\documentclass[12pt]{article}
\pdfoutput=1
\usepackage{titling}
\usepackage{amsmath}
\usepackage{slashed}
\usepackage{amssymb}
\usepackage{epsfig}
\usepackage{graphicx}
\usepackage{multirow}
\usepackage{color}
\usepackage[normal]{subfigure}
\usepackage{rotating}
\usepackage{hyperref}
\usepackage[margin=1.0in]{geometry}
\usepackage[table]{xcolor}
\usepackage{enumitem}
\usepackage[utf8x]{inputenc}
\usepackage[compress,numbers,sort]{natbib}
\usepackage{authblk}
\usepackage{colortbl}
\usepackage{pdflscape}
\usepackage{color}
\usepackage{hepnames}
\usepackage{numprint}
\usepackage{booktabs}
\usepackage{xparse}
\usepackage{mathrsfs}
\usepackage{multirow}
\usepackage{bbold}
\usepackage{siunitx}
\graphicspath{{../figures/}}
\definecolor{nicered}{rgb}{0.6,0.1,0.1}
\definecolor{nicegreen}{rgb}{0.1,0.5,0.1}
\definecolor{mediumcandyapplered}{rgb}{0.99, 0.12, 0.07}
\definecolor{red}{rgb}{1.0, 0, 0}
\hypersetup{colorlinks,citecolor= nicegreen,linkcolor= nicered}

\renewcommand{\bar}{\overline}

\newcommand{\SM}{\text{SM}}
\definecolor{myv}{HTML}{885B8D}
\definecolor{darkpastelgreen}{rgb}{0.01, 0.75, 0.24}

\definecolor{LightCyan}{rgb}{0.88,1,1}
\definecolor{piggypink}{rgb}{0.99, 0.87, 0.9}
\definecolor{applegreen}{rgb}{0.55, 0.71, 0.0}
\definecolor{green-yellow}{rgb}{0.68, 1.0, 0.18}
\newcommand{\GeV}{\si{\GeV}}
\newcommand{\TeV}{\si{\TeV}}

\newcommand{\femtobarn}{\si{\femto\barn}}

\newcommand{\published}[1]{%
\gdef\puB{#1}}
\newcommand{\puB}{}

\ExplSyntaxOn%
\newcommand\eff[2]
    { \fp_to_decimal:n { (#1)/(#2)) } }
\ExplSyntaxOff
\ExplSyntaxOn%
\newcommand\poiserr[2]
    { \fp_to_decimal:n { (#1)/(#2)* (1.0/(#1)+1.0/(#2))^0.5} }
\ExplSyntaxOff

\title{\bf{Probing Higgs couplings to light quarks via Higgs pair production}}

\author[1]
{Lina Alasfar\thanks{lina.alasfar@physik.hu-berlin.de}}
\author[2]
{Roberto Corral Lopez\thanks{robco23@correo.ugr.es}}
\author[1,3]
{Ramona Gr\"{o}ber\thanks{ramona.groeber@pd.infn.it}}

\affil[1]{\emph{\normalsize Humboldt-Universit\"at zu Berlin, Institut f\"ur Physik,
Newtonstr.~15, 12489 Berlin,  Germany.}}
\affil[2]{\emph{\normalsize CAFPE and Departamento de F\`isica T\'eorica y del Cosmos, Universidad de Granada, E18071 Granada, Spain.}}
\affil[3]{\emph{\normalsize Dipartimento di Fisica e Astronomia ``G. Galilei'', Universit\`a di Padova, Italy and Istituto Nazionale di Fisica Nucleare, Sezione di Padova, I-35131 Padova, Italy.}}

\date{}
\published{\flushright \vskip-0.5cm HU-EP-19/25}

\begin{document}

\maketitle
\begin{abstract}
\normalsize
We consider the potential of the Higgs boson pair production process to probe the light quark Yukawa couplings. 
We show within an effective theory description that the prospects of constraining enhanced first generation light quark Yukawa couplings in Higgs pair
production are similar to other methods and channels, due to a coupling of two Higgs bosons to two fermions. 
Higgs pair production can hence also probe if the Higgs sector couples non-linearly to the light quark generations. 
For the second generation, we show that by employing charm tagging
for the Higgs boson pair decaying to $c\bar{c}\gamma\gamma$, we can obtain similarly good prospects
for measuring the charm Yukawa coupling as in other direct probes.
\end{abstract}

\clearpage

%\tableofcontents

\clearpage

\section{Introduction}
After the Higgs boson discovery the era of precision measurements of Higgs properties has begun. While the Higgs boson couplings to vector bosons and third generation fermions have been measured at the LHC and agree with their Standard Model (SM) prediction at the level of 10\% -- 20\% \cite{Aad:2019mbh}, the situation for the Higgs self-couplings and couplings to first and second generation fermions is quite different.
Current bounds on the trilinear Higgs self-coupling  range from $-5.0< \lambda_{hhh}/\lambda_{hhh}^{SM}< 12.0 $  \cite{Aad:2019uzh} and are still above the limits of perturbative unitarity \cite{DiLuzio:2017tfn} or vacuum stability \cite{Falkowski:2019tft}. The quartic Higgs self-coupling is out of reach of the high-luminosity-LHC (HL-LHC) \cite{Plehn:2005nk, Binoth:2006ym}.
Upper limits on the Higgs boson decays to muons are $g_{h\bar{\mu}\mu}/g_{h\bar{\mu}\mu}^{SM}< 1.53$ \cite{Aad:2019mbh}, while current bounds on the Higgs coupling to electrons, $g_{h\bar{e}e}/g_{h\bar{e}e}^{SM}< 611$, are far away from the SM \cite{Khachatryan:2014aep}.
\par
For the Yukawa couplings to the first and second generation quarks, henceforth denoted as light quark Yukawa couplings, the current best limits are obtained from a global fit to Higgs data \cite{Kagan:2014ila, Perez:2015aoa}.  For instance for the HL-LHC, ref.~\cite{deBlas:2019rxi} obtained for a projection on the coupling strength modification, $\kappa_i=g_{h\bar{q}_iq_i}/g_{h\bar{q}_iq_i}^{SM}$, where $g_{h\bar{q}_iq_i}$ denotes the $i=u,d,s,c$ Higgs Yukawa coupling to quarks, in a global fit
\begin{equation}
|\kappa_u|< 570\,, \hspace*{0.5cm} |\kappa_d|< 270\,, \hspace*{0.5cm}|\kappa_s|< 13\,, \hspace*{0.5cm} |\kappa_c |< 1.2\,. \label{eq:fitbounds1}
\end{equation}
The determination of the light quark Yukawa couplings in a global fit is plagued by the fact  that the Higgs boson width can only be measured at the LHC under certain assumptions.\footnote{The width determination due to on- and off-shell measurements  \cite{Aaboud:2018puo, Sirunyan:2019twz} of Higgs boson couplings \cite{Caola:2013yja} is for instance made under the assumption that the couplings do not depend on the energy scale \cite{Englert:2014aca}.} The global fit can therefore not be considered to be completely model-independent. A more direct way of constraining the light Yukawa couplings is hence welcome.
\par
Searches for exclusive decays of the Higgs boson to a vector meson and a photon $h\to X\gamma $ with $X=\rho, \omega, \phi, J/\psi $, \footnote{In addition, $h\to \Upsilon
\gamma$ allows to probe the bottom Yukawa coupling \cite{Koenig:2015pha}.}
as a probe of light Yukawa couplings have been proposed in \cite{Bodwin:2013gca}
and can be even used to probe flavour-off-diagonal Yukawa couplings \cite{Kagan:2014ila}
for instance in Higgs boson rare decays such as $h\to M W^{\pm} $ or  $h\to M Z $, with $M$
denoting generically a scalar or pseudoscalar vector meson. From the experimental side,
ATLAS and CMS have reported upper bounds on the decays $h\to \rho \gamma$, $h\to \phi \gamma$ in \cite{Aaboud:2017xnb} and to $h \to J/\psi \gamma $ in \cite{Aaboud:2018txb, Sirunyan:2018fmm}.
The charm Yukawa coupling can also be constrained to a factor of a few times its SM value at the HL-LHC making use of charm tagging in $pp\to W/Zh $ with subsequent decay of the Higgs boson to $c \bar{c}$ \cite{Perez:2015lra} (see \cite{ATL-PHYS-PUB-2018-016, CMS-PAS-HIG-18-031} for first experimental results) or in $pp\to hc$ \cite{Brivio:2015fxa}.
\par
Another possibility for constraining the light quark Yukawa couplings is from Higgs kinematics.  If the Higgs boson is produced with an associated jet, the transverse momentum distribution changes with respect to the SM one in the presence of enhanced quark Yukawa couplings of the second and first generation. For the second generation quarks, the main effect stems from log-enhanced contributions due to interference between top and light quark loop diagrams. This allows to set a bound on $\kappa_c \in [-0.6, 3.0]$  at 95\% C.L. at the HL-LHC \cite{Bishara:2016jga}. Instead in the presence of significantly enhanced first generation quark Yukawa couplings the Higgs boson can be directly produced from initial state quarks, which again would alter the Higgs $p_{T}$-distribution \cite{Soreq:2016rae}. For non-collider probes of the light Yukawa couplings see ref.~\cite{Delaunay:2016brc}.
\par
In this paper, we will study the potential to constrain light quark Yukawa couplings from Higgs pair production. 
We note that though a measurement of the SM Higgs pair production process is quite challenging due to
the small signal cross section and large backgrounds, the prospects for several di-Higgs final states are quite 
promising, in particular for the $b\bar{b}\gamma\gamma$ \cite{Baur:2003gp, Baglio:2012np, Azatov:2015oxa, Kling:2016lay,
Barger:2013jfa, Lu:2015jza,  Adhikary:2017jtu, Alves:2017ued, Chang:2018uwu}, $b\bar{b}\tau^+\tau^-$ \cite{Dolan:2012rv, Baglio:2012np, Goertz:2014qta, Adhikary:2017jtu} and 
$b\bar{b}b\bar{b}$ \cite{Dolan:2012rv, deLima:2014dta, Behr:2015oqq, Wardrope:2014kya} final states. Experimental studies find that at the HL-LHC a 95\% C.L. bound of 
$\sigma/\sigma_{SM}< 1.1$ can be set on the Higgs pair production cross section \cite{ATL-PHYS-PUB-2018-053, CMS:2018ccd}.
\par 
As for Higgs plus jet production, we can make use of kinematical information in Higgs pair production. We will mainly consider the case in which the modifications of the light Yukawa couplings can be described by a dimension six effective operator, denoted schematically by
\begin{equation}
\mathcal{O}_f=(\phi^{\dagger}\phi) (\bar{Q}_L \phi q_R)\,. \label{eq:effop}
\end{equation}
The left-handed quark $SU(2)$ doublet has been denoted by $Q_L$,the right-handed quark $SU(2)$ singlet by $q_R$, while $\phi$ is the scalar Higgs doublet field. In the presence of such an operator, both a shift in the Yukawa coupling to one Higgs boson as well as a new coupling of two Higgs bosons to two fermions modifies the Higgs pair production cross section. In the case of the top quark it was shown that such a new coupling can lead to large enhancements of the double Higgs production process \cite{Dib:2005re, Grober:2010yv, Contino:2012xk, Grober:2016wmf}. For the light quark Yukawa couplings this was shown in \cite{Bar-Shalom:2018rjs} under the assumption of universally enhanced light Yukawa couplings.
We will consider more general scenarios and will show that indeed such an operator can also be constrained in di-Higgs production for the light generations of quarks. Under the assumption of linearly realised electroweak symmetry breaking we can then obtain a bound on the light quark Yukawa couplings which is competitive with the above mentioned ways of constraining them. We will also investigate how our bounds are modified if we allow for a modification of the trilinear Higgs self-coupling. Furthermore, we will discuss the possibility of charm tagging for di-Higgs final states, which will allow us to set bounds on the second generation quark Yukawa couplings.
\par
The paper is structured as follows: in sect.~\ref{sec:EFT} we will introduce our notation and point out under which circumstances scenarios considered in our analysis can be realised. In sect.~\ref{sec:HH} we present how the di-Higgs production process and the Higgs boson decays are modified in the presence of enhanced light quark Yukawa couplings. In sect.~\ref{sec:pheno} we present the results of our analysis both in the presence of enhanced first and second generation Yukawa couplings. We also consider the potential reach of the HL-LHC by employing charm tagging.  We conclude in sect.~\ref{sec:conclusion}.
%%%%%%%%%%%%%%%%%%%%%%%%%%%%%%%%%%%%%%%%%%%%%%%
\section{Effective Field Theory of light Yukawa couplings\label{sec:EFT}}
Within the SM, the Higgs couplings to quarks are described by the Lagrangian
\begin{equation}
\mathcal{L}_{y}=-y^u_{ij} \bar{Q}_L^i \tilde{\phi} u_R^j - y^d_{ij} \bar{Q}_L^i \phi d_R^j +h.c.\,,
\end{equation}
with $\tilde{\phi}=i \sigma_2 \phi^*$, $\sigma_2$ is the second Pauli matrix, $\phi$ denotes the Higgs doublet, $Q_L^i$ the left-handed $SU(2)$ quark doublet of the $i$-th generation and $u_R^j$ and $d_R^j$ the right-handed up- and down-type fields of the $j$-th generation, respectively.
Modifications of the SM from high-scale new physics can be described in a model-independent way by means of the SM effective field theory (SMEFT), in terms of higher dimensional operators.  In particular, the couplings of the quarks to the fermions are modified by the operator
\begin{equation}
\Delta \mathcal{L}_{y}=\frac{\phi^{\dagger}\phi}{\Lambda^2}\left( c^u_{ij} \bar{Q}_L^i \tilde{\phi} u_R^j + c^d_{ij} \bar{Q}_L^i \phi d_R^j +h.c.\right)\,,
\label{eq:EFTop}
\end{equation}
where $\Lambda$ denotes the cut-off of the effective field theory (EFT).  The mass matrices of the up-type and down-type quarks are
\begin{align}
M^u_{ij} =& \frac{v}{\sqrt{2}} \left( y^u_{ij}-\frac{1}{2} c^u_{ij}\frac{v^2}{\Lambda^2}\right)\,,\\
M^d_{ij} =& \frac{v}{\sqrt{2}} \left( y^d_{ij}-\frac{1}{2} c^d_{ij}\frac{v^2}{\Lambda^2}\right)\,.
\end{align}
They can be diagonalised by means of a bi-unitary transformation
\begin{equation}
m_{q_i}=\left((V_{L}^{u/d})^{\dagger} M^{u/d} V_R^{u/d}\right)_{ii}\,,
\label{eq:Diagonalize}
\end{equation}
while the CKM matrix is defined as $V_{CKM}=(V_L^u)^{\dagger} V_L^d$.
By defining
\begin{equation}
\tilde{c}^{q}_{ij}= \left(V_{L}^{q}\right)^*_{ni}c_{nm}^{q}\left(V_R^{q}\right)_{mj}\, , \; \; \; \;  \text{with } \;\;\;\;\; q = u,d\,,
\end{equation}
we can write the couplings of one and two Higgs boson to fermions with 
\begin{equation}
\mathcal{L}\supset g_{h\bar{q}_i q_j}\bar{q}_i q_j h + g_{hh\bar{q}_i q_j}\bar{q}_i q_j h^2
\end{equation} 
as
\begin{equation}
g_{h\bar{q}_i q_j} : \quad \frac{m_{q_i}}{v}\delta_{ij}-\frac{v^2}{\Lambda^2} \frac{\tilde{c}^q_{ij}}{\sqrt{2}}\,, \quad \quad \quad \quad \quad g_{h h\bar{q}_i q_j} : \quad -\frac{3}{2\sqrt{2}}\frac{v}{\Lambda^2}\tilde{c}^q_{ij}\,. \label{eq:couplingsEFT}
\end{equation}
 In the following, we will also use for the diagonal couplings alternatively  the notation
\begin{equation}
g_{h\bar{q}_i q_i} =\kappa_q g_{h\bar{q}_i q_i}^{\text{SM}} \,, \quad \quad \quad \quad \quad g_{h h\bar{q}_i q_i}= - \frac{3}{2}\frac{1-\kappa_q}{v}g_{h\bar{q}_i q_i}^{\text{SM}} \,,
\label{eq:def_kappa}
\end{equation}
in a slight abuse of language of the $\kappa$-framework used often in experimental analyses.
\par
Flavour-changing Higgs couplings are strongly constrained  from low-energy flavour observables, such as meson-antimeson mixing. The bounds are of order $|\tilde{c}_{uc/ds}| \lesssim 10^{-5} \Lambda^2/v^2$ and $|\tilde{c}_{db/sb}| \lesssim 10^{-4} \Lambda^2/v^2$ \cite{Blankenburg:2012ex}. Given that, a common assumption for the Wilson coefficients in eq. \eqref{eq:EFTop} is that of minimal flavour violation (MFV) \cite{DAmbrosio:2002vsn}, where
\begin{equation}
c^u_{ij} = \bar{c}_u \, y^u_{ij}\,, \quad\quad\quad \quad c^d_{ij} = \bar{c}_d \,y^d_{ij}\,,
\end{equation}
with flavour universal $\bar{c}_u$ and $\bar{c}_d$. Hence, under the assumption of MFV the Yukawa matrices $y_u$ ($y_d$) and the Wilson coefficients  $c^u$ ($c^d$) are simultaneously diagonalisable and no flavour changing Higgs interactions with quarks exist.
We refrain though from making the assumption of MFV, due to the reason that with the Wilson coefficients being proportional to the Yukawa couplings, we introduce a strong hierarchy into the Higgs couplings to quarks. Since we want to describe modifications of the order of the ones in eq.~\eqref{eq:fitbounds1} we would need to assume very low values of the new physics scale $\Lambda$ and/or large Wilson coefficients, rendering the validity of the EFT questionable and in potentially conflict with measurements of the third generation couplings to the Higgs boson.
\par
Instead, we will consider the case in which the $\tilde{c}^q_{ij}$ are diagonal, though not proportional to the Yukawa matrices.
This can be realised by appropriate choice of the parameters. For instance, $V_{L/R}^{u}=\mathbb{1}\, , \; V_{R}^d=\mathbb{1}\,, \; \text{and} \; V^d_L=V_{CKM}$, which keeps $\tilde{c}^u$ flavour-diagonal if $c^u$ is chosen flavour-diagonal. Flavour violation then originates only from the CKM matrix. We will refer to this as flavour alignment. However, from a UV-perspective there is no obvious symmetry argument to enforce this at low-energy.
\par
A possible way of keeping $\tilde{c}$ flavour-diagonal with symmetry arguments could be realised for flavour universal $c^{u/d}$ and a left-right symmetry rendering $V_L=V_R$.
Then by setting universal $\tilde{c}^{u/d}/\Lambda^2\approx1/(3\text{ TeV})^2$ we get for instance a modification of the up-quark coupling to the Higgs boson of a factor of 500, but only a  modification of the top Yukawa coupling by 1\%, which is still consistent with the current limits on the top Yukawa coupling \cite{ATLAS-CONF-2018-031, CERN-EP-2019-119}. Note that doing so for the down-type quarks would of course be more difficult, as it would imply a larger deviation in the bottom quark Yukawa coupling, due to its smaller mass.
Alternatively, one can chose $\tilde{c}^f$ flavour-diagonal (or with strongly suppressed flavour-off-diagonal elements) by choosing horizontal symmetries. We refer to~\cite{Bar-Shalom:2018rjs} for a model with vector-like quarks and strongly enhanced light quark Yukawa couplings.
Another realisation of large first and second generation Yukawa couplings without tree-level flavour-changing neutral currents (FCNCs) has been discussed in~\cite{Egana-Ugrinovic:2018znw}, and is referred to as spontaneous flavour violation.  The basic idea is to achieve this by breaking the quark family number symmetry via the RH up-type or down-type quark wave function renormalisation, leading to either enhanced up- or down-type quark Yukawa couplings. A concrete realisation of this idea for a two-Higgs doublet model was discussed in~\cite {Egana-Ugrinovic:2019dqu}.
\par
\begin{figure}[!t]
\centering
  \includegraphics[width = 0.25\textwidth]{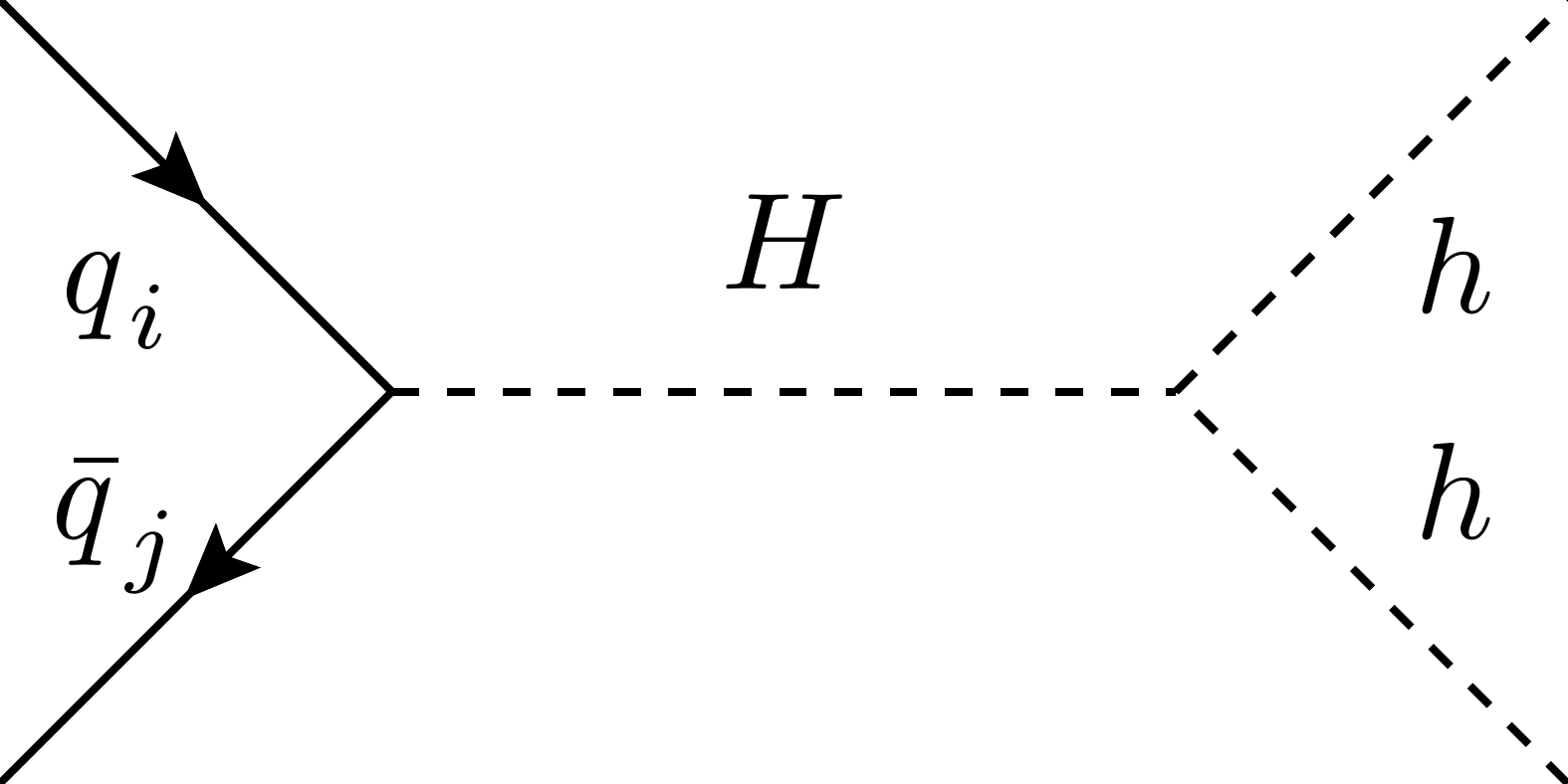}
   \hspace{0.5 cm}
  \includegraphics[width = 0.2\textwidth]{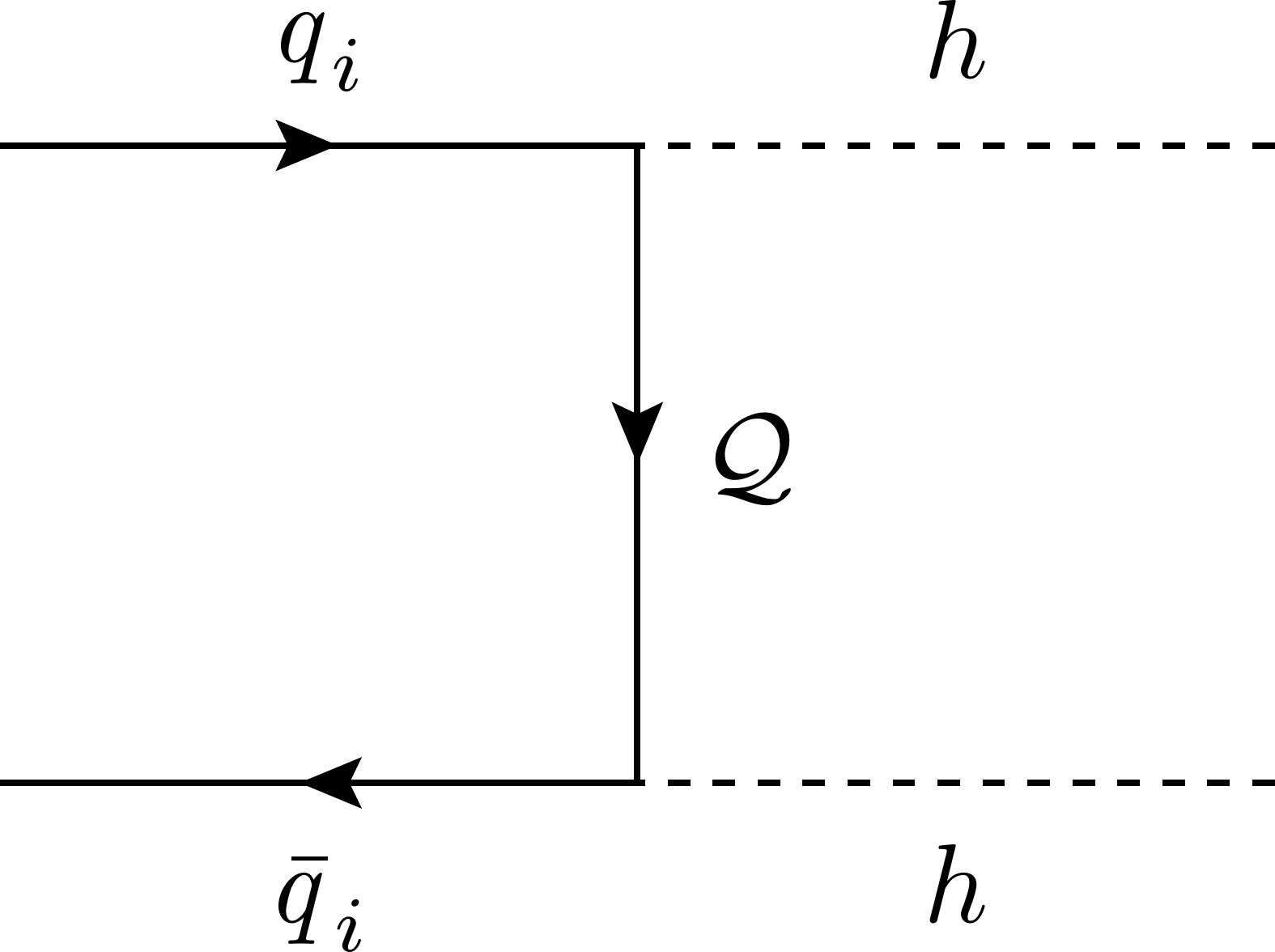}
  \caption{Examples of potential concrete models leading to a  $hh  \bar{q}  q$ coupling. The left Feynman diagram shows a heavy Higgs $H$, the right diagram a vector-like quark $\mathcal Q $.} % verify the notation
  \label{fig_uv_qqhh}
\end{figure}

We would also like to stress that from a UV perspective it makes sense to assume that if there is a modification in the light quark Yukawa couplings with respect to the SM, deviations in the di-Higgs production process can be expected, which in the limit of heavy new physics can be traced back to a coupling of two Higgs boson to two fermions. We show this schematically in fig.~\ref{fig_uv_qqhh} for a heavy new scalar and a heavy new vector-like fermion. The coupling of the SM-like Higgs boson in the models extended by a heavy new Higgs boson or a heavy new vector-like quark as shown in fig.~\ref{fig_uv_qqhh}  is modified due to a mixing with either the new Higgs boson, if it acquires a vacuum expectation value (VEV), or by the mixing between the quark and the new vector-like fermion.
For the case of the heavy new scalar, the effective coupling of two SM-like Higgs bosons to fermions in the limit of $m_H \gg E$, with $E$ denoting the energy scale of the process and $m_H$ the Higgs mass of the heavy Higgs boson,
can be written as
\begin{equation}
g_{hh\bar{q}q} \to -i \frac{g_{H\bar{q}q} g_{Hhh}}{m_H^2}\,.
\end{equation}
A coupling $g_{Hhh}$ always exist, if both of the Higgs fields acquire a VEV, since a portal term in the Lagrangian, $(\phi^{\dagger} \phi) (\Phi^{\dagger}\Phi)$, is always allowed  by the symmetries. We denoted here the new Higgs multiplet by $\Phi$ with neutral component $H$.
\par
In the presence of new vector-like quarks that mix with the SM quarks, the coupling of two Higgs bosons to two fermions comes from $\hat{t}/\hat{u}$ channel diagrams. If the mass of the new vector-like quark $m_Q$ is $m_Q\gg E$ one obtains for the coupling\footnote{In Composite Higgs Models with vector-like quarks there is also a contribution from the non-linearities of the model.}
\begin{equation}
g_{hh\bar{q}q} \to -i \frac{g_{h\bar{q}\mathcal Q}g_{h\bar{\mathcal Q}q}}{m_\mathcal Q}\,.
\end{equation}
A more explicit consideration of models that realise large light Yukawa couplings is beyond the scope of this paper and we refer to existing work~\cite{Bar-Shalom:2018rjs, Egana-Ugrinovic:2019dqu}.

\par
We finally note that an alternative way of describing model-independent deviations from the SM Higgs couplings is by a non-linear effective Lagrangian (alternatively referred to as electroweak chiral Lagrangian) \cite{Coleman:1969sm, Callan:1969sn}.  While in SMEFT the Higgs boson is assumed to be part of an $SU(2)$ doublet and the expansion is
organised in terms of dimensionality of the operator, in the chiral Lagrangian the Higgs boson is assumed to be a singlet and the expansion is organised in terms
of chiral dimension, where bosonic fields are assigned chiral dimension 0 and derivatives and fermion bi-linears chiral dimension 1.
The Lagrangian responsible for a potential modification of the Yukawa couplings can be written as \cite{Contino:2010mh}
\begin{equation}
\mathcal{L}=-\frac{v}{\sqrt{2}}(\bar{u}_L^i, \bar{d}_L^i)\Sigma \left( y_{q,ij}+ k_{q,ij} \frac{h}{v}+k_{2q,ij}\frac{h^2}{v^2}+...\right)\left(\begin{array}{c} u_R^j \\ d_R^j\end{array}\right)
\label{chirallag}
\end{equation}
with
\begin{equation}
\Sigma=e^{i\sigma^a \pi^a(x)/v}\,,
\end{equation}
in terms of the Pauli matrices $\sigma^a$ and the Goldstone bosons $\pi^a$ with $a=1,2,3$. The field $\Sigma$ transforms linearly under the custodial symmetry $SU(2)_L\times SU(2)_R$. We note again as for the SMEFT that off-diagonal elements of $k_{q}$ are strongly constrained. Compared to SMEFT the couplings of one or two Higgs boson to fermions are now uncorrelated, leading to different coefficients $k_{q}$ and $k_{2q}$.
In principle, the coefficients of the light fermion couplings to two Higgs bosons are yet unconstrained and di-Higgs production is \textit{the} place to test if there exists a correlation among those and hence whether a linear or non-linear EFT prescription is to be preferred.
While in the following we will mainly concentrate on the case of SMEFT we shall shortly comment also on the case of non-linear EFT.
%%%%%%%%%%%%%%%%%%%%%%%%%%%%%%%%%%%%%%%%%%%%%%%%%%
\section{Higgs pair production and Higgs decays with modified light Yukawa couplings \label{sec:HH}}
In this section we will describe how the Higgs pair production process for modified light quark Yukawa couplings is affected.
While in the SM Higgs pair production is dominantly mediated by gluons fusing into a heavy quark loop coupling to the Higgs boson, for large first and second generation quark Yukawa couplings
also quark annihilation becomes relevant.
For a phenomenological analysis we also need to take into account the Higgs boson decays, which we describe in the last part of the section.
\subsection{Higgs pair production via gluon fusion \label{sec:ggF}}
\begin{figure}[!t]
\centering
  \includegraphics[width = 0.35\textwidth]{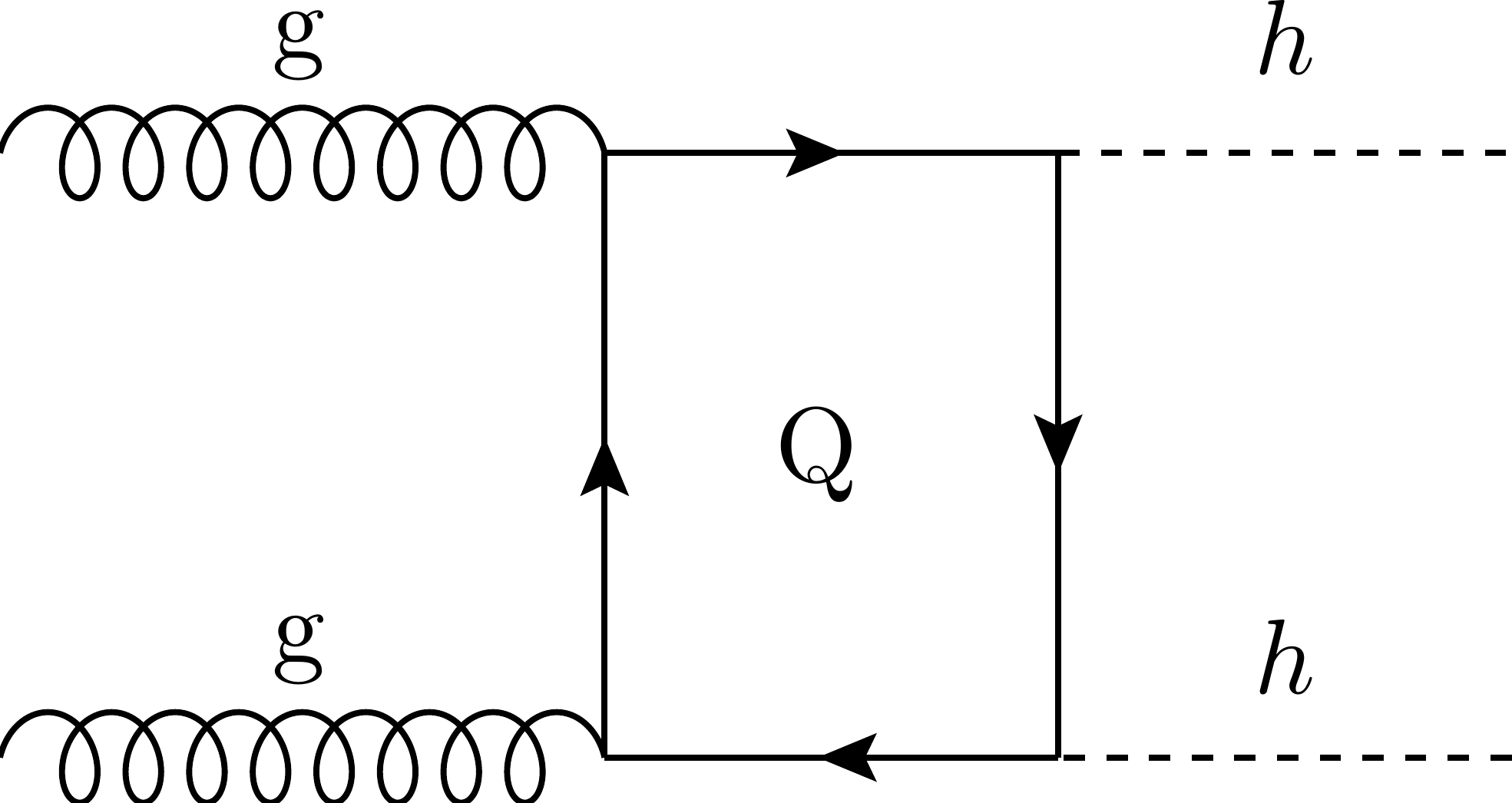}
 \hspace{0.3 cm}
  \includegraphics[width = 0.4\textwidth]{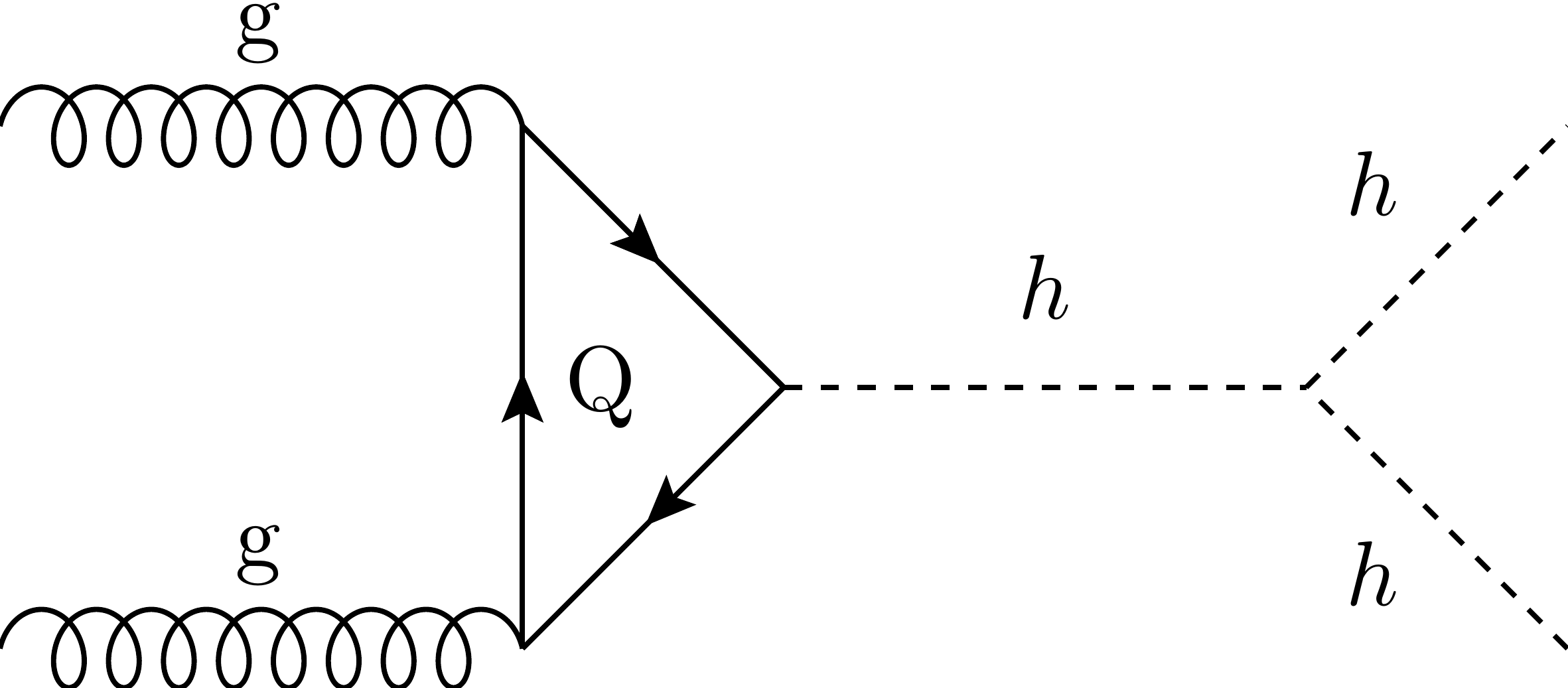}
  \caption{Feynman diagrams for the ggF process of Higgs pair production in the SM.}
  \label{fig_ggf_sm}
\end{figure}

The dominant process for Higgs pair production at the LHC in the SM is the gluon fusion process~(ggF) via a heavy quark loop~$Q$, where $Q$ stands mainly  for the top quark. The bottom quark contributes with less than $1\%$. We show the Feynman diagrams for the process in  fig.~\ref{fig_ggf_sm}.
 The process has been known since long at leading order~(LO) in full mass dependence \cite{EBOLI1987269,GLOVER1988282,DICUS1988457,Plehn:1996wb}. The next-to-leading order~(NLO) in the strong coupling constant was initially computed using the infinite top mass limit~($m_t \to \infty$) and reweighted with the full LO results \cite{Dawson:1998py}. However, this approximation is only valid up to the top quark threshold. More recently, the NLO QCD corrections have been computed in full top mass dependence, showing that the infinite top mass limit overestimates the full result by 14\%~\cite{Borowka:2016ypz,Borowka:2016ehy,Baglio:2018lrj}.\footnote{ The numerical NLO QCD results for the virtual corrections were cross-checked by employing different analytic expansions \cite{Bonciani:2018omm, Grober:2017uho, Davies:2018qvx}.} For distributions, the approximation of infinite top mass is even worse.  At next-to-next-to leading order (NNLO) results are available in the infinite top mass limit \cite{Grigo:2014jma, deFlorian:2013jea} and by including top mass effects for the double real radiation \cite{Grazzini:2018bsd}. First steps towards an inclusion of top mass effects for the virtual corrections (for the triangle only) have been made in~\cite{Davies:2019nhm, Davies:2019djw} and for the light fermion triangle contributions the NNLO has been computed in \cite{Harlander:2019ioe}.
 \par
\begin{figure}[!hb]
\centering
  \includegraphics[width = 0.25\textwidth]{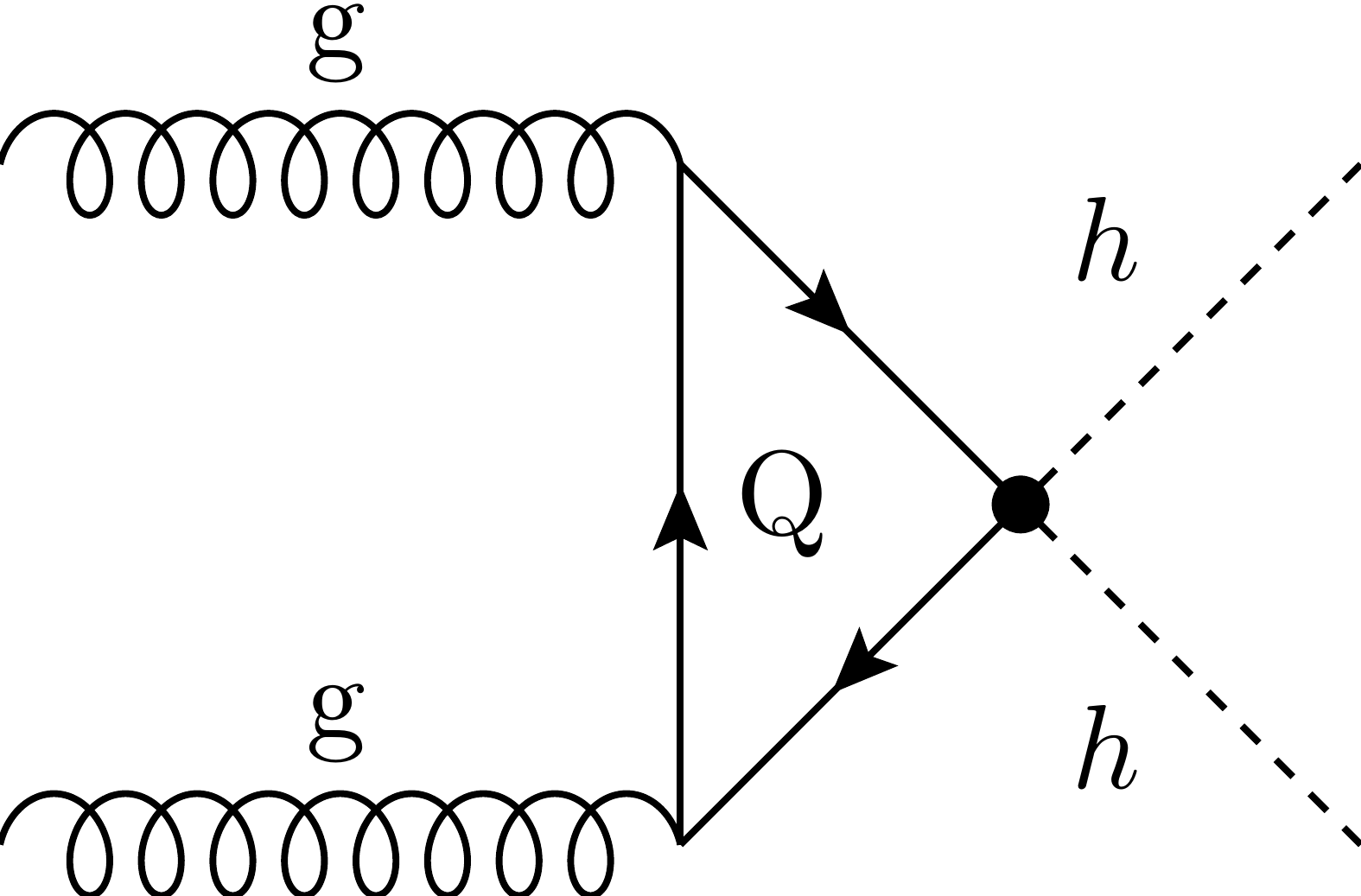}
  \caption{The new diagram for ggF emerging from the $hh q \bar q$ coupling stemming from an effective dim-6 operator.}
  \label{fig_ggf_diag}
\end{figure}

{}
For our analysis, we have calculated the $ \sqrt{s} = 14$ \text{TeV}\xspace LO ggF inclusive cross section and distributions with modified light Yukawa couplings by including the light quark loops and the coupling $hh q \bar q$ shown in fig.~\ref{fig_ggf_diag}.
 The calculation was carried out using a private FORTRAN implementation of the LO cross section utilising the \texttt{VEGAS} integration algorithm, and NNPDF30 parton distribution functions~(PDF's)\cite{Ball:2017nwa} implemented via the \texttt{LHAPDF-6} package \cite{Buckley:2014ana}. For the one-loop integrals appearing in the form factors of the box and triangle diagrams, we have used the \textsc{Collier} library~\cite{Denner:2014gla} to ensure numerical stability of the loop integral calculation for massless quarks inside the loops.
   A $K$-factor for the NNLO correction was used following the recommendations by the Higgs cross section working group~\cite{deFlorian:2016spz}
\begin{equation}
  K = \frac{\sigma_{NNLO}}{\sigma_{LO}}, \;\;\;\;\; K_{14\, \mathrm{ TeV}} = 1.72.
\end{equation}
For differential distributions in the invariant mass of the Higgs boson pair, $M_{hh}$, we extract a differential $K$-factor from \cite{Grazzini:2018bsd}.
As a reference cross section at NNLO \cite{Grazzini:2018bsd}  for the analysis in sect.~\ref{sec:pheno} we use
\begin{equation}
\sigma^{\text{SM}}_{NNLO}=36.69^{+1.99}_{-2.57} \text{ fb}\,.
\end{equation}
The uncertainty stems from the scale choice, the PDF+$\alpha_s$ error and the uncertainty associated to the usage of the infinite top mass limit in parts of the calculation.
Since we found that the cross section does not change much once the effects of the modified light Yukawa couplings are included, we use the same NNLO $K$-factor for all values of the scalings.
The renormalisation, $\mu_R$, and factorisation scales, ~$ \mu_F$, are set to $\mu_0 =M_{hh}/2$ as has been pointed out as an optimal choice in ref.~\cite{deFlorian:2015moa}, and $\alpha_s(M_Z) = 0.118$.
\subsubsection{Results}
For comparison of the results with modified Yukawa couplings with the SM results, we define as a benchmark point the case where all first and second generation quark Yukawa couplings are scaled to the SM bottom Yukawa coupling, which we will refer to in plots and tables as $g_{hq \bar q} = g_{h b \bar b}^{\SM}$. This means we scale the Yukawa couplings by $\kappa_q=g_{h\bar{q}q}/g_{h\bar q q}^{\SM}$ with
\begin{equation}
\kappa_u=  1879\,, \hspace*{0.5cm} \kappa_d= 889 \,, \hspace*{0.5cm}\kappa_s= 44\,, \hspace*{0.5cm} \kappa_c =3.3\,, \label{eq:fitbounds}
\end{equation}
and use only flavour-diagonal modifications of the quark Yukawa couplings.
This benchmark is inspired by ref.~\cite{Bar-Shalom:2018rjs}.
\par
\begin{figure}[!t]
\centering
\begin{picture}(200,200)
\put(-140,0){\includegraphics[scale =0.28]{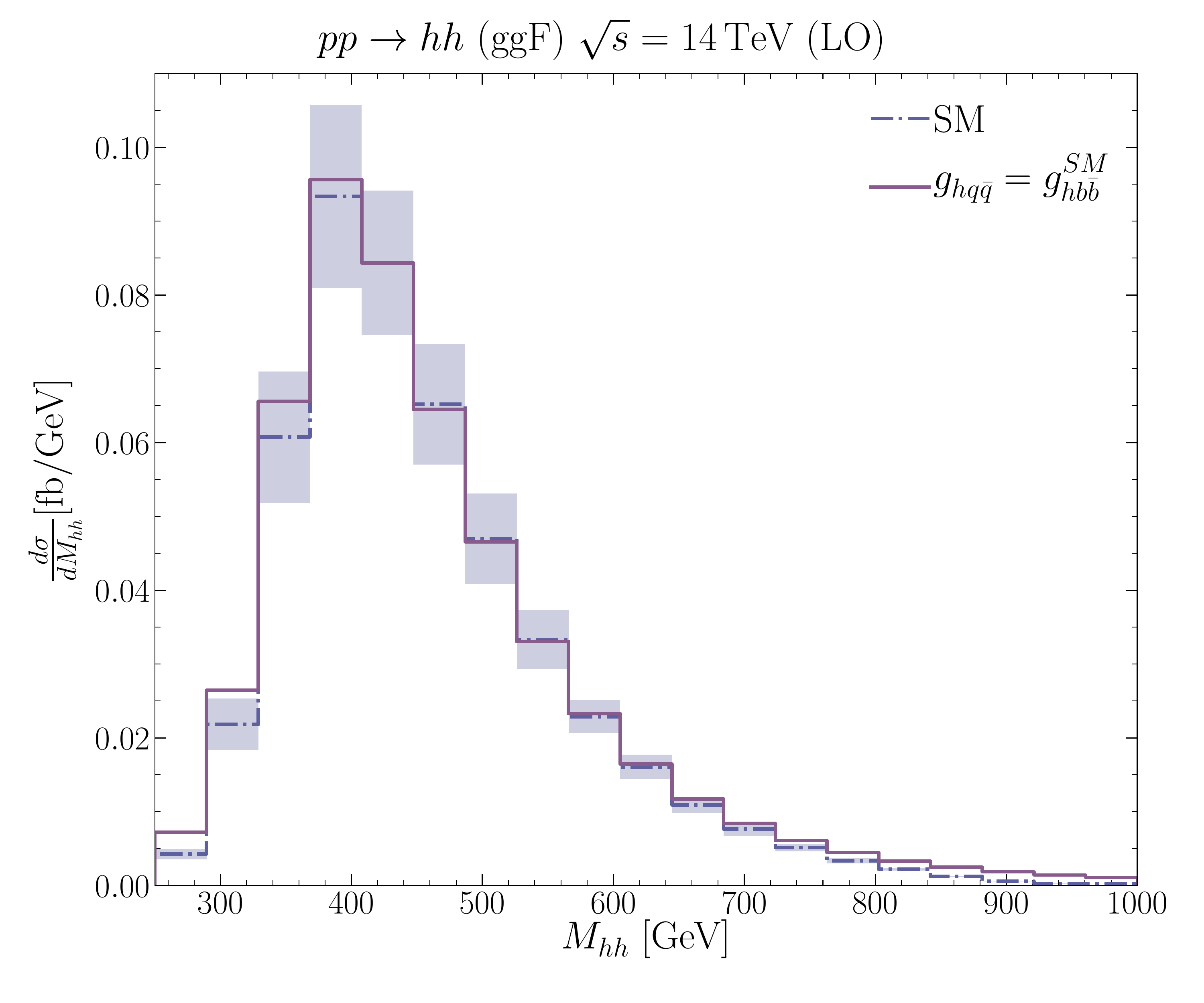}}
\put(100,0){\includegraphics[scale = 0.28]{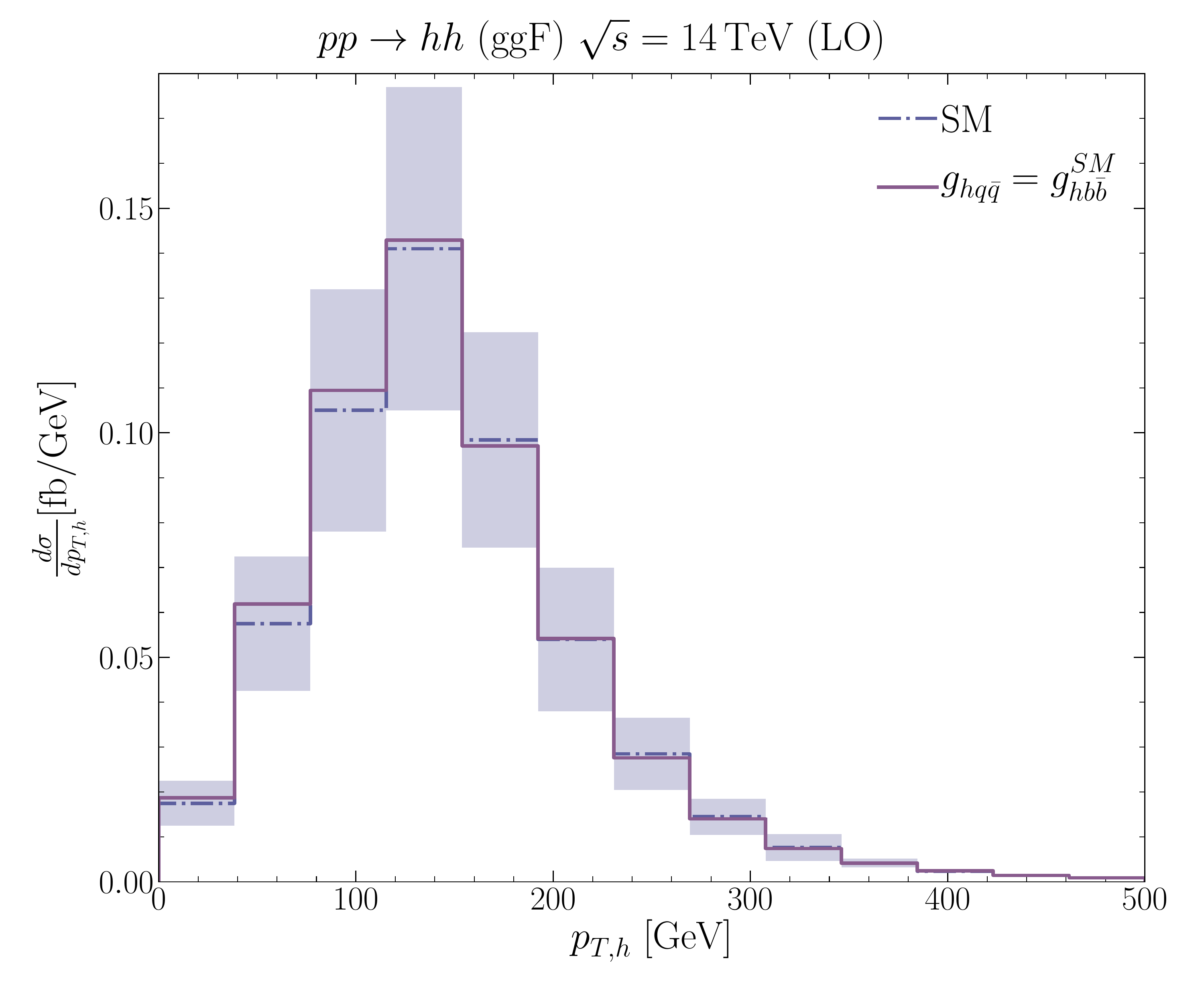}}
\end{picture}
  \caption{\textit{Left:} The di-Higgs invariant mass differential cross section $d \sigma/dM_{hh}$ for the SM at LO and the benchmark point toy. The error boxes denote the total scale, PDF and $\alpha_s$ uncertainties. \textit{Right:} The same but for the Higgs transverse momentum $p_{T,h}$ distribution. }
 \label{dsig_dmhh_ggf}
\end{figure}
Figure~\ref{dsig_dmhh_ggf} shows the di-Higgs invariant mass~$M_{hh}$- and the~$p_{T,h}$-distributions for the computed LO process.
From the distributions it is evident, that the change of the ggF process in the presence of enhanced light Yukawa couplings is quite small. The reason is that the box contribution which is the major part of the cross section has two fermion coupling insertions and hence is strongly suppressed for all the light quarks with respect to the top quark loop diagrams. The bottom quark contribution to the ggF process in the SM is less than 1\% and comes mainly from the triangle diagram, so adding several contributions from similar size does not change the cross section by much. Also the new diagrams (\textit{cf.}~fig.~\ref{fig_ggf_diag}) are suppressed compared to the box diagrams of the top quark.
In the presence of enhanced light quark Yukawa couplings the Higgs boson pair can though be directly produced by quark annihilation. We turn to discuss this process in the next part. In the meanwhile we can conclude that for the ggF process we can improve on the LO predictions by using SM $K$-factors and that the effects of light Yukawa coupling modifications for the ggF process are small for the still allowed modifications.
%%%%%%%%%%%%%%%%%%%%%%%%%%%%%%%
\subsection{Higgs pair production via quark anti-quark annihilation}
\begin{figure}[!tb]
\centering
\begin{picture}(180,200)
\put(-120,120){\includegraphics[scale =0.25]{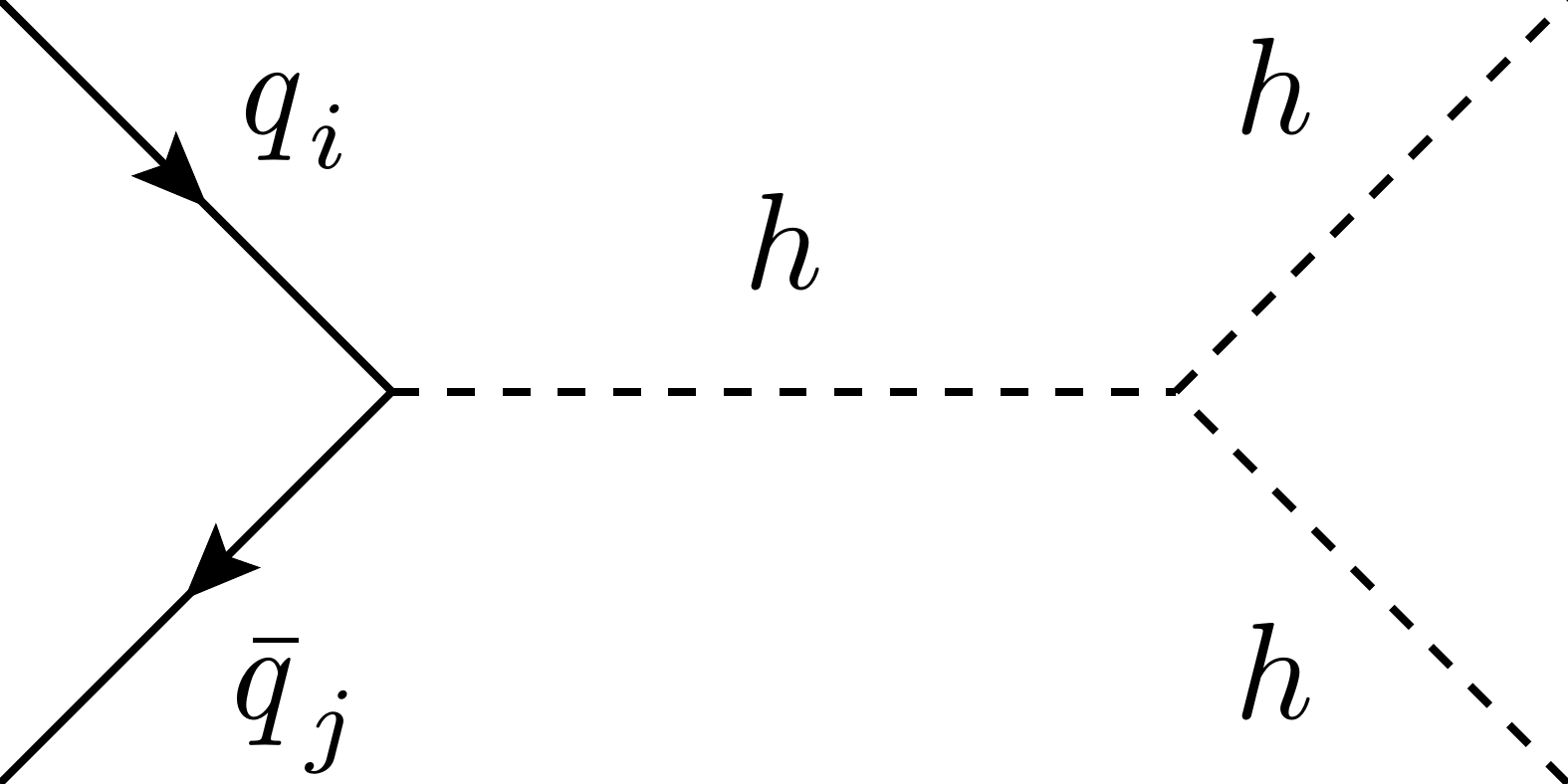}}
\put(20,120){\includegraphics[scale = 0.25]{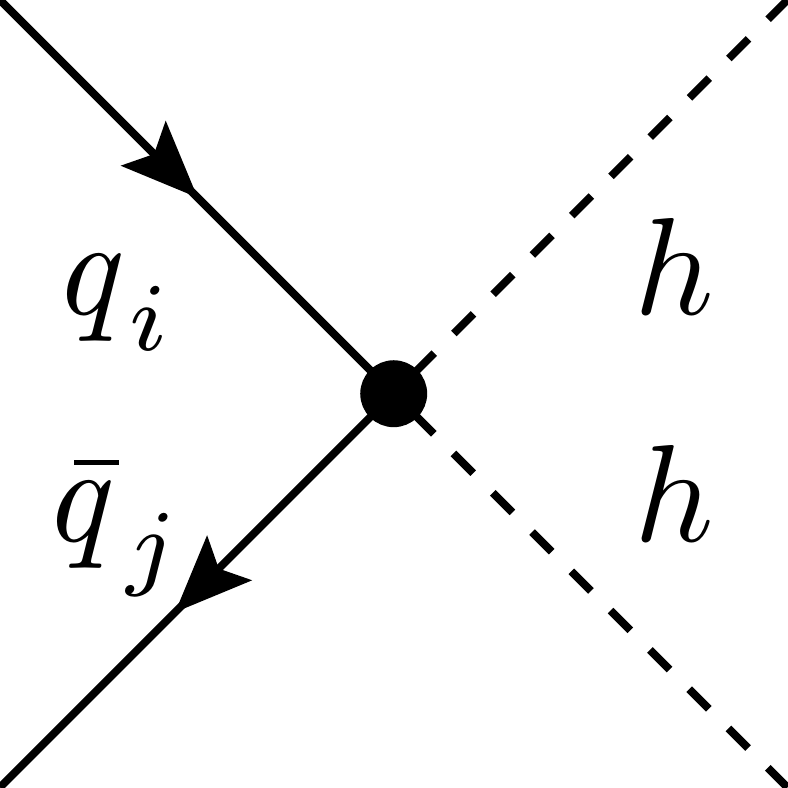}}
\put(110,110){\includegraphics[scale =0.21]{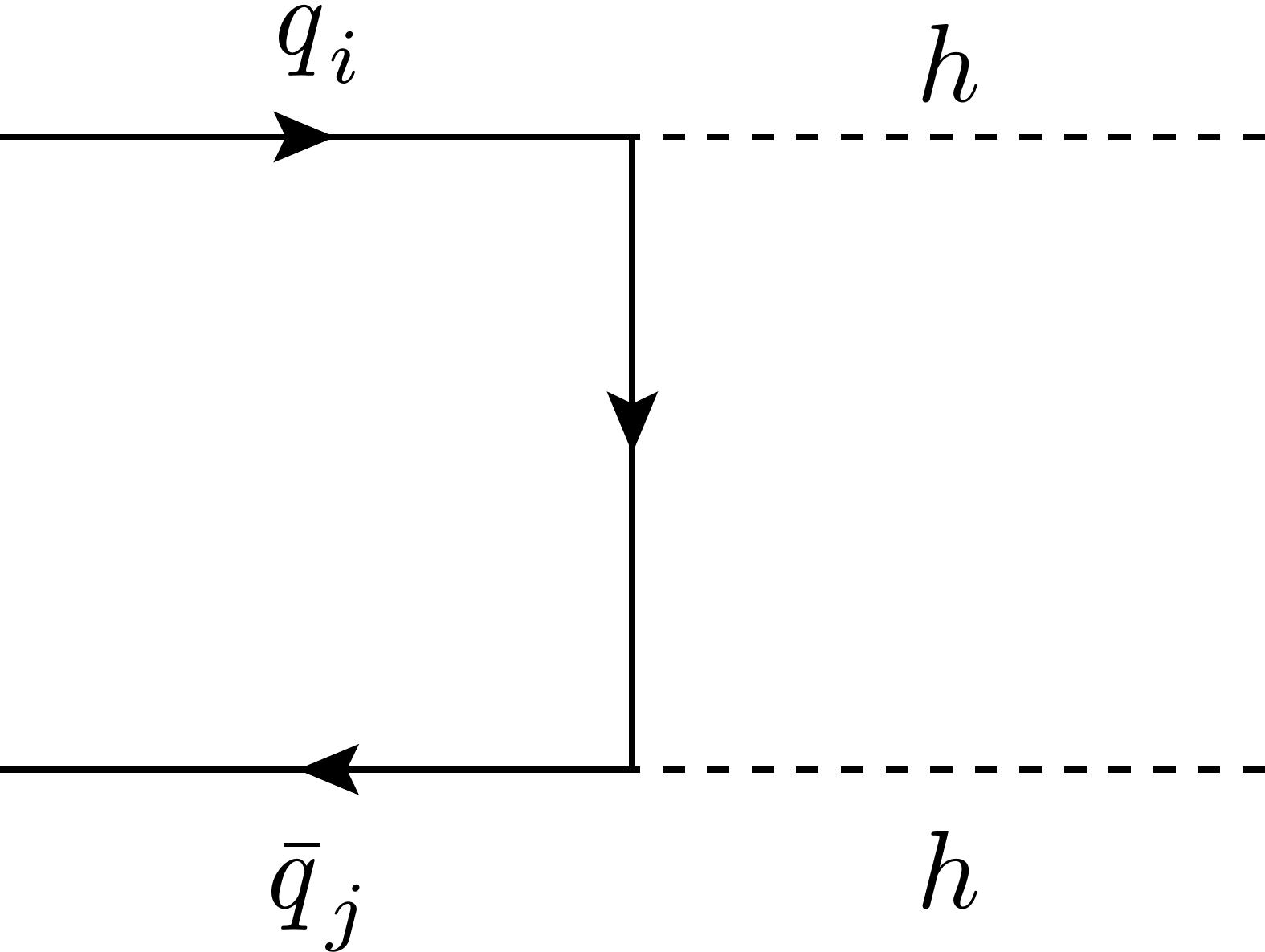}}
\put(230,145 ){{\large+ crossed} }
\end{picture}
\vspace*{-4cm}
  \caption{ Feynman diagrams for the qqA Higgs pair production.}
  \label{qqA_fd}
\end{figure}

If the Yukawa couplings of the light quark generations are sufficiently increased, the Higgs bosons will be produced directly from the constituents of the proton with a sizeable rate. The quark anti-quark annihilation (qqA) process becomes then relevant for Higgs pair production.
The qqA process has four Feynman diagrams shown in the fig.~\ref{qqA_fd}.

The differential cross section given by
\begin{align}
\frac{d \hat \sigma_{q_i\bar{q}_j}}{d \hat t} &= \frac{1}{16 \pi}\, \frac{1}{12  \hat{s}} \bigg[ \left| 2  g_{hh q_i \bar q_j} + \frac{g_{hhh}\, g_{h q_i \bar q_j}}{\hat{s}-m_h^2-im_h\Gamma_h}\right|^2+ \mathcal{O}(g_{h q_i \bar q_j}^4) \bigg].
\label{sigmaqqa}
\end{align}

We neglect here the $\hat{t}$ and $\hat{u}$ channel diagrams, as their contribution is $\sim 0.1 \%$ of the total cross section, as they are suppressed by $g_{h q_i \bar q_j}^4$ and only interfere with each other.

The hadronic cross section is then obtained by
\begin{equation}
  \sigma_{\mathrm{hadronic}} =  \int_{\tau_0}^1 d\tau \int_{\hat{t}_-}^{\hat{t}_+} d\hat{t} \sum_{i,j} \frac{d\mathcal{L}^{q_i\bar{q}_j}}{d\tau}\frac{ d\hat \sigma_{q_i\bar{q}_j}}{d \hat t}\,, \label{eq:sigmahadron}
\end{equation}
with $ \tau_0= 4\, m_h^2/s$, $\hat{s}=\tau s$ and
\begin{equation}
\hat{t}_{\pm}=m_h^2-\frac{\hat{s}(1\mp \beta)}{2} \quad\quad \text{and}\quad \quad \beta=\sqrt{1-\frac{4 m_h^2}{\hat{s}}}\,.
\end{equation}
The parton luminosity is given by
\begin{equation}
  \frac{d{\cal L}^{q_i \bar q_j}}{d\tau} = \int_\tau ^1 \frac{dx}{x} \,\left[  f_{q_i}(x/\tau,\mu_F^2) f_{\bar{q}_j}(x,\mu_F^2) + \,f_{\bar{q}_j}(x/\tau,\mu_F^2) f_{q_i}(x,\mu_F^2)\right]\,.
\end{equation}
We neglected all the kinematical masses in accordance with the 5-flavour scheme of the PDFs while the coupling of the Higgs boson to the light quarks (for flavour diagonal couplings) is
\begin{equation}
g_{hq_i\bar{q}_j}=\frac{m^{\bar{MS}}_q(\mu_R)}{v}  \kappa_q \delta_{ij}\,,
\end{equation}
and analogously for the $g_{hhq_i\bar{q}_j}$ coupling.\footnote{We note that there is no inconsistency with such an assumption since in scenarios of modified Yukawa couplings, the masses of the quarks need not to be generated by electroweak symmetry breaking.}
\subsubsection{NLO QCD correction \label{sec:qqA_NLO}}
\begin{figure}[!t]
\centering
  \includegraphics[width = 0.8\textwidth, angle = 0]{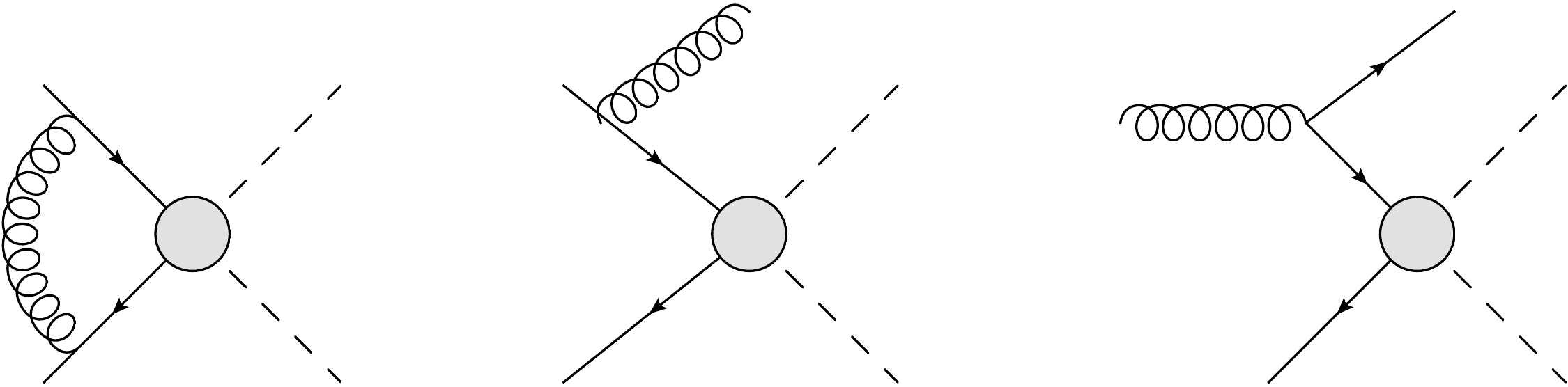}
  \caption{Generic form of the QCD corrections of order $\mathcal O(\alpha_s)$ to the qqA Higgs pair production. }
  \label{qqA_nlo}
\end{figure}

Since NLO QCD corrections are sizeable, we will take them into account in our analysis. For this purpose, we will detail here how we obtained them.
Since the $\hat{t}$ and $\hat{u}$ channel diagrams are strongly suppressed we can take the NLO QCD corrections over from $ b \bar b \to h$ in the 5-flavour scheme~\cite{Dicus:1998hs, Balazs:1998sb, Harlander:2003ai}\footnote{Note that the NLO and NNLO QCD corrections for $b\bar{b}hh$ have been given in \cite{Dawson:2006dm,  H:2018hqz}. It was found that the $b\bar{b}hh$ specific contributions of $\hat{t}$ and $\hat{u}$ channel diagrams are, as stated before at tree-level, also negligible at (N)NLO QCD.}  by some adjustments taking into account the modified LO cross section and the different kinematics of the process.
The Feynman diagrams at NLO QCD are shown in fig.~\ref{qqA_nlo}.
For convenience and for making our adjustments explicit we report here the formulae from \cite{Spira:2016ztx}
\begin{subequations}
\begin{eqnarray}
\sigma(q\bar q\to h) & = & \sigma_{LO} + \Delta\sigma_{q\bar q} +
\Delta\sigma_{qg}  \\
\Delta\sigma_{q\bar q} & = & \frac{\alpha_s(\mu_R)}{\pi} \int_{\tau_0}^1
d\tau \sum_q \frac{d{\cal L}^{q\bar q}}{d\tau} ~\int_\tau^1 dz~\hat{\sigma}_{LO}(Q^2=z\tau s)~\omega_{q\bar
q}(z)  \\
\Delta\sigma_{qg} & = & \frac{\alpha_s(\mu_R)}{\pi} \int_{\tau_0}^1 d\tau
\sum_{q,\bar q} \frac{d{\cal L}^{qg}}{d\tau}~\int_\tau^1 dz~\hat{\sigma}_{LO}(Q^2=z\tau s)~\omega_{qg}(z)
\end{eqnarray}
\end{subequations}
and
\begin{equation}
\hat{\sigma}_{LO}(Q^2)= \int_{\hat{t}_-}^{\hat{t}_+} \frac{ d\hat \sigma_{q_i\bar{q}_j}}{d \hat t}
\end{equation}
 with $z=\tau_0/\tau$, $\sigma_{LO}=\sigma_{\mathrm{hadronic}}$ of eq.~\eqref{eq:sigmahadron}, and the $\omega$ factors are given by \\
\begin{subequations}
 \begin{eqnarray}
 \omega_{q\bar q}(z) & = & -P_{qq}(z) \ln \frac{\mu_F^2}{\tau s}
 + \frac{4}{3}\left\{ \left(2\zeta_2-1 +
 \frac{3}{2}\ln\frac{\mu_R^2}{M_{hh}^2} \right)\delta(1-z)  \right. \\ &+ & \left.  (1+z^2) \left[
 2 {\cal D}_1(z) - \frac{\ln z}{1-z} \right] + 1-z \right\} \nonumber \,, \\
 \omega_{qg}(z) & = & -\frac{1}{2} P_{qg}(z) \ln \left(
 \frac{\mu_F^2}{(1-z)^2 \tau s} \right) - \frac{1}{8}(1-z)(3-7z)\,,
 \end{eqnarray}
\end{subequations}
  with $ \zeta_2 = \frac{\pi^2}{6}$.
The Altarelli Parisi splitting functions $ P_{qq}(z)$ and $ P_{qg}(z)$~\cite{gribov1972deep,Altarelli:1977zs,Dokshitzer:1977sg} are given by
 \begin{subequations}
 \begin{align}
 P_{qq}(z) & = \frac{4}{3} \, \left[2{\cal D}_0(z)-1-z+\frac{3}{2}\, \delta(1-z)\right],  \\
 P_{qg} &= \frac{1}{2}\, \left[  z^2+(1-z)^2\right],
\end{align}
\end{subequations}
and the `plus' distribution is
 \begin{equation}
   {\cal D}_n(z) := \left(\frac{\ln(1-z)^n}{1-z} \right)_+.
 \end{equation}
We have chosen the renormalisation scale $ \mu_R = M_{hh}$ and the factorisation scale $ \mu_F= M_{hh}/4$, as central values.
We define the NLO $K$-factor as 
\begin{equation}
K_{NLO}=\frac{\sigma_{NLO}}{\sigma_{LO}} = 1.28 \pm 0.02 \pm 0.17,
\end{equation}
with the first error denoting the uncertainty from varying the various $\kappa_q$ and the second error is propagated from LO and NLO scale and PDFs+$\alpha_s$ uncertainty.
 The $K$-factor does not depend on the scaling of the couplings, nor the flavour of the initial $q \bar q$ since the LO cross section factors out (with exception of the different integration in the real contributions). We finally note that at NNLO the qqA process interferes with diagrams with top quark loops, which contribute to ggF also at NLO. These contributions can in the SM limit be rather large, i.e. of similar order or even larger than the tree-level qqA process depending on the flavour considered.\footnote{We thank M.~Spira for pointing this out to us.} Due to the fact that the modifications of the Yukawa couplings that we consider in our analysis are rather large and that we are mostly interested in the case where qqA is of similar size or large than the ggF process we can safely neglect these contributions.

\subsubsection{Results}
While in the SM, the contribution from quark annihilation to a Higgs boson pair is below $0.11$ fb at NLO, it scales like $ \sim \kappa_q^2 m_q^2/v^4$, dominated by the $hh \bar q q$ diagram as can be seen from eq.~\eqref{sigmaqqa}, hence showing significant enhancement for enhanced Yukawa couplings.
For our benchmark scenario $(g_{hq \bar q} = g_{h b \bar b}^{\SM})$ we find for the cross section
\begin{equation}
\sigma^{qqA}_{NLO}= 284 \pm 25 \text{ fb}\,,
\end{equation}
and therefore a significantly larger cross section as for the ggF process.
\begin{figure}[!t]
\centering
  \includegraphics[width = 0.75\textwidth]{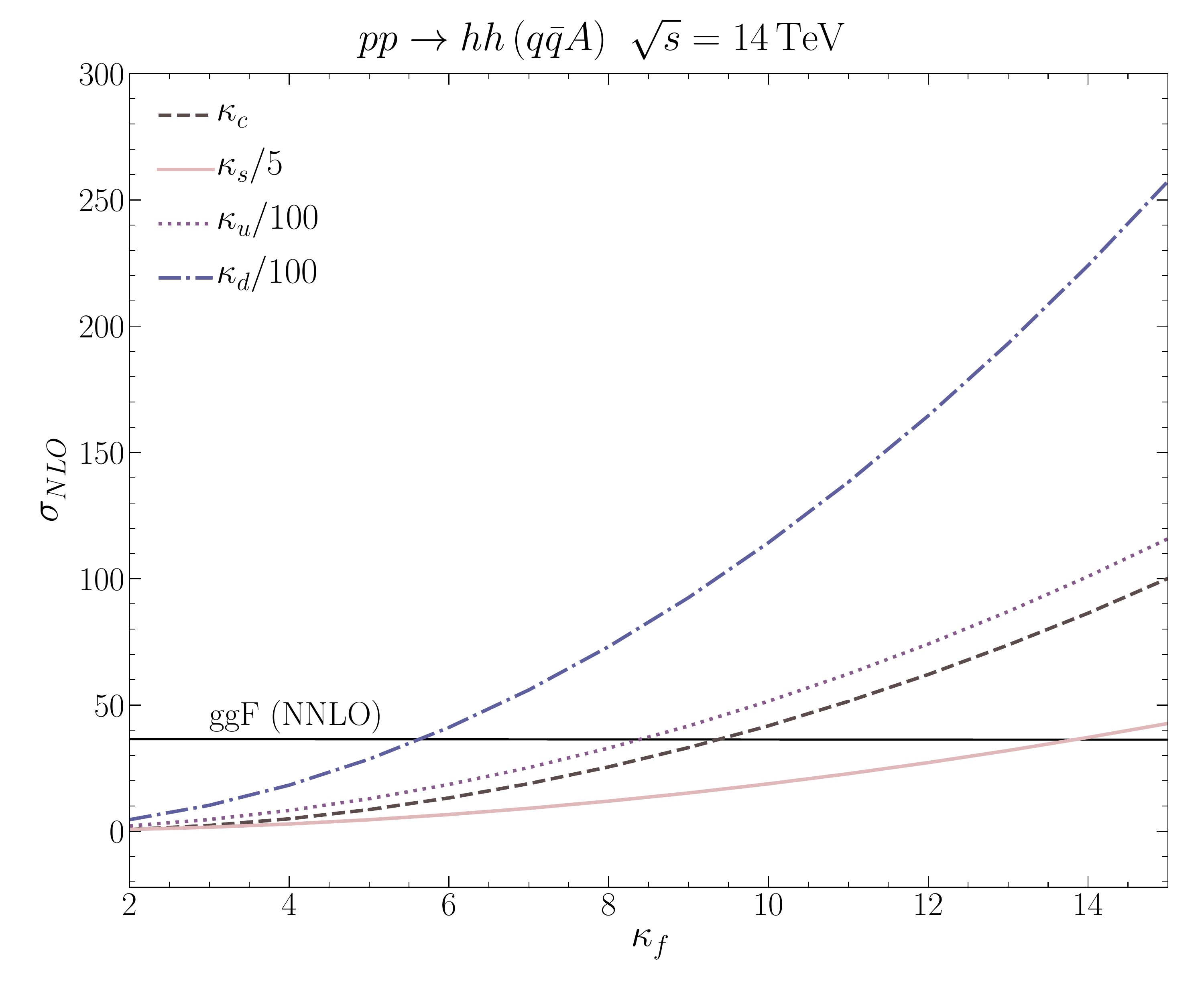}
  \caption{The NLO cross section for the qqA process for different scalings of the quark Yukawa couplings. The solid black line shows the NNLO ggF process width rescaled charm Yukawa coupling, whose effect though is unrecognisable in the plot. }
  \label{qqA_ggF}
\end{figure}
In fig.~\ref{qqA_ggF} we compare the ggF process (black line) for rescaled charm coupling to the Higgs boson(s) with the qqA process for different scalings of the light quark Yukawa couplings  (different coloured, dashed, dotted solid and dashed dotted lines).
We find that for sufficiently large scaling of the Yukawa couplings still allowed by current data, qqA can be even  the dominant di-Higgs production channel. Note that in the figure we scale the Yukawa couplings for the different quark mass eigenstates differently. For the up and down quark Yukawa coupling the scaling is the same, hence the effect from rescaling the down Yukawa coupling is larger even though the up quark is more abundant in the proton. The plot shows nicely for which values of the coupling modifications the qqA process surpasses ggF. \\
We would also like to give a qualitative argument for the dominance of qqA for large $\kappa_q$.
The dominant term for the qqA comes from the $hh q \bar q$ vertex diagram, such that the qqA cross section behaves for large values of $\kappa$ as (assuming that $ \sigma^{qqA}_{SM}\sim 0 $)
\begin{equation}
 (\sigma^{qqA}-\sigma^{qqA}_{SM}) \sim  g_{hh q \bar q}^2 \sim v^{-4}\,{m_q^2\,\kappa_q^2}.
 \end{equation}
 The ggF cross section instead gets contributions from light quark loops from  the diagram in fig.~\ref{fig_ggf_diag} interfering with top quark loops in the triangle SM diagram,  leading to a scaling of
 \begin{equation}
 (\sigma^{ggF} - \sigma^{ggF}_{SM} ) \sim  \kappa_q\, \frac{m_q^2}{ v^2\,M_{hh}^2}\,\ln^2{\left(\frac{M_{hh}}{m_q}\right)}\,.
 \end{equation}
 Taking the ratio we get
\begin{equation}
    \frac{(\sigma^{qqA}-\sigma^{qqA}_{SM})}{(\sigma^{ggF} - \sigma^{ggF}_{SM} )} \sim  \frac{\kappa_q}{ v^2\left(\frac{
   \ln^2{\left(\frac{M_{hh}}{m_q}\right)}}{M_{hh}^2}\right)}\,.
\end{equation}
This ratio approaches one (neglecting effects from different PDFs) when
\begin{equation}
    \kappa_q^{qqA = ggF} \sim  \frac{v^2\,\ln^2{\left(\frac{M_{hh}}{m_q}\right)}}{M_{hh}^2}\,.
\end{equation}
  Using this order of magnitude estimate, we see that  the two cross sections are roughly equal if $\kappa_c^{qqA = ggF} \sim 1$, $\kappa_s^{qqA = ggF} \sim 10$ and $\kappa_u^{qqA = ggF} \sim \kappa_d^{qqA = ggF} \sim 10^3$.
  The actual values of $\kappa_q^{qqA = ggF}$ can be read from fig.~\ref{qqA_ggF}. We observe that $ \kappa_q^{qqA = ggF}$ values  are not yet excluded, particularly for the first family.
\par
\begin{figure}[!b]
\centering
  \includegraphics[width = 0.75\textwidth]{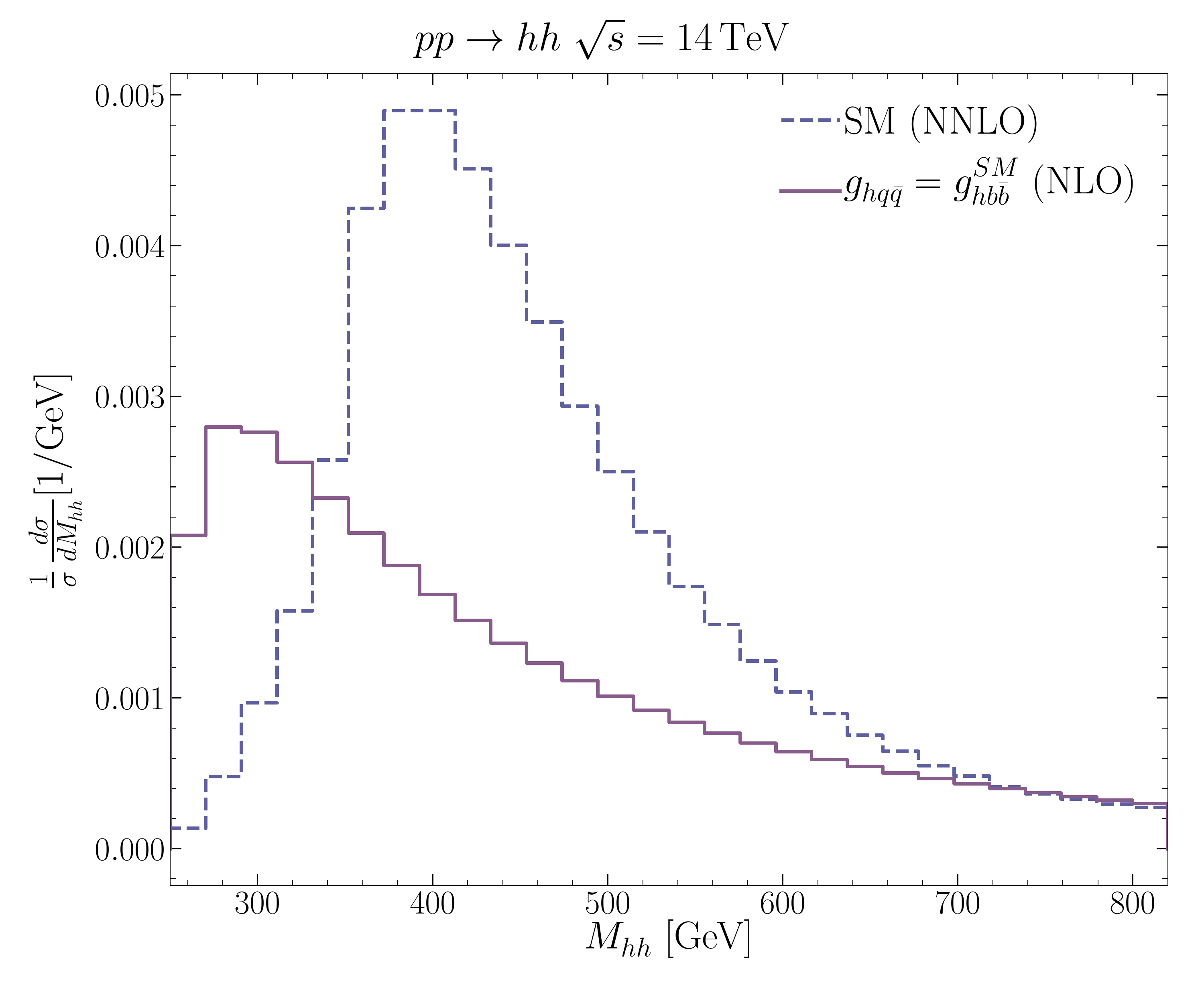}
  \caption{The qqA normalised NLO invariant mass differential cross section distribution for the benchmark point ($g_{hq \bar q} = g_{h b \bar b}^{\SM}$) (solid line) and the NNLO SM ggF cross section obtained from~\cite{Grazzini:2018bsd} (dashed line).
  }
  \label{qqA_dsigdmhh}
\end{figure}

In fig.~\ref{qqA_dsigdmhh} we show the di-Higgs invariant mass  normalised differential cross section distributions for the $g_{hq \bar q} = g_{h b \bar b}^{\SM}$ benchmark point at NLO compared to the NNLO SM ggF cross section extracted from~\cite{Grazzini:2018bsd}. We notice a considerable shape difference, with shifted peak to the left, and a larger tail. This will allow us later on to use kinematical information to extract the light quark Yukawa couplings.

\subsection{Higgs decays \label{sec:Hdecay}}
The light fermion decay channels will no longer be negligible for enhanced light Yukawa couplings. The decay channels $ h \to gg$, $h \to \gamma \gamma$  and $h \to Z \gamma$ containing fermion loops will get modified, but similarly to the production, the modification is ~$\sim 2\,\kappa_q\,(m_q^2/m_h^2) \,\ln^2(m_q/m_h)$. Thus, the main effect on the Higgs boson branching ratios and width is the `opening' of the new light fermion channels. \\ In order to compute the Higgs partial widths and branching ratios (BR) at higher orders in QCD, we have modified the FORTRAN programme \texttt{HDECAY}~\cite{Djouadi:1997yw,Djouadi:2018xqq} to include the light fermion decay channels and loops in the above-mentioned decays. In the SM, light fermion BRs are of order $\mathcal{O}(10^{-4})$ for $ h \to c \bar{c} $,  $\mathcal{O}(10^{-6})$ for $ h \to s \bar{s} $  and  $<\mathcal{O}(10^{-9})$ for the first generation quarks \cite{deFlorian:2016spz}. In our benchmark point ($g_{hq \bar q} = g_{h b \bar b}^{\SM}$) these would increase to $ \sim 18 \%$. Correspondingly, the BRs for  $ h \to b \bar b/VV/\tau^+\tau^-$ decrease due to the increased Higgs width in the model.
\par
\begin{figure}[!t]
\centering
  \includegraphics[width = 0.49\textwidth]{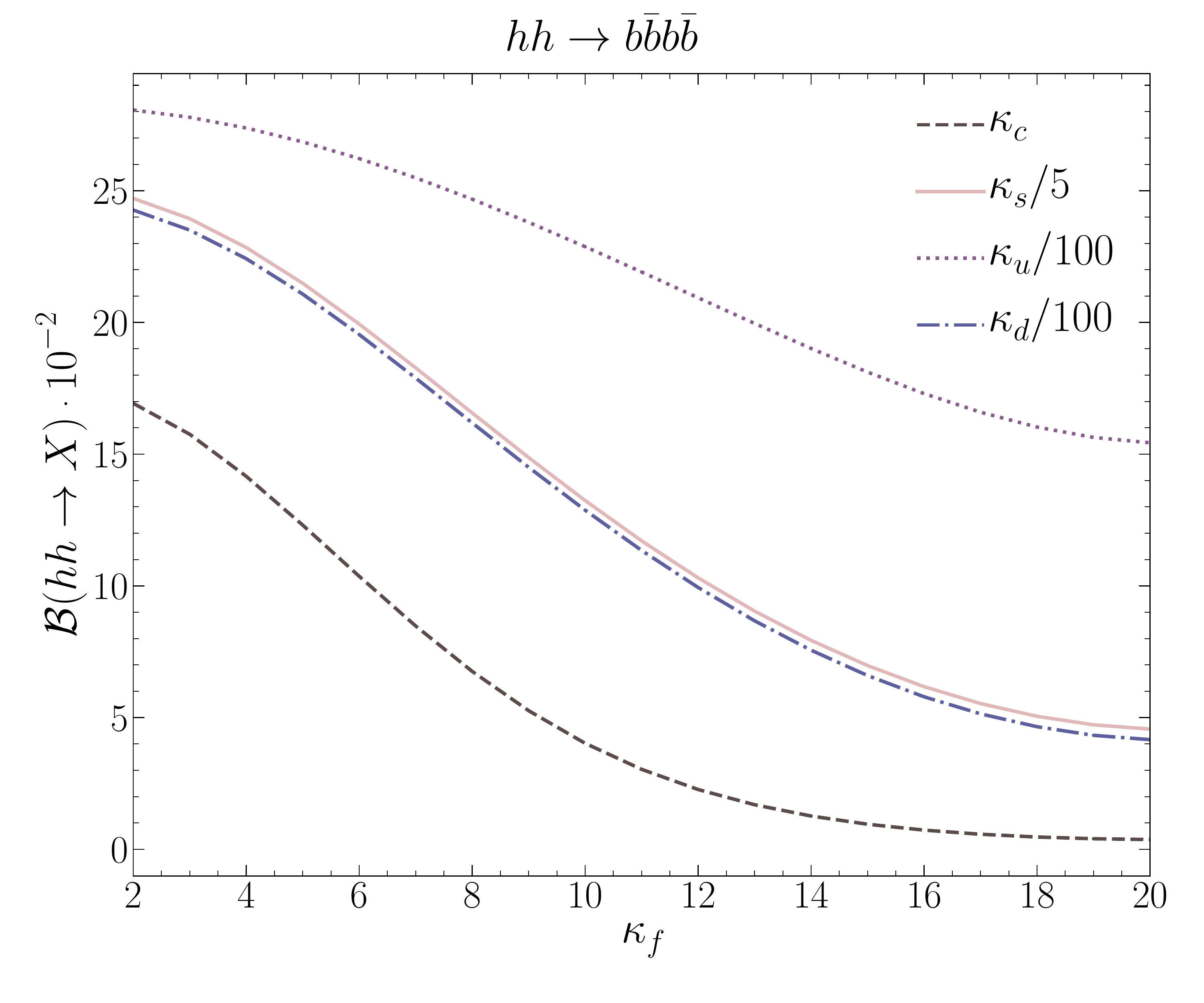}
  \includegraphics[width = 0.49\textwidth]{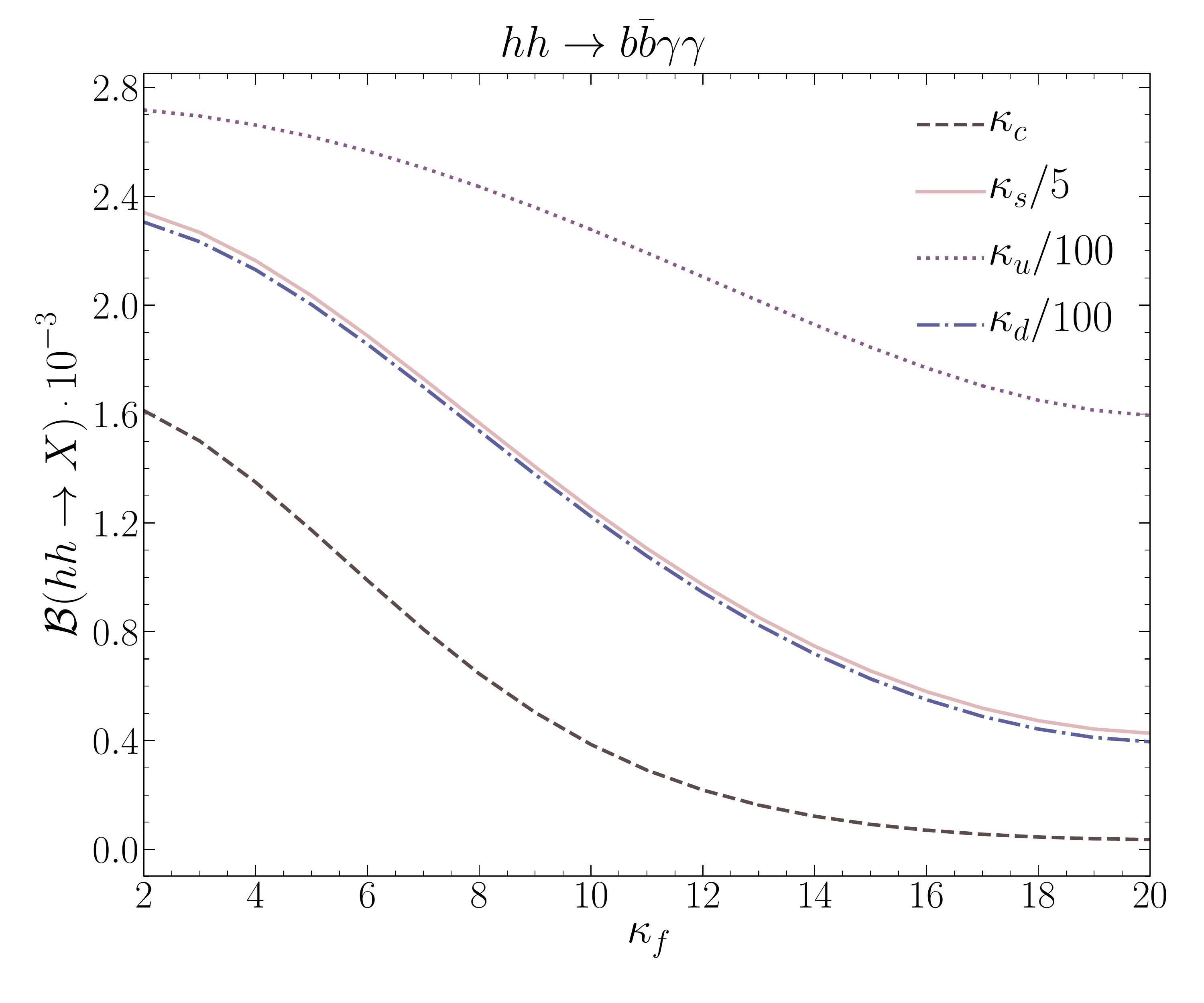}
  \\
  \includegraphics[width = 0.49\textwidth]{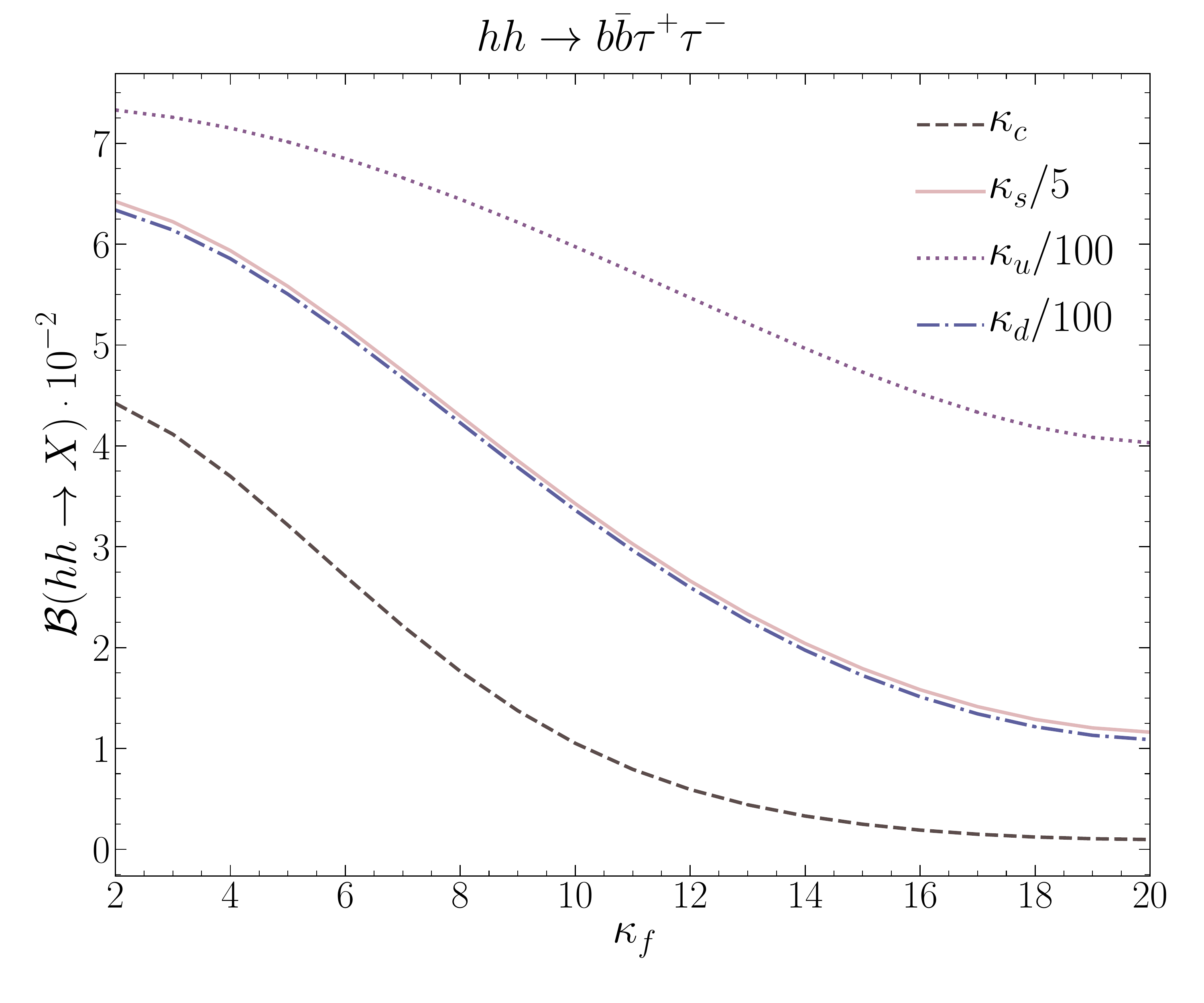}
  \includegraphics[width = 0.49\textwidth]{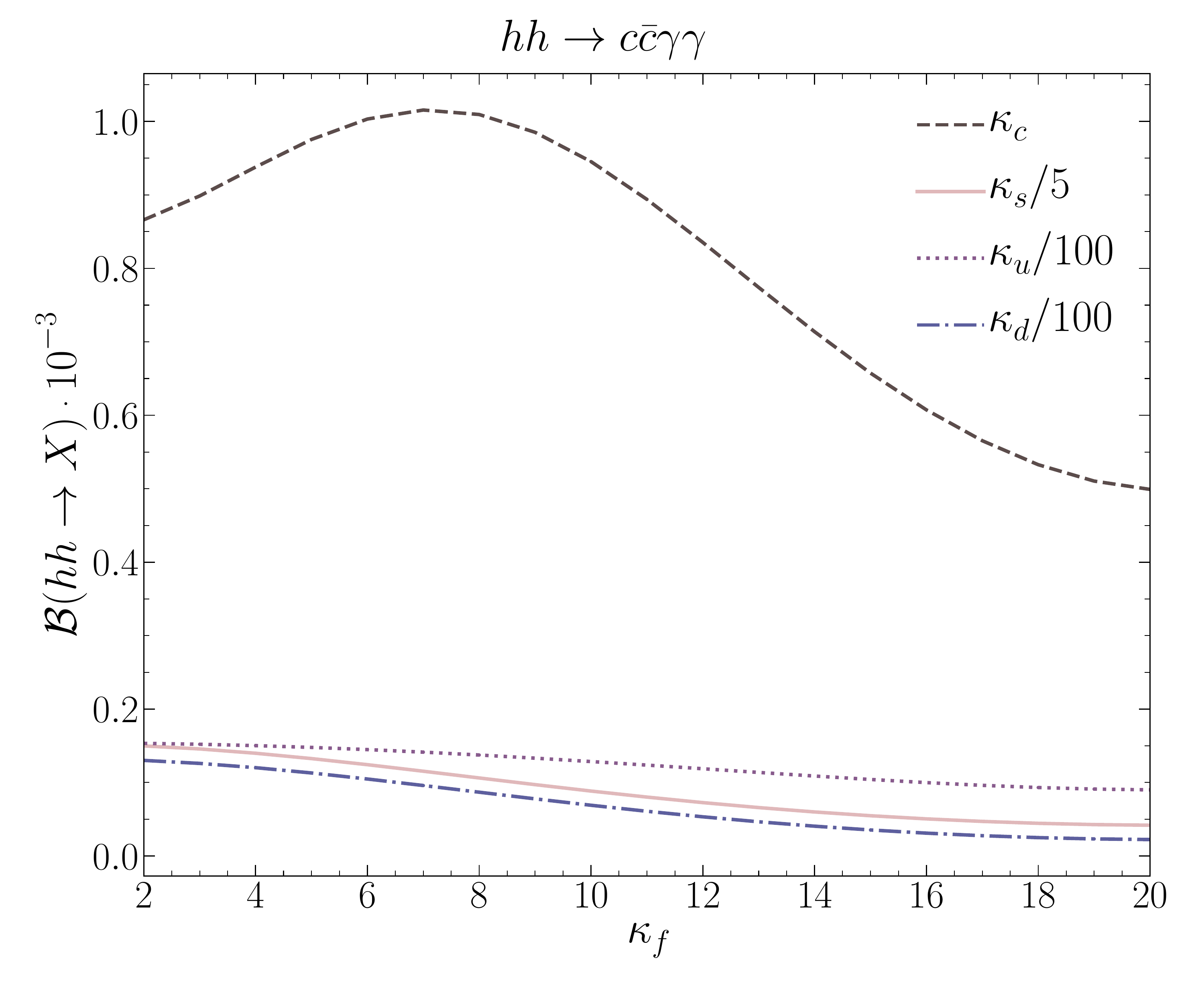}
  \caption{Different Higgs pair final states BRs including state-of-the-art QCD corrections as functions of the coupling modification factors $\kappa_f$. \textit{Top left}: $ hh \to b \bar b b \bar b$. \textit{Top right}: $ hh \to b \bar b \gamma \gamma$. \textit{Bottom left}: $ hh \to b \bar b \tau^+ \tau^-$. \textit{Bottom right}: $ hh \to c \bar c \gamma \gamma$. }
  \label{brs}
\end{figure}

In fig.~\ref{brs} we show the BRs, denoted by $\mathcal{B}$ in the following,  of the Higgs boson pair with the best prospects for discovering Higgs pair production, $hh\to b\bar{b}b\bar{b}$, $hh\to b\bar{b}\gamma\gamma$ and $hh\to b\bar{b}\tau^+\tau^-$ \cite{Aad:2019uzh}, and in addition we show for later purpose also  $hh\to c\bar{c}\gamma\gamma$. Once we increase the light quark Yukawa couplings (shown for the different quarks by the different coloured lines) the BRs to $b\bar{b}b\bar{b}$, $b\bar{b}\gamma\gamma$ and $b\bar{b}\tau^+\tau^-$ decrease due to the increased Higgs width. Instead the $\mathcal{B}(hh\to c\bar{c}\gamma\gamma)$ first increases with increasing $\kappa_c$, but starts decreasing after reaching a maximum around $\kappa_c\approx 8$, where the $\mathcal{B}(h\to c\bar{c})$ asymptotically reaches 1 while the $\mathcal{B}(h\to \gamma \gamma)$ continues decreasing.

\begin{figure}[!t]
\centering
\includegraphics[width = 0.49\textwidth]{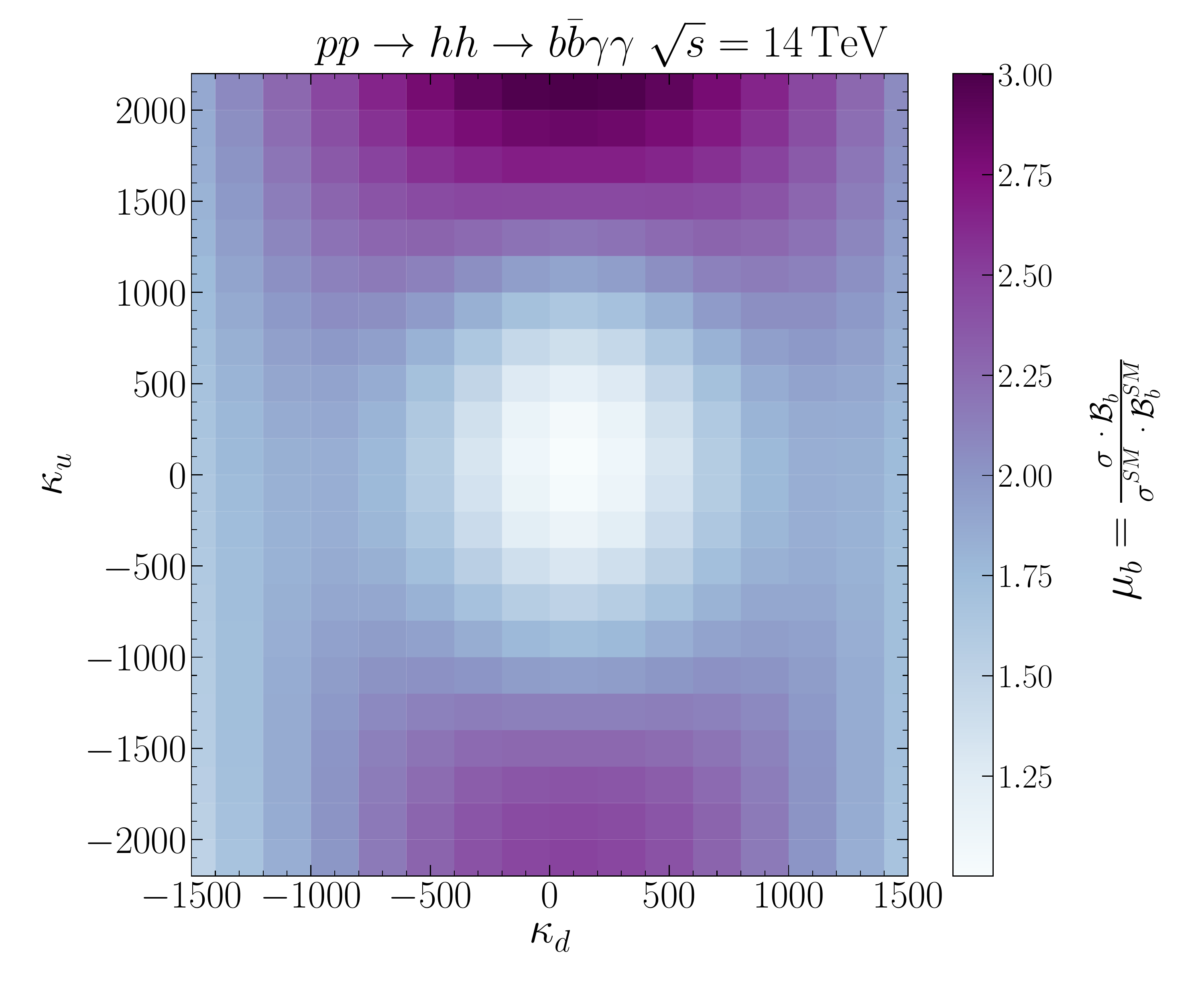}
\includegraphics[width = 0.49\textwidth]{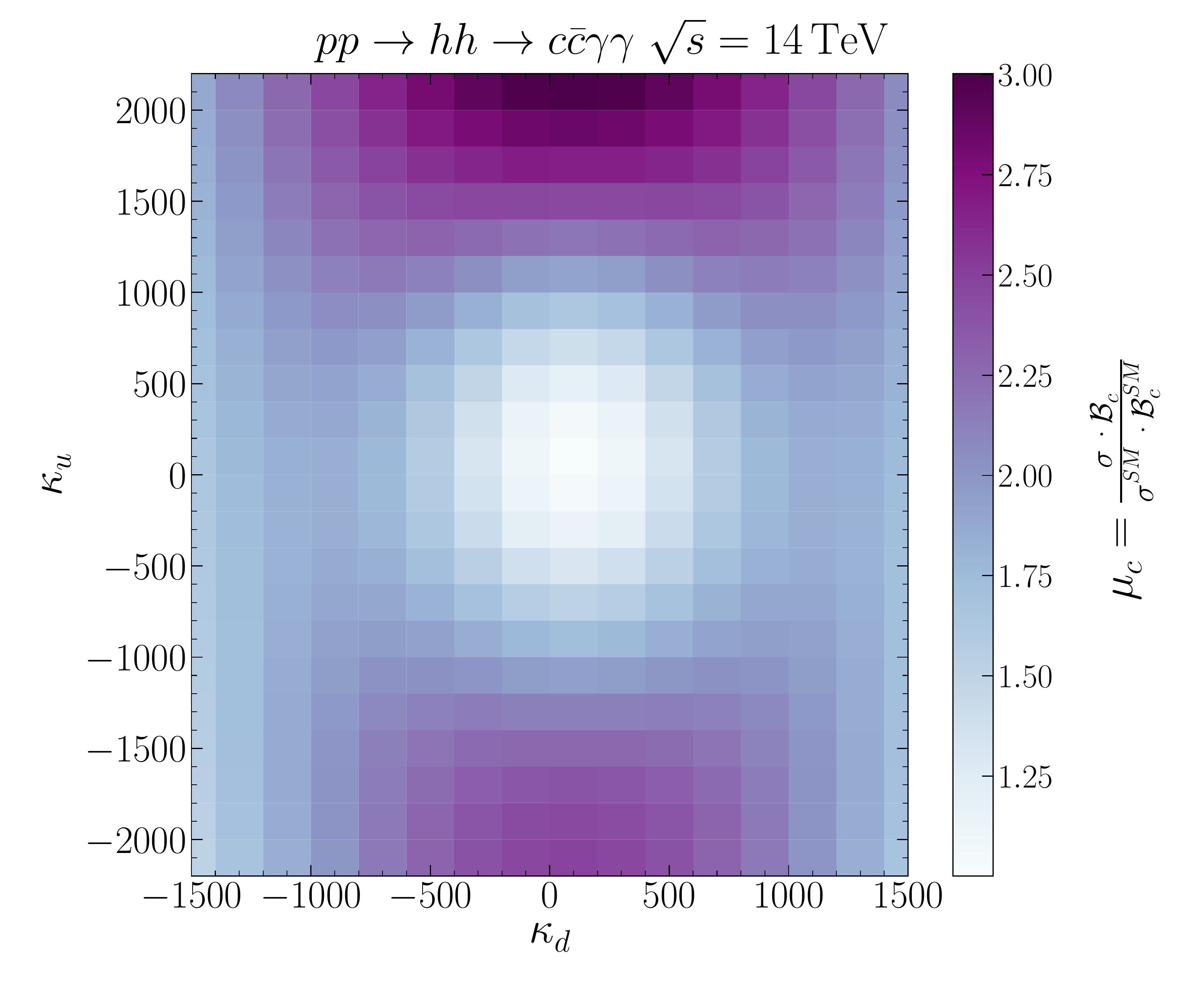}
\\
\includegraphics[width = 0.49\textwidth]{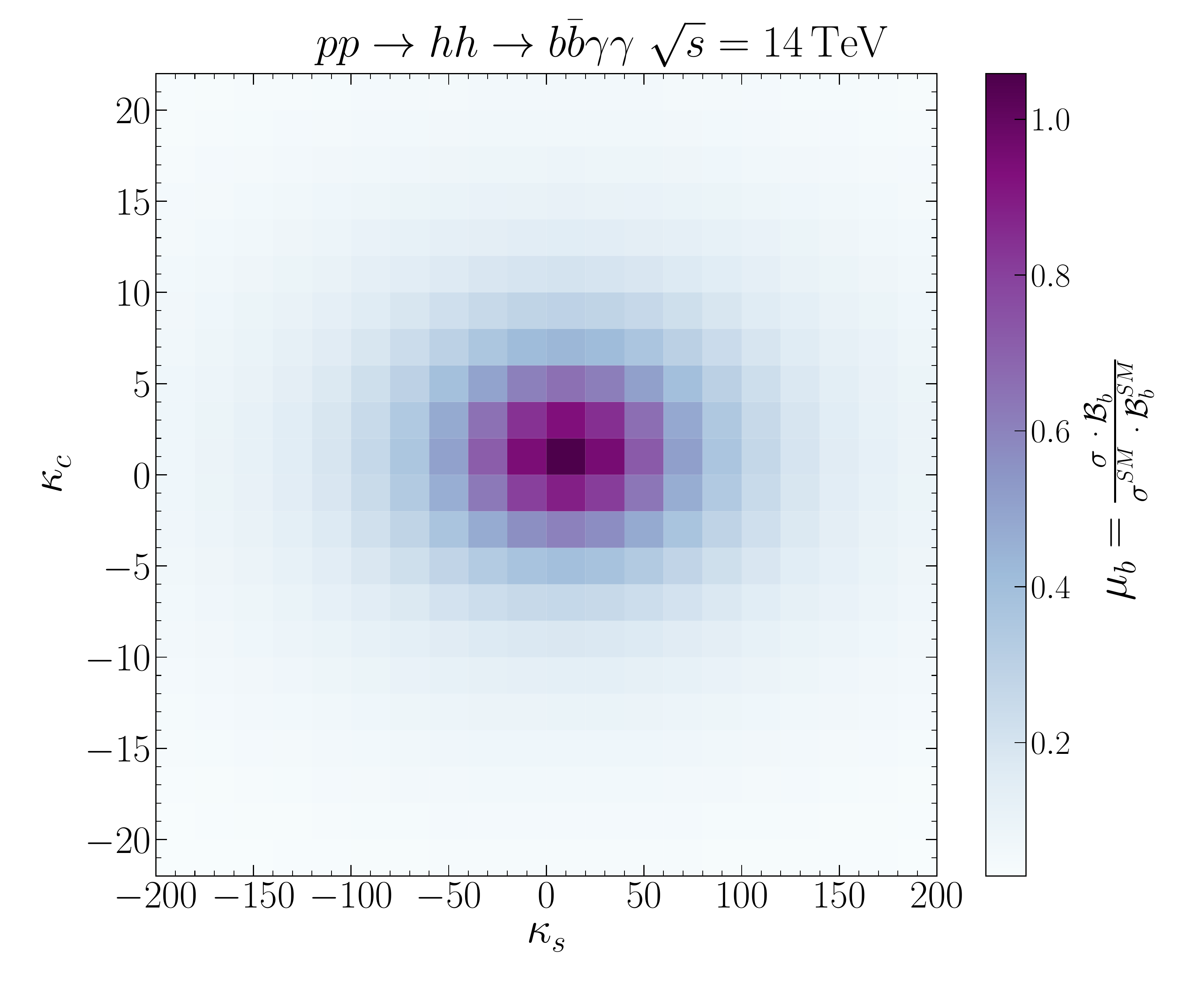}
\includegraphics[width = 0.49\textwidth]{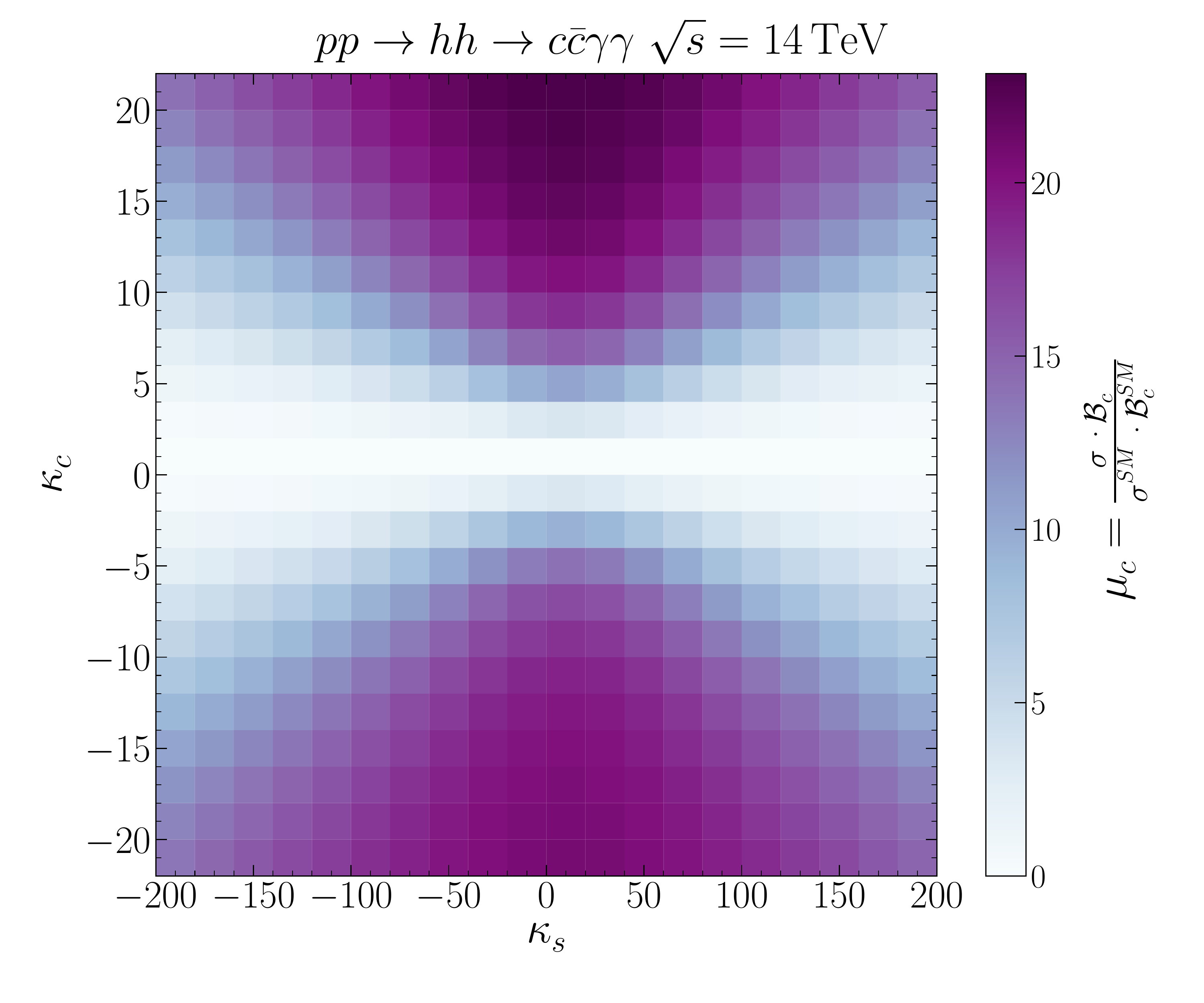}
  \caption{Signal strength modifier $\mu = \sigma \,{\cal B}(hh \to X) / (\sigma^{SM} \,{\cal B}^{SM}(hh \to X) )$  fits for bottom quark (left plots) and charm quark (right plots)  final states for  first~(upper row) and second~(lower row) generations quark Yukawa modifications.}
  \label{kcks_mu_contors}
\end{figure}

In fig.~\ref{kcks_mu_contors} we show the signal strength modifier defined  here as\\

\begin{equation}
\mu_i : = \frac{\sigma \, {\cal B}_i } {\sigma^{\SM} \, {\cal B}_i^{\SM}}\quad \quad (i=b,c),
\end{equation}
for final states with bottom (left hand side) and charm quarks (right hand side) for first generation (plots in the upper row) and second generation (plots in the lower row) modified Yukawa couplings. For the first generation, we obtain enhancement of both of the signal strengths $\mu_c$ and $\mu_b$, as seen  plots in the top of fig.~\ref{kcks_mu_contors}. The  second generation  signal strength is instead reduced with respect to the SM for the channels with bottom quarks in the final state $\mu_b : = \sigma \, {\cal B}_b/ \sigma ^{\SM}\, {\cal B}_b^{\SM}$ when scaling the charm and strange Yukawa couplings, as seen in the lower left plot of fig.~\ref{kcks_mu_contors}. Nevertheless, when considering channels with charm quarks in the final state the signal strength $\mu_c : = \sigma \, {\cal B}_c /\sigma^{\SM} \, {\cal B}_c^{\SM} $ is enhanced due to both enhancements from the cross section and BRs.
The increased cross section in the presence of enhanced light quark Yukawa couplings has to compete with the decreased BRs for the standard search channels for di-Higgs production.
We shall notice however, that while the increase of the cross section comes mainly from the $q\bar{q}hh$ vertex diagram, the decrease of the BRs stems from the increased width which would be in good approximation (for flavour-diagonal couplings)
\begin{equation}
\Gamma_H \approx \Gamma_{\text{SM}}+\sum_{q=c,s,u,d}\frac{g_{h \bar{q}_i q_i}^2}{(g_{h \bar{q}_i q_i}^{\text{SM}})^2}\Gamma_{q}\,,
\end{equation}
where $\Gamma_q$ stands generically for the partial width of the Higgs boson decaying to light quarks.
In a non-linear EFT as briefly discussed in sect.~\ref{sec:EFT}, the couplings of one Higgs boson to quarks and two Higgs bosons to quarks are uncorrelated.
So an increase of the cross section for $hh$ production in the presence of modified light quark Yukawa couplings does not need to go hand in hand with a decrease of the BRs in the final states with bottom quarks (or at least the decrease could be in-proportional).

%%%%%%%%%%%%%%%%%%%%%%%%%%%%
\section{Phenomenological analysis \label{sec:pheno}}
In this section we will investigate whether enhanced light quark Yukawa couplings can be measured in Higgs pair production. As we have seen in the previous section,
we can get an enhancement in the signal strengths for first generation quarks from the enhanced cross sections while BRs in the standard di-Higgs search channels decrease. We have also seen that final states with charm quarks might be worth studying further for enhanced second generation Yukawa couplings.
Here in this section, we will perform a phenomenological analysis to see if the HL-LHC has potential to constrain the light quark Yukawa couplings in di-Higgs channels.
The first part of the section is devoted to the analysis strategy, before we discuss the bounds from final states with bottom quarks. We will be  focussing in particular on the $b\bar{b}\gamma\gamma$ final state as it holds promising prospects \cite{Azatov:2015oxa, Baur:2003gp, Baglio:2012np, Kling:2016lay, Barger:2013jfa, Adhikary:2017jtu} despite the low BR of 0.27\% in the SM for the Higgs boson pair. At the end of the section we take a closer look at the $c\bar{c}\gamma \gamma$ final state, which is in particular interesting for enhanced charm Yukawa couplings.

For our phenomenological analysis we do not assume that the efficiency is constant for the new physics hypothesis with respect to the SM efficiency.  Hence, we use the full definition of the signal strength $\mu$ as the ratio of the number of events measured or expected given the new physics hypothesis over the number of events expected by the SM~(null) hypothesis
\begin{equation}
    \mu = \frac{ N_{expec}}{ N^{SM}_{expec}}.
\end{equation}
The number of expected events $ N_{expec}$ at a hadron collider with integrated luminosity $L$ and selection efficiency $\epsilon_{SEL}$  in the narrow width approximation for a process $pp\to R$ with subsequent decay of $R\to X$ is given by the formula
\begin{equation}
  N_{expec} = \sigma(pp\to R) \, \mathcal B(R \to X)\, L \, \epsilon_{\mathrm{SEL}}.
\end{equation}
The selection efficiency can be written in terms of several factors by
\begin{equation}
  \epsilon_{\mathrm{SEL}} = \epsilon_{\mathrm{Acc}} \cdot \epsilon_{\mathrm{Rec}} \cdot \epsilon_{\mathrm{Trig}}\cdot \epsilon_{\mathrm{cut}},
\end{equation}
with $ \epsilon_{\mathrm{Acc}}$ being the detector acceptance efficiency, $ \epsilon_{\mathrm{Rec}}$ the efficiency from reconstruction,  $\epsilon_{\mathrm{Trig}}$ the trigger efficiency and $\epsilon_{\mathrm{cut}}$ the efficiency obtained from the applied kinematical cuts on the signal. For the ATLAS and CMS experiments, the acceptance for the Higgs pair production is close to $100 \%$ due to the complete coverage of the pseudorapidity range of~$2.5< |\eta |< 5$, so we use $\epsilon_{\mathrm{Acc}}=1$.
The other efficiencies will be discussed in more detail in subsect.~\ref{sec:analysisstrategy}.
%%%%%%%%%%%%%%%%%%%%%%%%%%%%%%%%%%%%%%%%%%%%%%%%%%%%%
\subsection{Event generation}
The parton showering and hadronisation of the process $pp \to hh \to b \bar b \gamma \gamma$ has been simulated using \texttt{Pythia} 6.4~\cite{Sjostrand:2006za} with the settings detailed in appendix~\ref{app:pythia}. The cross section of the Higgs pair production (ggF and qqA both at LO multiplied by a $K$-factor as described in subsect.~\ref{sec:ggF} and \ref{sec:qqA_NLO}) is fed to \texttt{Pythia} which decays the two Higgs bosons and then performs the parton showering. We have accounted for the correct BRs by using the values obtained as described in subsect.~\ref{sec:Hdecay} from {\tt HDECAY}. We have turned on initial and final state QCD and QED radiation and multiple interactions.  \\
The generated events were written to a ROOT file via \texttt{RootTuple} tool~\cite{roottuple} for further analysis.
%%%%%%%%%%%%%%%%%%%%%%%%%%%%%%
\subsection{Analysis strategy \label{sec:analysisstrategy}}
The analysis strategy follows the one performed in~\cite{Azatov:2015oxa} allowing us to use their backgrounds. Note that the analysis was based on the SM  simulated events, meaning that the significances could be potentially improved performing a dedicated new physics analysis.
In order to satisfy the minimal reconstruction requirements of the LHC we select only events with
\begin{equation}
  p_T(\gamma/j) > 25 \, \GeV\,, \; \; \; \; |\eta(\gamma/j)| < 2.5\,.
\end{equation}
Moreover, we veto events with hard leptons
 \begin{equation}
   p_T(\ell) > 20 \, \GeV, \; \; \; \; |\eta(\ell)| < 2.5\,,
 \end{equation}
corresponding of an expected $ \epsilon_{\mathrm{Trig}}=0.9$.
Jets were clustered using \texttt{fastjet}~\cite{Cacciari:2011ma}  with the anti-kt algorithm with a radius parameter of $R=0.5$. \\
We have used a $b$-tagging efficiency of $ \epsilon_b = 0.7$ \footnote{We have explicitly cross checked the number by doing a mass-drop tagger analysis \cite{Butterworth:2008iy}.}.
The contamination probability of $\epsilon_{j \to b} < 1 \%$  is found to be consistent with ATLAS and CMS performance~\cite{Chatrchyan:2012dk,CMS:2013vea,ATL-PHYS-PUB-2013-009}. For the photon reconstruction efficiency  we used $ \epsilon_\gamma = 0.8$ as reported by ATLAS and CMS in~\cite{ATL-PHYS-PUB-2013-009,CMS:2013aoa}.
The selection cuts we used are the same ones as in~\cite{Azatov:2015oxa}, starting with the cuts of the transverse momentum $p_T$ of the photons and $b$-tagged jets. The two hardest photons/$b$-tagged jets,  with transverse momentum $p_{T>}$, and the softer ones with $p_{T<}$ are selected to satisfy
\begin{equation}
    p_{T>}(b/ \gamma) > 50 \, \GeV, \quad \text{and} \quad   p_{T<}(b/ \gamma) > 30 \, \GeV\,.
    \label{cut1}
\end{equation}
In order to ensure well-separation of the photons and $b$-jets, we require the following cuts on the jet radius,
\begin{equation}
    \Delta R(b,b) < 2  ,\; \; \; \; \Delta R(\gamma,\gamma) < 2, \; \;  \; \; \Delta R(b,\gamma) > 1.5\,.
    \label{cut2}
\end{equation}
While the majority of the signal lies within this region, these cuts significantly reduce the backgrounds.

We choose a wide $m_{\gamma \gamma}$  window (see eq.~\eqref{cut3}) corresponding to 2-3 times the photon resolution of ATLAS and CMS~\cite{ATL-PHYS-PUB-2013-009,CMS:2013aoa} which does not cause any significant loss. As for the Higgs  mass window reconstructed from 2 $b$-jets $ m_{b\bar b}$, the mass window chosen in eq.~\eqref{cut3} corresponds to the given $b$-tagging efficiency. The mass windows used are then

\begin{equation}
   105\,\GeV < m_{b \bar b} < 145 \, \GeV, \; \; \; \;123\, \GeV < m_{\gamma \gamma} < 130 \, \GeV\,.
    \label{cut3}
\end{equation}
 The selection cuts are summarised in table~\ref{cuts_eff} with their corresponding efficiency. In table~\ref{eff} we summarise all the efficiencies used in the analysis.
\npdecimalsign{.}
\nprounddigits{2}
\begin{table}[!t]
    \centering
    \begin{tabular}{l n{5}{2} n{5}{2} }
\toprule
  cut  & $\epsilon_{\mathrm{cut}}$  &  $ \delta \epsilon_{\mathrm{cut}}$  \\
  \midrule
  $p_T$ cuts in eq.~\eqref{cut1} & 0.35 & 0.07\\
   $\Delta R$ cuts  in eq.~\eqref{cut2} & 0.69 & 0.21 \\
  \hline
  total    & 0.16 & 0.05 \\
    \bottomrule
    \end{tabular}
    \caption{The cuts used in the analysis with their efficiency $\epsilon_{\mathrm{cut}}$ and uncertainties on these efficiencies $ \delta \epsilon_{\mathrm{cut}} = \sqrt{\epsilon(1-\epsilon)\,N}$, where $N$ is the total number of events. The analysis was performed on 100K SM simulated events.}
    \label{cuts_eff}
\end{table}
\npnoround
\begin{table}[!t]
    \centering
    \begin{tabular}{ c c }
\toprule
    Type & efficiency \\
  \midrule
$\epsilon_{\mathrm{Acc}}$  & $\sim 1$ \\
$\epsilon_{\mathrm{Rec}}$ & 0.31 \\
$\epsilon_{\mathrm{Trig}}$ & 0.90 \\
$\epsilon_{\mathrm{Cut}}$ & 0.16 \\
  \hline
  total    &  0.044\\
    \bottomrule
    \end{tabular}
    \caption{Values of the efficiencies calculated/used in this analysis. }
    \label{eff}
\end{table}
\par
\begin{figure}[!t]
\centering
  \includegraphics[width = 0.75\textwidth]{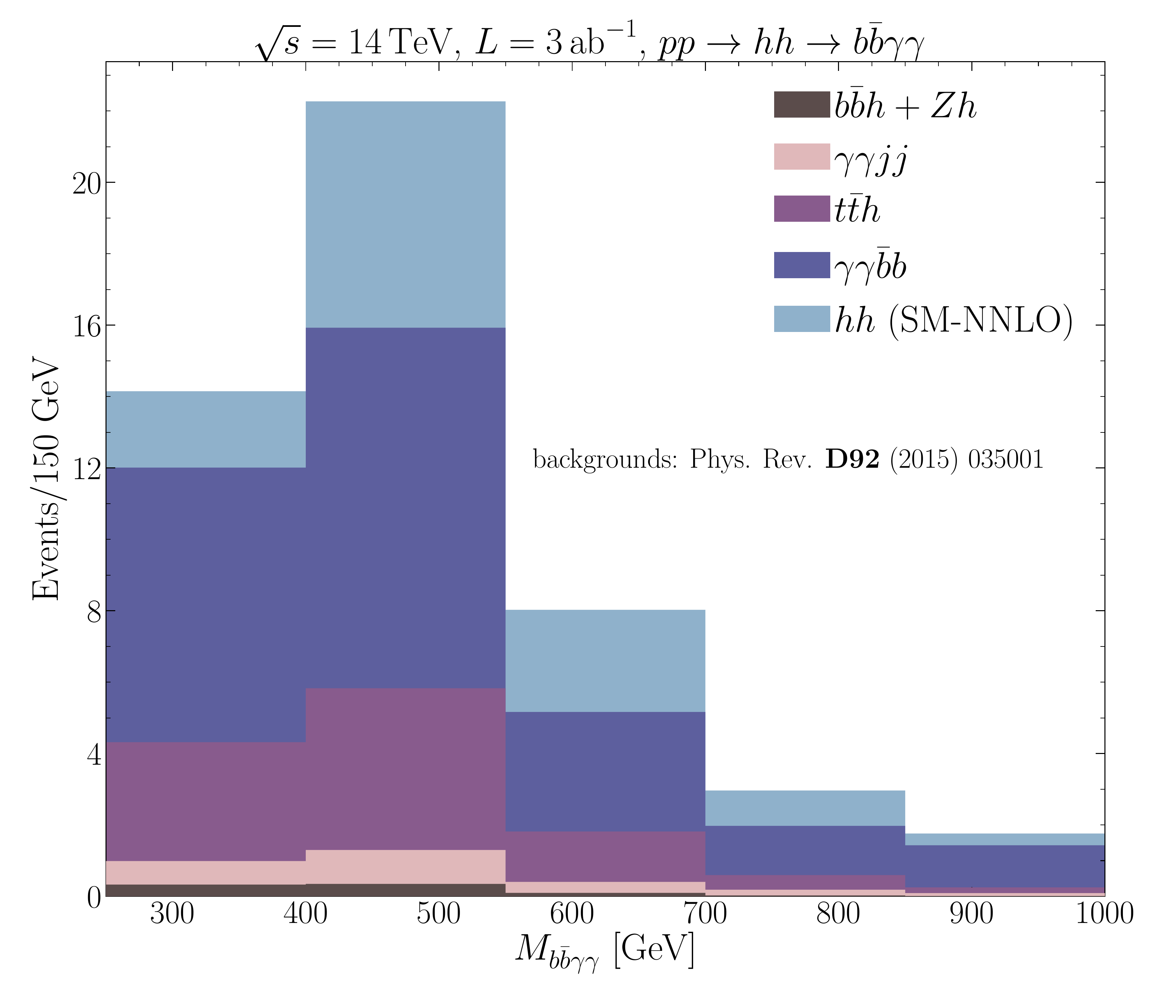}
  \caption{The number of expected events in SM $hh$ signal with the most relevant backgrounds as estimated by~\cite{Azatov:2015oxa}.}
  \label{sm_hh_and_bkg}
\end{figure}

The major backgrounds for the considered final state are the $b\bar{b}\gamma\gamma$ continuum background, $\gamma\gamma jj$ with two mistagged jets, $t\bar{t}h$, $Zh$ and $b\bar{b}h$ in the order of importance after the cuts in eq.~\eqref{cut1}. The number of background events (surviving the cuts) is taken from~\cite{Azatov:2015oxa}. The backgrounds are illustrated in the fig.~\ref{sm_hh_and_bkg} in which we show the number of events for the SM Higgs pair signal in light blue and the most relevant backgrounds in other colors. It should be noted that the background $h(\to \gamma \gamma) Z(\to b \bar b)$  is modified in the presence of enhanced light quark Yukawa couplings. We checked though explicitly that scaling the Yukawa couplings to the values of our benchmark point only changes the NLO cross section by less than 1\%, making this effect negligible.

\begin{figure}[!t]
\centering
\begin{picture}(200,250)
\put(-180,0){\includegraphics[scale =0.30]{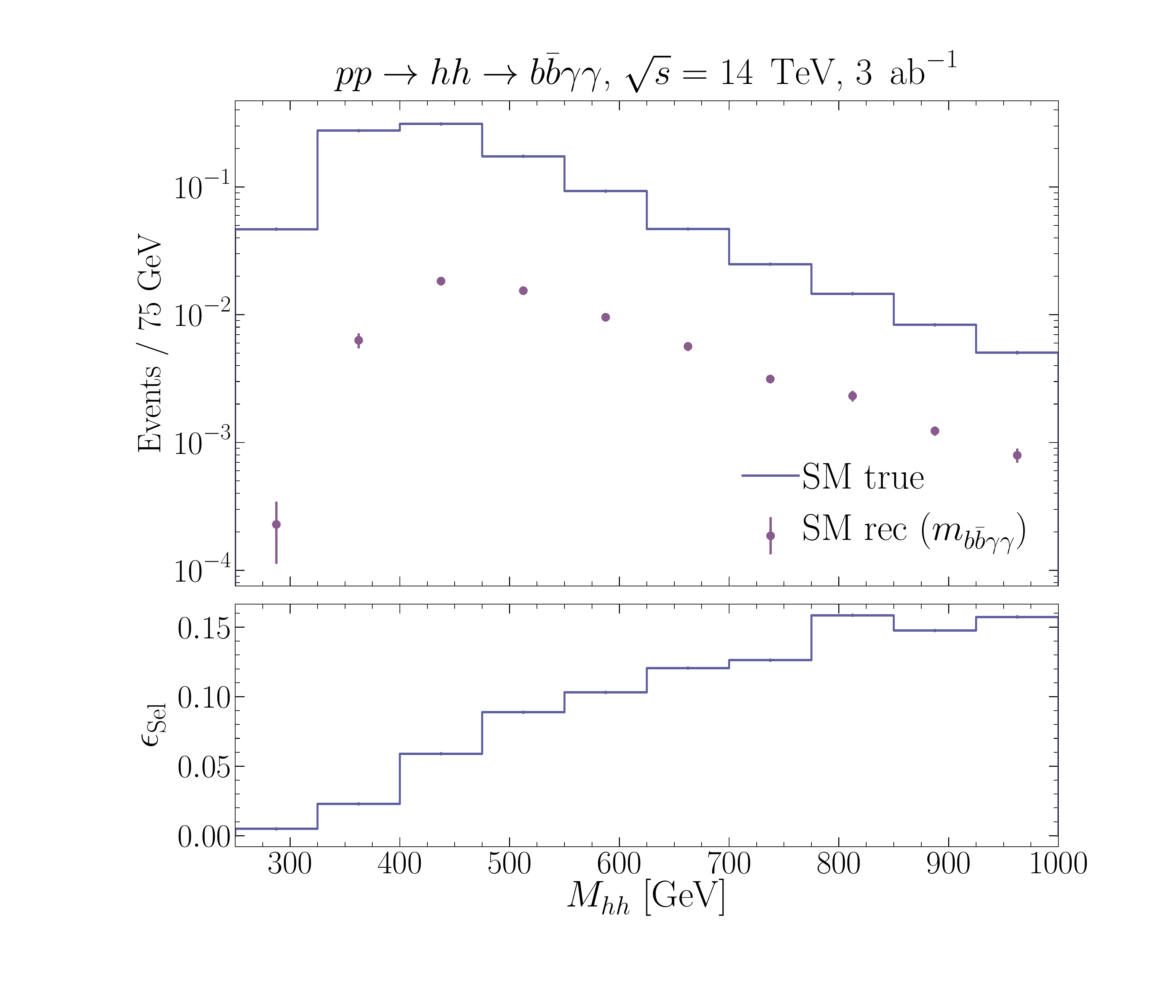}}
\put(80,0){\includegraphics[scale = 0.30]{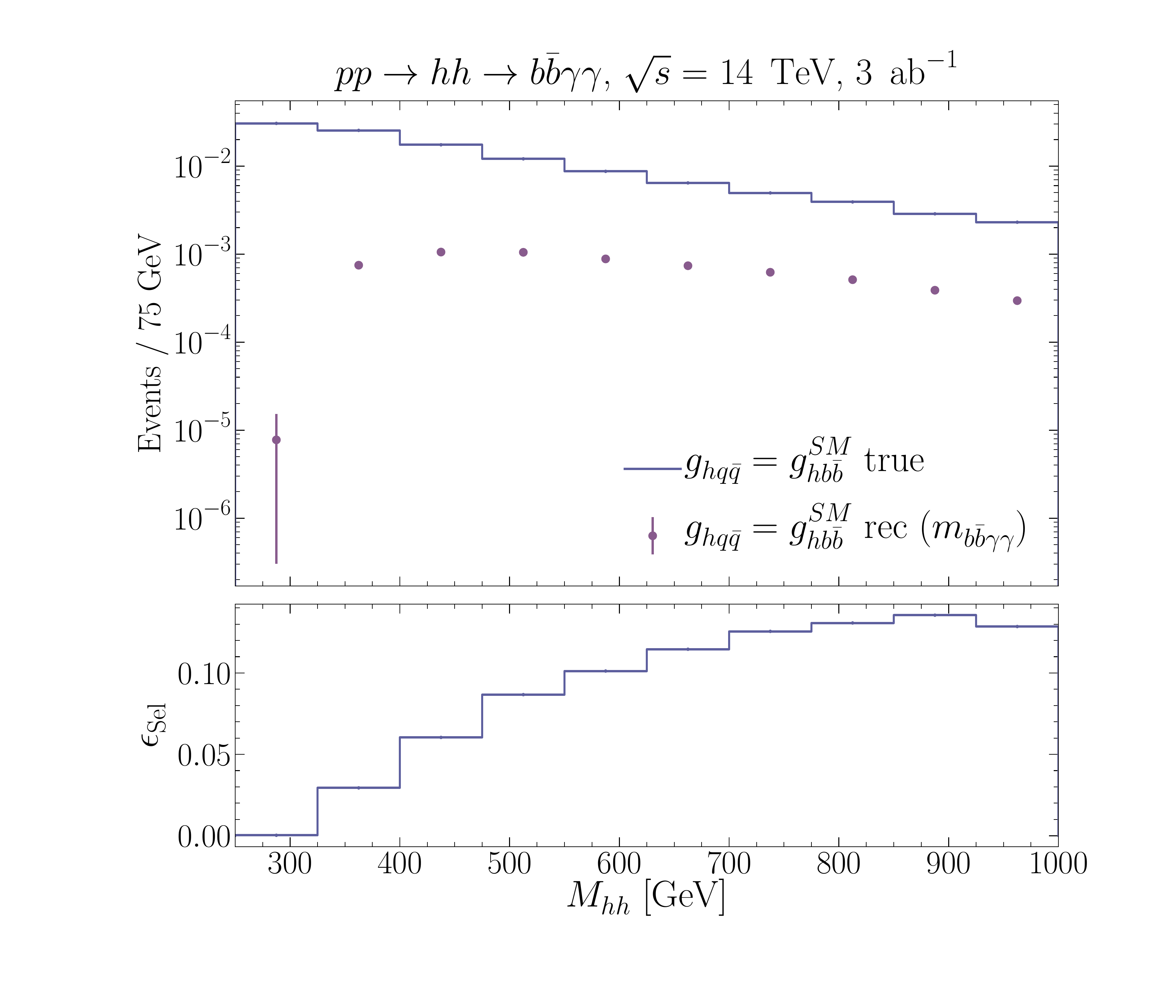}}
\end{picture}
\vspace*{-1.cm}
  \caption{Truth-level (no cuts) vs reconstructed $M_{hh}$ distributions for the SM (left) and the benchmark point (right). In the lower part of the plot we show the ratio between truth-level and reconstructed distribution, which is equivalent to the efficiency.}
  \label{eff_ratio}
\end{figure}

The analysis was carried out for varying values of $\kappa_f$ for the different flavours. Due to the change in the kinematical distributions (cf.~fig.~\ref{eff_ratio}) resulting from the PDFs of the different flavours, the efficiencies depend on the flavour of the quarks.
For $\kappa_f\gg 1$ the $\kappa_f$ dependence factors out of the cross section such that for the values considered in the analysis of the distributions no dependence on the concrete value of $\kappa_f$ is seen. The flavour-specific efficiency ratio~$\epsilon_f$ is given by
\begin{equation}
   \epsilon_f =  \frac{\sigma_{ggF} \, \epsilon_{ggF} + \sigma_{q\bar q}  \, \epsilon_{qq} }{\sigma_{gg} + \sigma_{q\bar q}}\,,
   \label{effrat}
\end{equation}
with $\sigma_{ggF}$ being the gluon fusion cross section, $\sigma_{q\bar q}$ the quark annihilation cross section and $\epsilon_{ggF} = 0.044$. We give the values for the qqA efficiency $\epsilon_{qq}$ in table~\ref{eps_vark}.
\begin{table}[!b]
   \centering
   \begin{tabular}{cc}
       \toprule
       $\delta \kappa$ 	& $ \epsilon_{qq}$	 \\
       \midrule
           $\kappa_{u}$   & $0.050$  \\
           $\kappa_{d}$   & $0.049$   \\
           $\kappa_{u}$ \& $\kappa_{d}$   & $0.053$  \\
           \hline
           $\kappa_{c}$   & $0.034$  \\
           $\kappa_{s}$   & $0.037$   \\
           $\kappa_{c}$ \& $\kappa_{s}$   & $0.039$  \\
       \bottomrule
   \end{tabular}
   \caption{The dependence of $\epsilon_{qq}$ on the flavour of the Yukawa couplings' scalings.}
   \label{eps_vark}
\end{table}

In fig.~\ref{eff_ratio} we show for the SM and for our benchmark point $g_{hq\bar{q}}=g_{hb\bar{b}}^{\text{SM}}$ the $M_{hh}$ distribution. The lower panels in the plot show the efficiencies. These plots illustrate how the efficiency depends on the shape of the distribution, and hence the flavour~$f$ that is scaled by $\kappa_f$.
\hspace{2cm}
\subsection{Statistical analysis}
We have used the likelihood ratio test statistic~$q_\mu$ in order to estimate the HL-LHC sensitivity, and set projected limits on the scalings of the light Yukawa couplings. A (log)--likelihood was constructed from the signal and background events in each bin of the histogram in fig.~\ref{sm_hh_and_bkg},
\begin{equation} \label{loglik}
    - \ln \mathscr L (\mu) = \sum_{i \in \mathrm{bins}} (N_{bi} + \mu\, N_{si}) - n_i\, \ln(N_{bi} + \mu\, N_{si}),
\end{equation}
with  $N_{bi}$ and $N_{si}$ being the number of background  and signal events in the $i$th $ M_{hh}$ distribution, respectively. In order to include the theoretical uncertainties on the expected number of signal events, the above likelihood was extended by a gaussian distribution for $N_{si}$ in which the mean equals to the central value of the bin values and standard deviation $\sigma$ equals to its theoretical uncertainty.
The signal strength $ \mu$ was then estimated by minimising $- \ln \mathscr L (\mu)$ to obtain the estimator for $\hat \mu$ by injecting SM signal + background events $n_i$. The test statistic is then given by
\begin{equation}
    q_\mu = 2 (\ln \mathscr L (\mu)- \ln \mathscr L ( \hat \mu) ),
\end{equation}
following the procedure described in~\cite{Cowan:2010js}.
\par
In order to set bounds on the scalings, we have fitted the signal strength inclusively by a function depending on the scaling of the Yukawa couplings
\begin{equation}
    \mu(\kappa_1,\kappa_2) = \Bigg\{ \frac{1}{Z}\,\left[ A_0\,
    \left(
    \kappa_1^2 \frac{m_{q1}^2}{M_{hh}^2} \, \ln^2\left(\frac{M_{hh}}{m_{q1}} \right) \right)+ A_1\,\left(\kappa_2^2 \frac{m_{q2}^2}{M_{hh}^2} \, \ln^2 \left(\frac{M_{hh}}{m_{q2}} \right)
    \right)
    \right] + B_2 \Bigg\}\, \epsilon_f,
    \label{modelmu}
\end{equation}
with
\begin{equation}
    Z = \frac{\kappa_1^2 \, m_{q1}^2 +\kappa_2^2 \, m_{q2}^2  + B_0}{m_{q1}^2+m_{q2}^2+ B_1}
\end{equation}
and $m_{q1}$ and $m_{q2}$ denoting the $\bar{\text{MS}}$ masses of the quarks.

Taking $M_{hh} \approx 300$ GeV, we could perform a fit for the signal strength for each of the quark generations scalings separately. Note that one could of course also extend the model to include the dependence of the signal strength on four Yukawa coupling modifications, taking into account the correlation between them when fitting the likelihood in eq.~\eqref{loglik}.

The expected HL-LHC sensitivity for the signal strength  at 95\% (68 \%) CL is found to be $\mu  = 2.1 (1.6)$.
\par
\subsection{Results for the $b\bar{b}\gamma\gamma$ final state}
\begin{figure}[!b]
\centering
  \includegraphics[width = 0.75\textwidth]{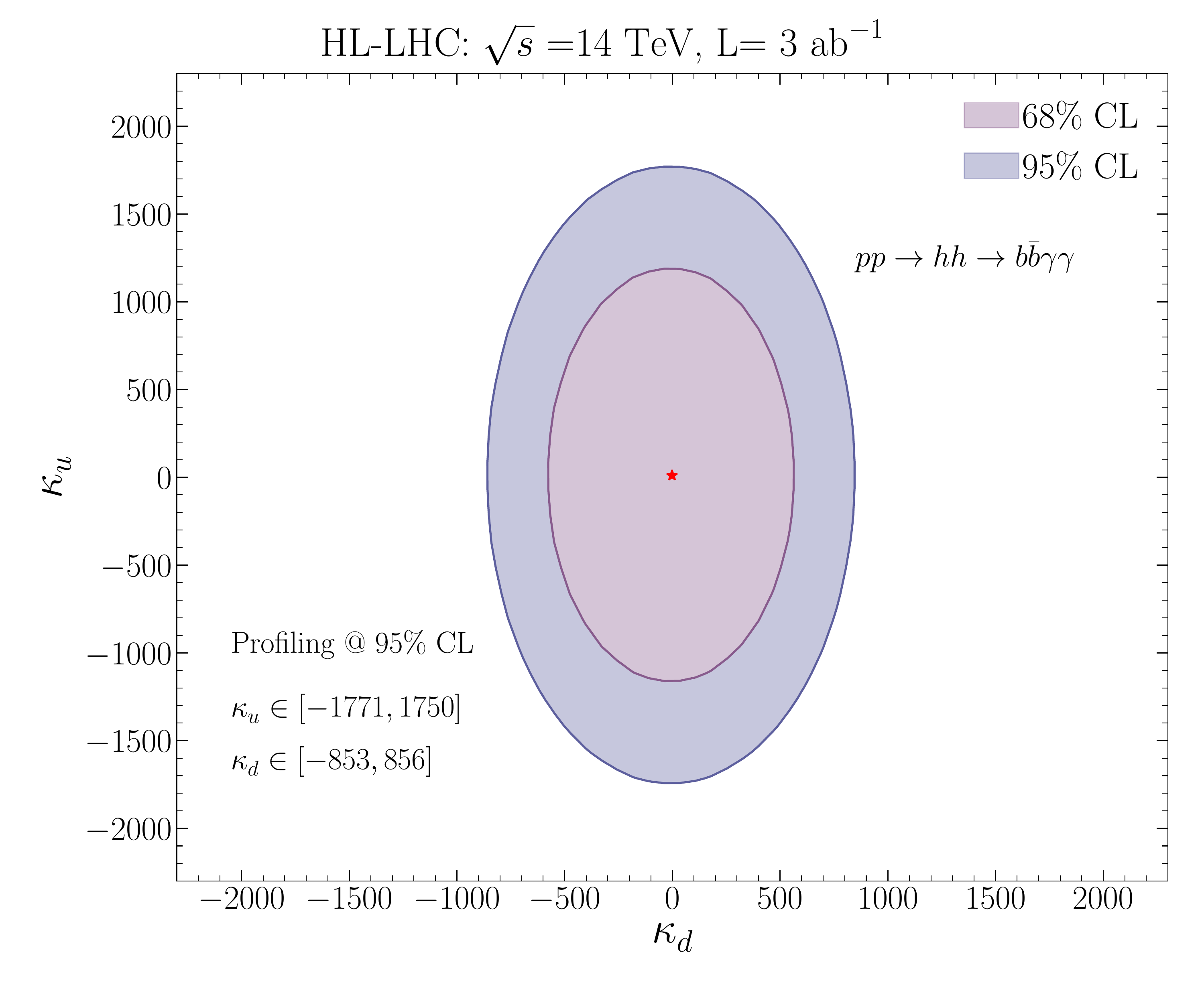}
  \caption{The expected sensitivity likelihood contours  at  68\% and 95\% CL of the HL-LHC for the first generation Yukawa coupling scalings. }
  \label{bounds_1stgen}
\end{figure}

We have performed a scan on the first generation Yukawa coupling scalings $ \kappa_u$ and $ \kappa_d$ in order to obtain exclusion limits, derived from the likelihood contours shown in fig.~\ref{bounds_1stgen}. The individual $\kappa_q$ expected upper bounds at 68\% and 95\% CL are obtained by profiling the likelihood over the other first generation $\kappa_q$.  Doing so, we obtain the following upper bounds for HL-LHC
\begin{equation}
   -571  < \kappa_d <  575, \;(\text{68\% CL}), \, \;\;\;\; \,
     -853 < \kappa_d <  856,\;(\text{95\% CL}),
\end{equation}
 and %
\begin{equation}
   -1192  < \kappa_u < 1170, \;(\text{68\% CL}), \, \;\;\;\; \,
    -1771 < \kappa_u <  1750,\;(\text{95\% CL}).
\end{equation}

\par
Note that these bounds are not directly comparable to the standard $\kappa$ formalism bounds since we relate with $\kappa$ the Yukawa couplings $g_{hq\bar{q}}$ and the new coupling  $g_{hhq\bar{q}}$.
For the second generation quarks we were not able to obtain similar bounds due to the reduction of ~$\mu /\mu_{\SM}$ with increasing $ \kappa_s$ and $\kappa_c$ away from the SM, which stems from the decrease of the branching ratio $\mathcal B (hh \to b \bar b \gamma \gamma)$ as new decay channels open, while the cross section is not as much enhanced as for up and down quarks due to the charm and strange quark being less abundant in the proton. This leads to signal strength modifiers $\mu/\mu_{\SM}< 1$ (\textit{cf}.~fig.~\ref{kcks_mu_contors}).  We will analyse the second generation Yukawa couplings instead for the final state $hh \to c \bar c \gamma \gamma$, in which we observe significant enhancement of the relative signal strength modifier $\mu/\mu_{\SM}$ (\textit{cf}.~fig.~\ref{kcks_mu_contors}). Before turning to a different final state though, we will reanalyse the $b\bar{b}\gamma\gamma$ final state under the point of view of a non-linear effective field theory, hence leaving the couplings $g_{hq\bar{q}}$ and $g_{hhq\bar{q}}$ independent.
%%%%%%%
\subsubsection{Results for non-linear EFT}
We will consider in this part a non-linear EFT as introduced in eq.~\eqref{chirallag}. By expanding in the chiral modes, taking the 0th mode and the flavour diagonal terms, we get
\begin{equation}
    -\mathcal L = \bar q_{L} \frac{m_q}{v} \left( v + c_{q}h + \frac{c_{qq}}{v} h^2 + \dots \right) q_R + h.c,
\end{equation}
where we rescaled the coefficients $k_{q}$ and $k_{2q}$ of eq.~\eqref{chirallag} as  $ k_{q,ii} =\sqrt{2}  c_{q} m_q/v $ and $ k_{2q,ii} =\sqrt{2}  c_{qq}m_q/v^2 $.  Unlike the linear EFT, the Wilson coefficients $c_{q}$ and $ c_{qq}$ are independent of each other leading to the coupling constants
	\begin{equation}
	g_{h\bar{q}_i q_i} =c_q g_{h\bar{q}_i q_i}^{\text{SM}} \,, \quad \quad \quad \quad \quad g_{h h\bar{q}_i q_i}= \frac{c_{qq}g_{h\bar{q}_i q_i}^{\text{SM}}}{v} \,.
	\label{eq:def_nolinear}
	\end{equation}
We can observe that compared to SMEFT (see eq.~\eqref{eq:couplingsEFT}) the interaction $hh \bar q q$ becomes independent of the Yukawa coupling $h \bar q q$, with the first contributing to the contact interaction diagram and the latter to the $\hat s$ channel Higgs exchange diagram and the $\hat t$ and $\hat u$ channel diagrams as shown in fig.~\ref{qqA_fd}. As we found already for SMEFT, the ggF process depends only very little on the modifications of the light quark coupling to the Higgs boson, hence barely changes for the considered values of the coefficients $c_q$ and $c_{qq}$.  The Higgs boson decays are only affected by a variation of $c_q$ but not $c_{qq}$, as the latter does not contribute to single Higgs interactions. We have observed that the shape of the differential $hh$ production distribution is dominated by a change of the $hh \bar q q$ coupling, hence the efficiency changed in a similar way to the linear EFT when changing $c_{qq}$ and remained almost constant when changing $c_q$ alone. 
Unlike the linear EFT case, we have two parameters to vary independently per quark flavour, making a total of eight Wilson coefficients when restricting ourselves to the first and second generation. 
\par
\begin{figure}[!ht]
\centering
 \includegraphics[width = 0.49\textwidth]{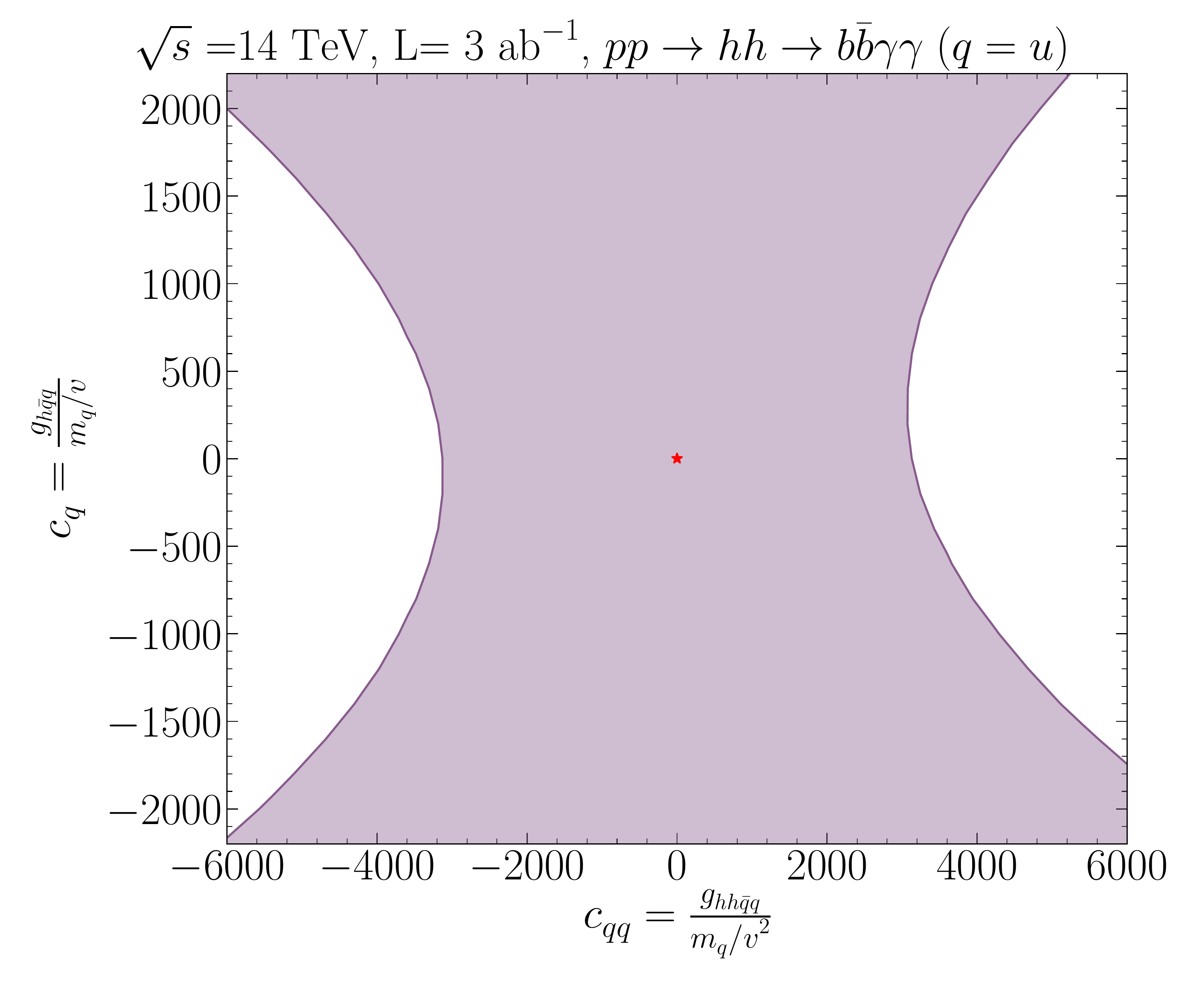}
 \includegraphics[width = 0.49\textwidth]{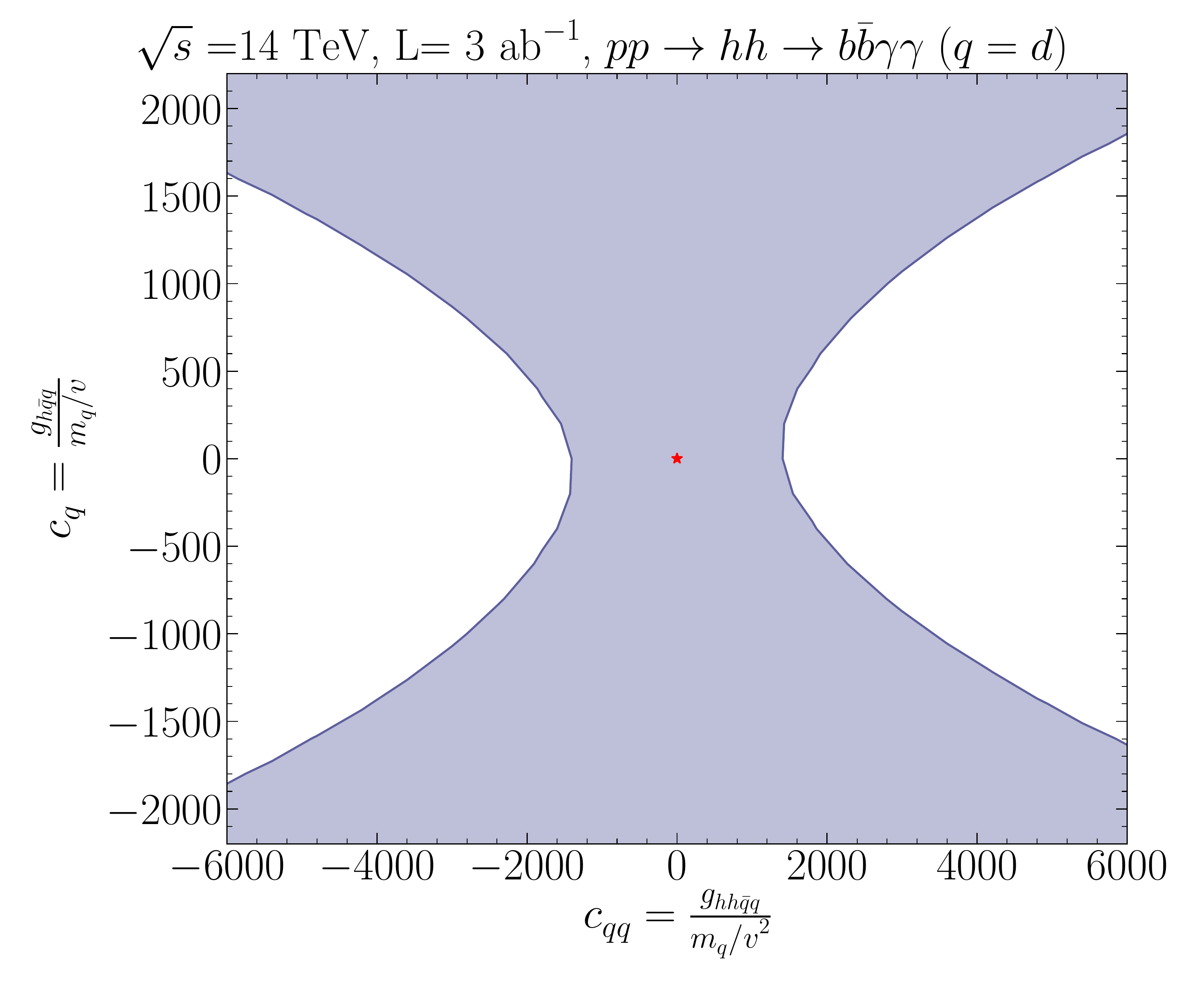}
 \\
 \includegraphics[width = 0.49\textwidth]{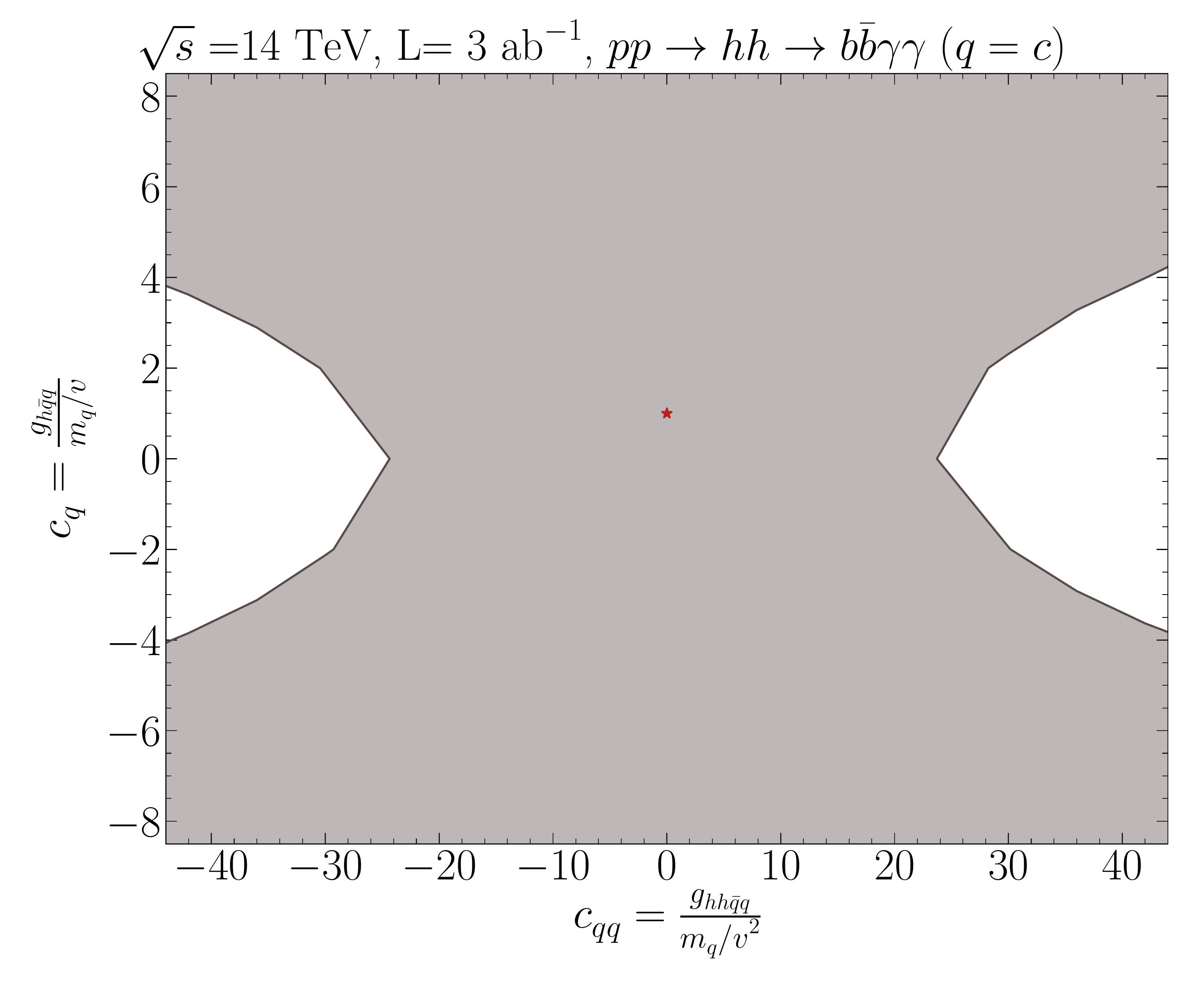}
 \includegraphics[width = 0.49\textwidth]{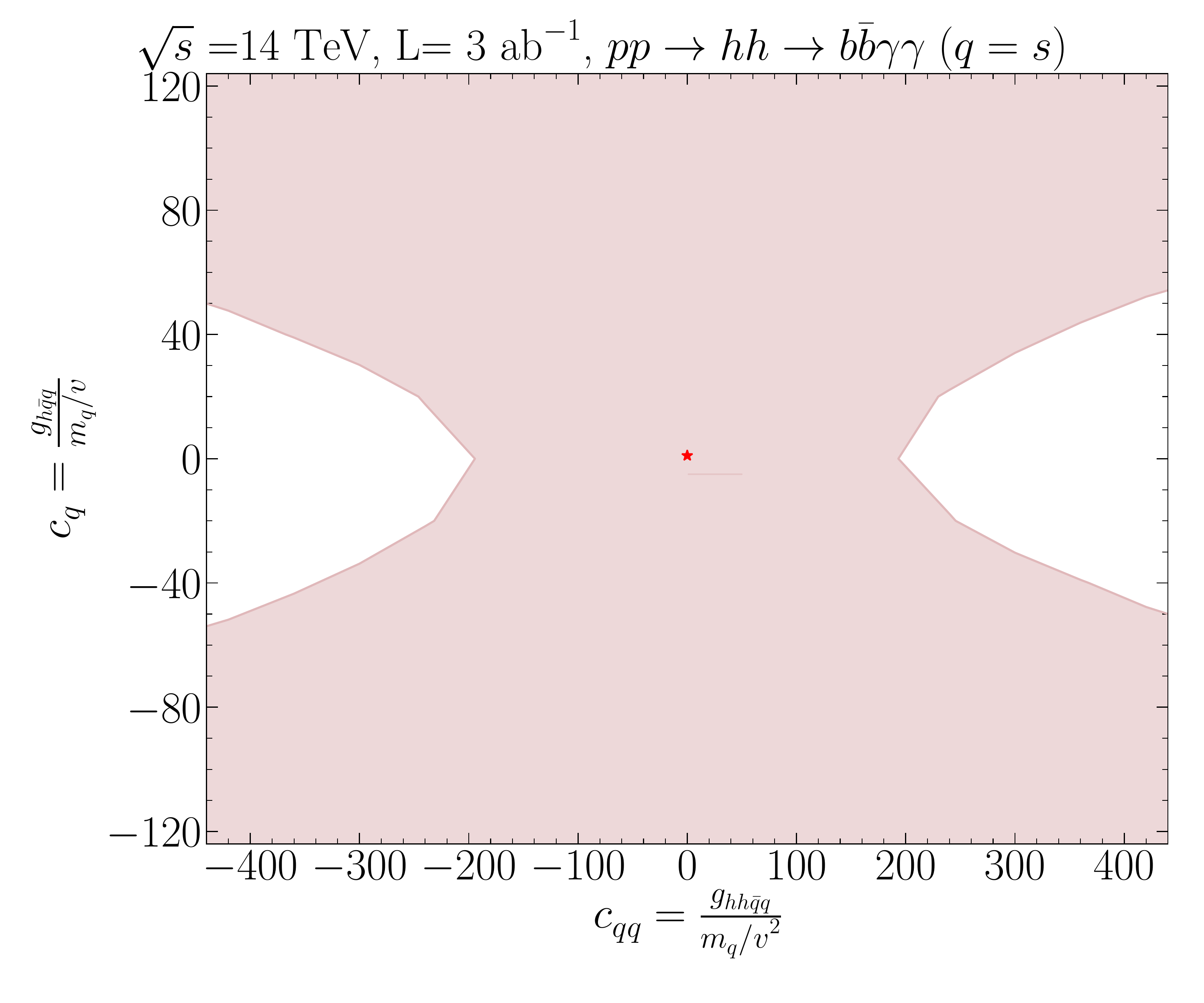}
 \caption{95\% CL likelihood contours for the non-linear EFT Wilson coefficients $c_{qq}$ and $ c_{q}$ for up  (\textit{upper left}), down  (\textit{upper right}), charm (\textit{lower left}) and strange quarks (\textit{lower right}).}
 \label{nleft}
\end{figure}

The analysis used is identical to the one of the linear EFT, with the same statistical technique, except here we have used spline functions to fit the signal strength $\mu$, as it yielded a better fit result than the simple model of eq.~\eqref{modelmu}, though the same test statistics was utilised as for the SMEFT case. The thus obtained sensitivity bounds are given in fig.~\ref{nleft}.  
We observe that without the $hh \bar q q$ interaction, one cannot set bounds on any of the light Yukawa couplings from Higgs pair production. We remark though that in case any deviation in the light Yukawa couplings is observed, the di-Higgs channel can distinguish whether electroweak symmetry breaking is realised linearly or non-linearly.
%%%%%%%%%%%%%%%%%%%%%%%%%%%%%%%%%%%%%%%%%%%%%%%%%%%%
\subsection{Charm-tagging and second generation bounds}
In order to set bounds on the second generation Yukawa couplings, we use the method developed in~\cite{Perez:2015aoa,Kim:2015oua} that re-analyses final states with $b$-quarks based on the mistagging of $c$-jets as $b$-jets in associated $VH$ production. The analysis relies on the current CMS~\cite{Chatrchyan:2013zna} and ATLAS~\cite{Aad:2014xzb} working points for $b$-tagging, as illustrated in the table~\ref{btag_wp}.
\begin{table}
    \centering
    \begin{tabular}{ccc}
        \toprule
        Detector	& Cuts (1st, 2nd) $b$-jets	& $\epsilon_{c/b}^{\rm{b-tag}\,2}$  \\
        \midrule
        CMS    & Med1-Med1   & $0.18$               \\
        CMS    & Med1-Loose  & $0.23$                \\
        \hline
        ATLAS  & Med-Med     & $8.2 \cdot 10^{-2}$ \\
        ATLAS  & Tight-Tight & $5.9 \cdot 10^{-3}$ \\
        \bottomrule
    \end{tabular}
    \caption{The $b$-tagging working points used in the analysis, for CMS~\cite{Chatrchyan:2013zna} and ATLAS~\cite{Aad:2014xzb}. }
    \label{btag_wp}
\end{table}
The signal strength estimator when considering the mistagging probability  of $b$-jets to $c$-jets~(i.e. $c$-jet contamination of b-tagged jets)~$\epsilon_{b\to c}$ is
\begin{equation}
    \hat \mu = \frac{\sigma_{hh}\, \mathcal B_b\ \,\epsilon_{b1}\,\epsilon_{b2} \,\epsilon_f+\sigma_{hh}\, \mathcal B_c \,\epsilon_{b\to c,1} \,\epsilon_{b\to c,2}\,\epsilon_f }{\sigma_{hh}^{\SM}\, \mathcal B_b^{\SM}\,  \epsilon_{b1}\,\epsilon_{b2}},
\end{equation}
 with~$\epsilon_f$ being the efficiency ratio in eq.~\eqref{effrat}. The above expression simplifies to
\begin{equation}
    \hat \mu = \mu_b\,\epsilon_f +0.05\cdot \left(\epsilon_{c/b}^{\text{$b$-tag}}\right)^2 \, \epsilon_f \cdot  \mu_c\,,
\end{equation}
for~$ \mathcal B_c^{\SM} / \mathcal B_b^{\SM} \approx 0.05$. The signal strength modifier of the $b\bar{b}\gamma\gamma$ final state is denoted by~$\mu_b$ and the one of the  $c\bar{c}\gamma\gamma$ final state by~$\mu_c$. 
The ratio of tagging efficiencies is defined as
\begin{equation}
   \left( \epsilon_{c/b}^{\text{$b$-tag}}\right)^2 = \frac{\epsilon_{b\to c,1}  \epsilon_{b\to c,2} }{\epsilon_{b1} \epsilon_{b2}}\,.
\end{equation}
\par
\begin{figure}[!ht]
\centering
  \includegraphics[width = 0.75\textwidth]{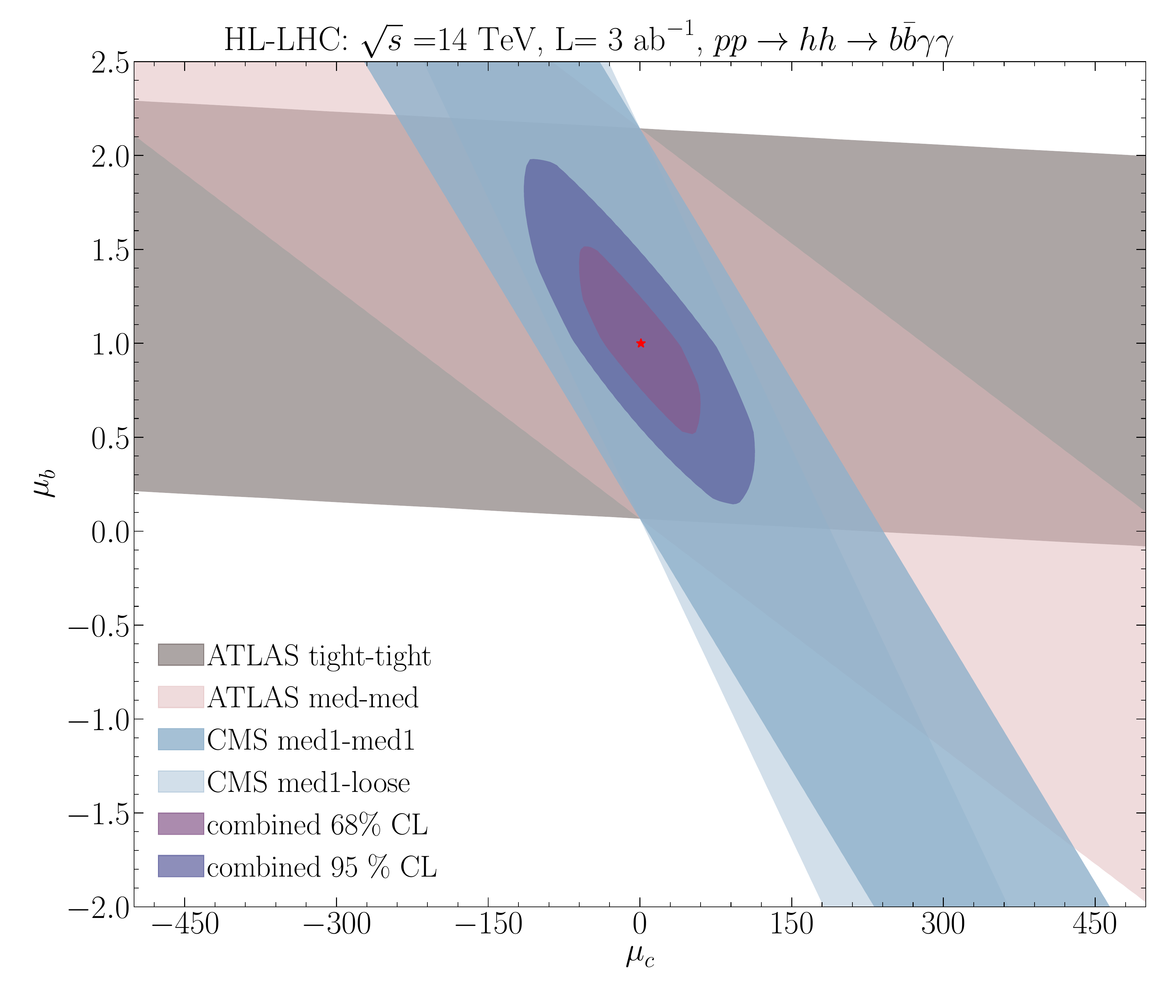}
  \caption{The  95 \% CL contours of $\mu_b$ vs $\mu_c$, obtained from fitting of the signal strength for several CMS and ATLAS $b$-tagging working points. Their combination with the 68\% and 95\% CL upper limits on $\mu_b$ and $\mu_c$ are shown.}
  \label{bworkingpoints}
\end{figure}
\begin{figure}[!b]
\centering
  \includegraphics[width = 0.49\textwidth]{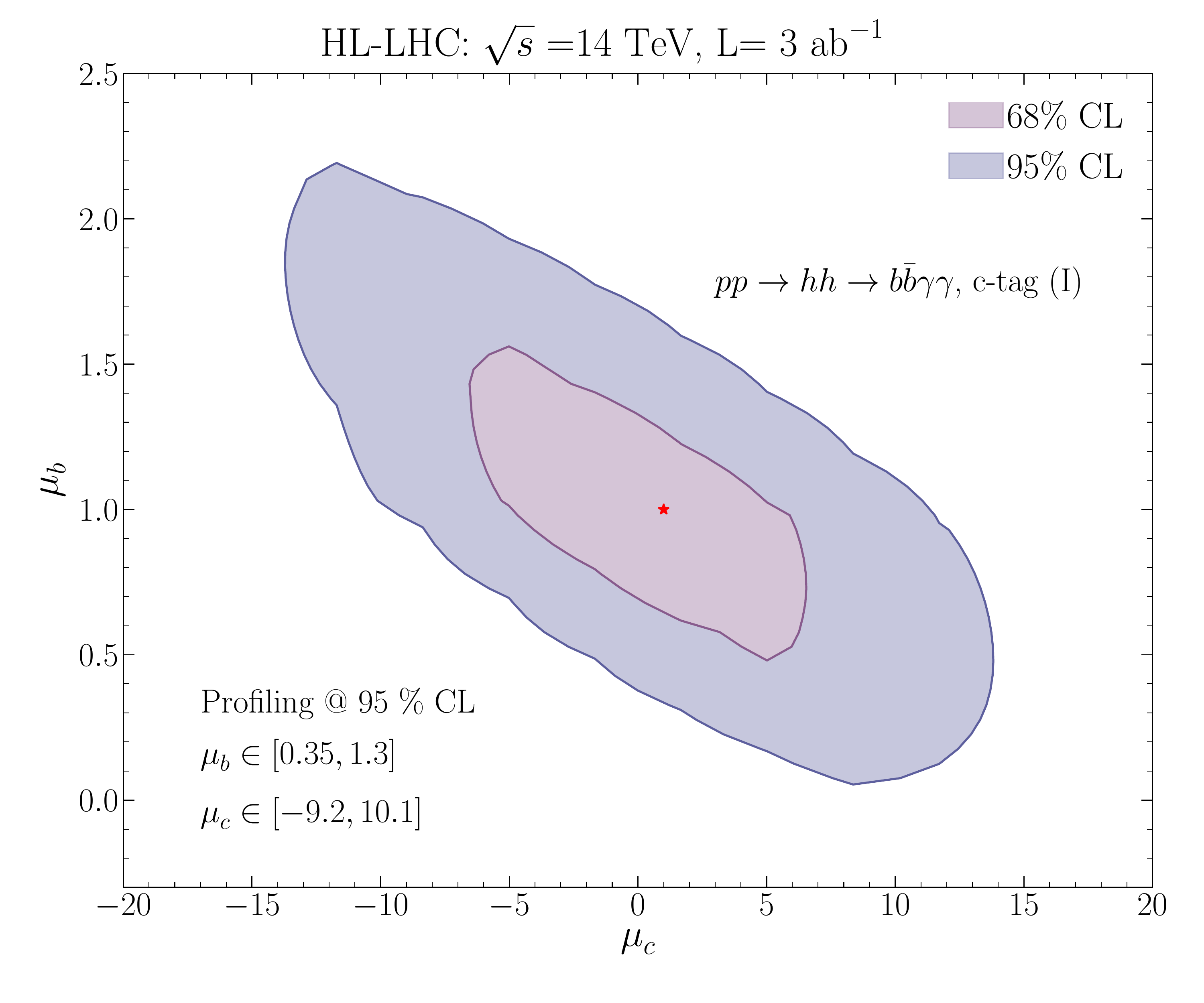}
   \includegraphics[width = 0.49\textwidth]{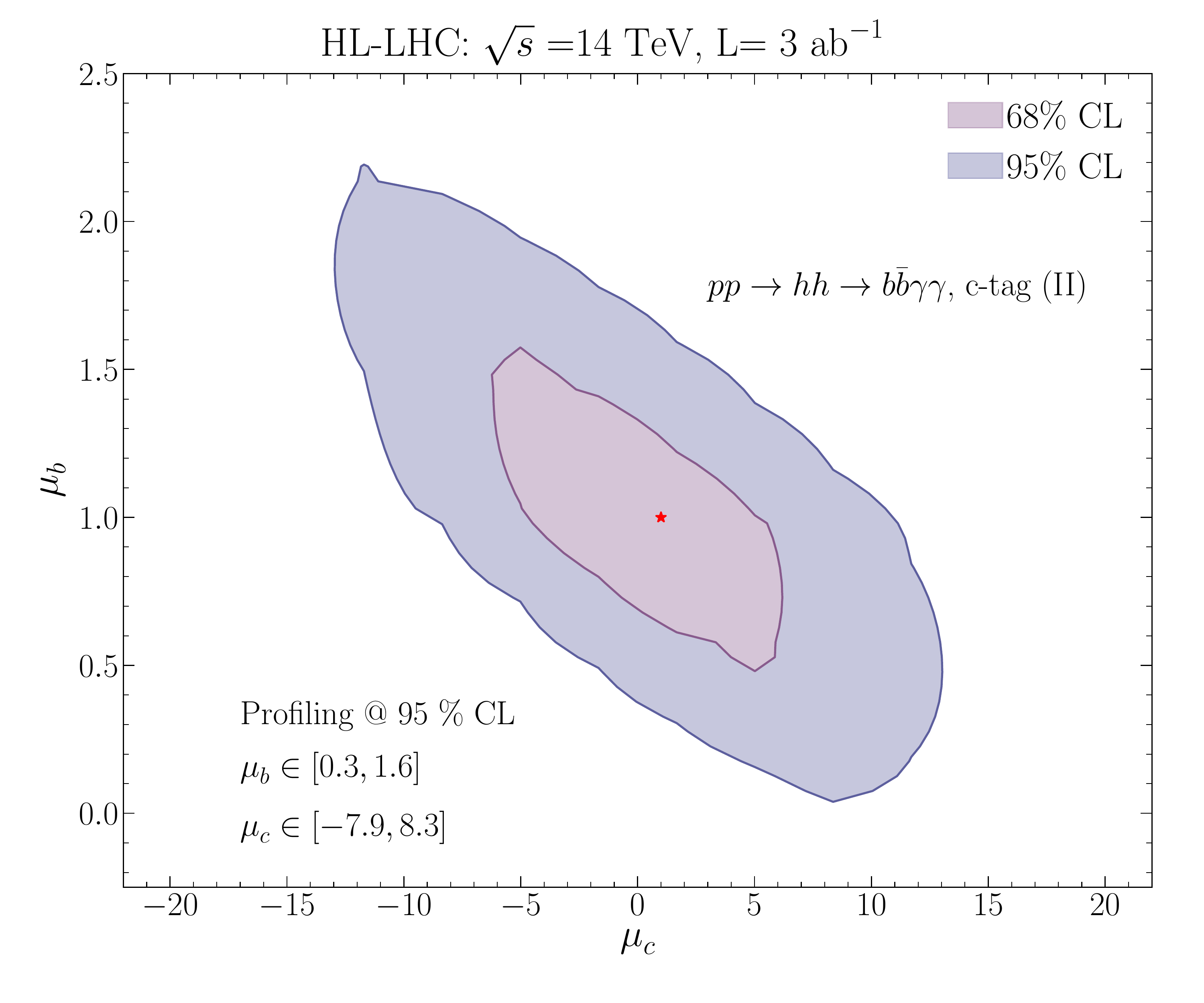}
   \centering
   \includegraphics[width = 0.49\textwidth]{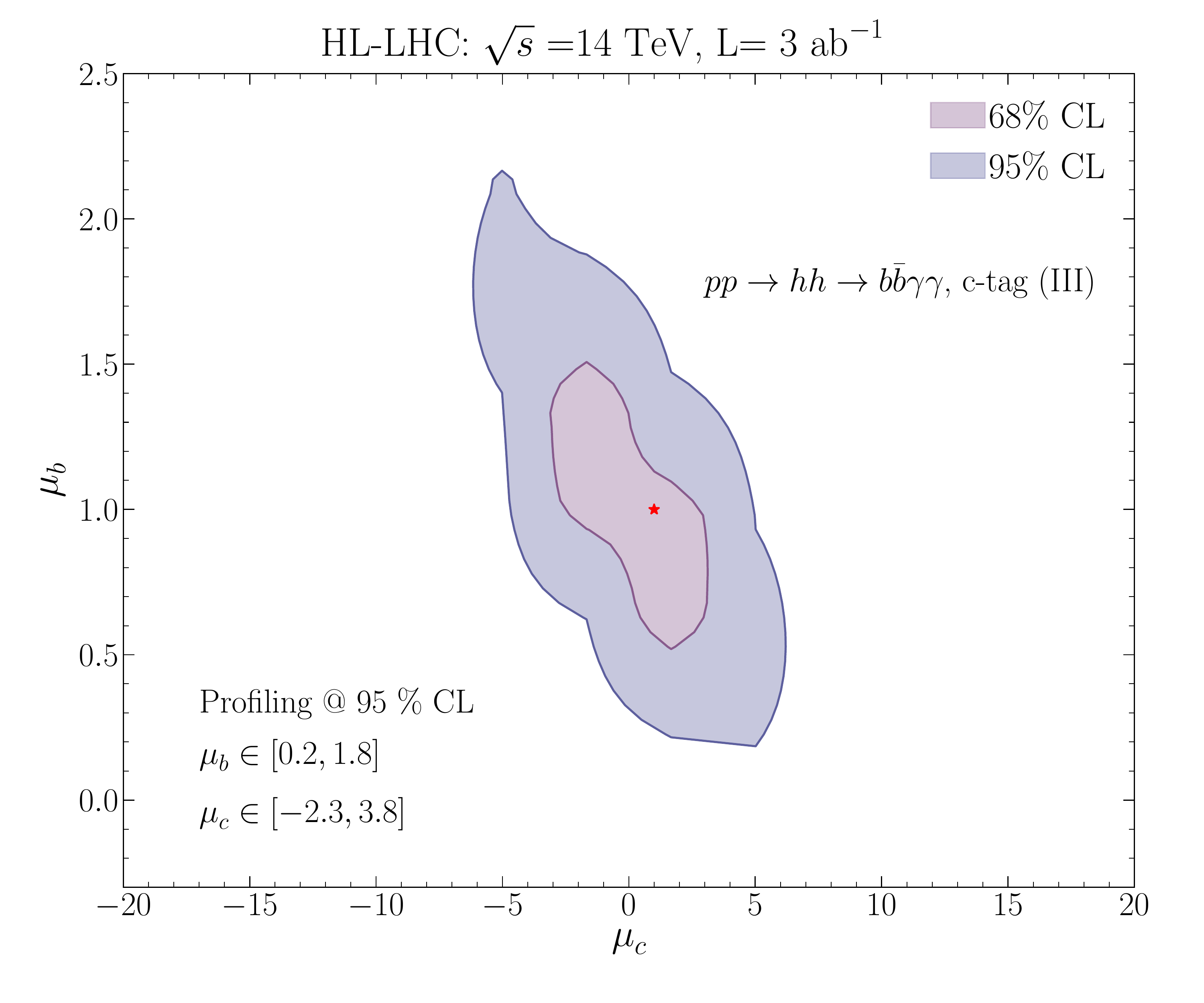}
  \caption{The expected sensitivity likelihood contours at 68\% CL and 95\% CL for an integrated luminosity $L=3000\text{ fb}^{-1}$ for the signal strengths $\mu_c$ and $\mu_b$, using the c-tagging I (\textit{upper pannel, left}), II (\textit{upper pannel, right}) and III (\textit{lower pannel}) working points combined with the ATLAS med $b$-tagging working point. }
  \label{bounds_2ndgen1_btag}
\end{figure}

One~$b$-tagging working point could only constrain either~$ \mu_b$ or~$\mu_c$. In order to resolve the flat direction several $b$-tagging working points~$\left(\epsilon_{c/b}^{\text{$b$-tag}} \right)^2$ are needed. This is illustrated in fig.~\ref{bworkingpoints}, where the working points fitting contours are combined  using Fisher's method~\cite{heard2018choosing}. We thus obtain an upper projected limit on the charm final state signal strength after profiling over~$\mu_b$,
\begin{equation}
\mu_c(\mathrm{up}) = 36.6 \;(\text{68\% CL})\,, \, \;\;\;\; \,  \mu_c(\mathrm{up}) = 74.8 \;(\text{95\% CL})\,.
\end{equation}
However, the obtained sensitivity is not sufficient to set any better limits at 95\% CL than the existing ones (or projected ones in other  channels) for the Yukawa coupling modifiers~$\kappa_c$, and~$\kappa_s$. Instead, we can improve on them by introducing $c$-tagging working points~$(\epsilon_{c/b}^{\text{$c$-tag}})^2$
\begin{equation}
   \left( \epsilon_{c/b}^{\text{$c$-tag}}\right)^2 = \frac{\epsilon_{c1}  \epsilon_{c2} }{\epsilon_{c \to b,1} \epsilon_{c \to b,2}}\,,
\end{equation}
mixed with the  $b$-tagging ones. We denoted  the contamination of $c$-jets with $b$-jets by $\epsilon_{c \to b}$.  For mixed tagging,  the signal strength estimator becomes
\begin{equation}
    \hat \mu = \frac{\sigma_{hh}\, \mathcal B_b \,\epsilon_{b1}\,\epsilon_{b2} \, \epsilon_f+\sigma_{hh}\, \mathcal B_c  \,\epsilon_{c1}\,\epsilon_{c2} \,\epsilon_f }{\sigma_{hh}^{\SM}\, \mathcal B_b^{\SM}\, \epsilon_{b1}\,\epsilon_{b2}
    +\sigma_{hh}^{\SM}\, \mathcal B_c^{\SM}\, \epsilon_{c1}\,\epsilon_{c2}
    } \,,
\end{equation}
where now $\epsilon_{b}$ is either $\epsilon_{b}$ or $\epsilon_{c\to b}$  and $\epsilon_{c}$ either $\epsilon_{c}$ or $\epsilon_{b\to c}$.
This simplifies to
\begin{equation}
\hat \mu  = \frac{\mu_b+0.05\,\epsilon_{c/b}^2\,\mu_c}{1+0.05\,\epsilon_{c/b}^2}\, \epsilon_f\,.
\end{equation}
 The working point~$\epsilon_{c/b}^2$ could be the $b$-tagging or $c$-tagging working point.
Assuming that $c$-tagging and $b$-tagging are uncorrelated, and working with the methods discussed in~\cite{Perez:2015aoa,Perez:2015lra}, i.e.~combining the ATLAS medium cuts (med.) for $b$-tagging with the $c$-tagging working points in order to break the degeneracy, we could improve the 95\% CL sensitivity on $\mu_c$.
We start by the $c$-tagging working point used by the ATLAS collaboration in Run I searches for top squarks decays to charm and neutralino~\cite{Aad:2015gna,ATLAS-CONF-2013-063}, which we refer to as $c$-tagging I.
Further $c$-tagging working points from the HL-LHC upgrade are used: with the expected insertable B-layer (IBL) sub-detector that is to be installed during the ATLAS HL-LHC upgrade~\cite{Capeans:1291633,ATL-PHYS-PUB-2015-018}, the new $c$-tagging II and III points, as illustrated in table~\ref{ctag_wp},
can be identified. In fig.~\ref{bounds_2ndgen1_btag}  we used them to obtain in combination with the ATLAS med $b$-tagging expected 95\% CL upper limits on~$\mu_c$ for the HL-LHC from an analysis of the final state~$b\bar{b}\gamma\gamma$.
 \begin{table}
     \centering
     \begin{tabular}{cccc}
         \toprule
         $c$-tagging working point	&$\epsilon_{c}$	& $\epsilon_{c \to b}$ & $\mu_c(\mathrm{up})$ 95\% CL  \\
         \midrule
            $c$-tag I ~\cite{Aad:2015gna,ATLAS-CONF-2013-063}
            & $19\%$ & $13 \%$ & $ 10.1$\\
            $c$-tag II ~\cite{Capeans:1291633,ATL-PHYS-PUB-2015-018}
            & $30\%$ & $20 \%$ & $ 8.2$\\
            $c$-tag III ~\cite{Capeans:1291633,ATL-PHYS-PUB-2015-018}
            & $50\%$ & $20 \%$ & $ 3.8$\\
         \bottomrule
     \end{tabular}
     \caption{The $c$-tagging working points with the expected 95\% CL upper limit~(sensitivity) of $\mu_c$ obtained after profiling over $\mu_b$. }
     \label{ctag_wp}
 \end{table}

\begin{figure}[!t]
\centering
  \includegraphics[width = 0.49\textwidth]{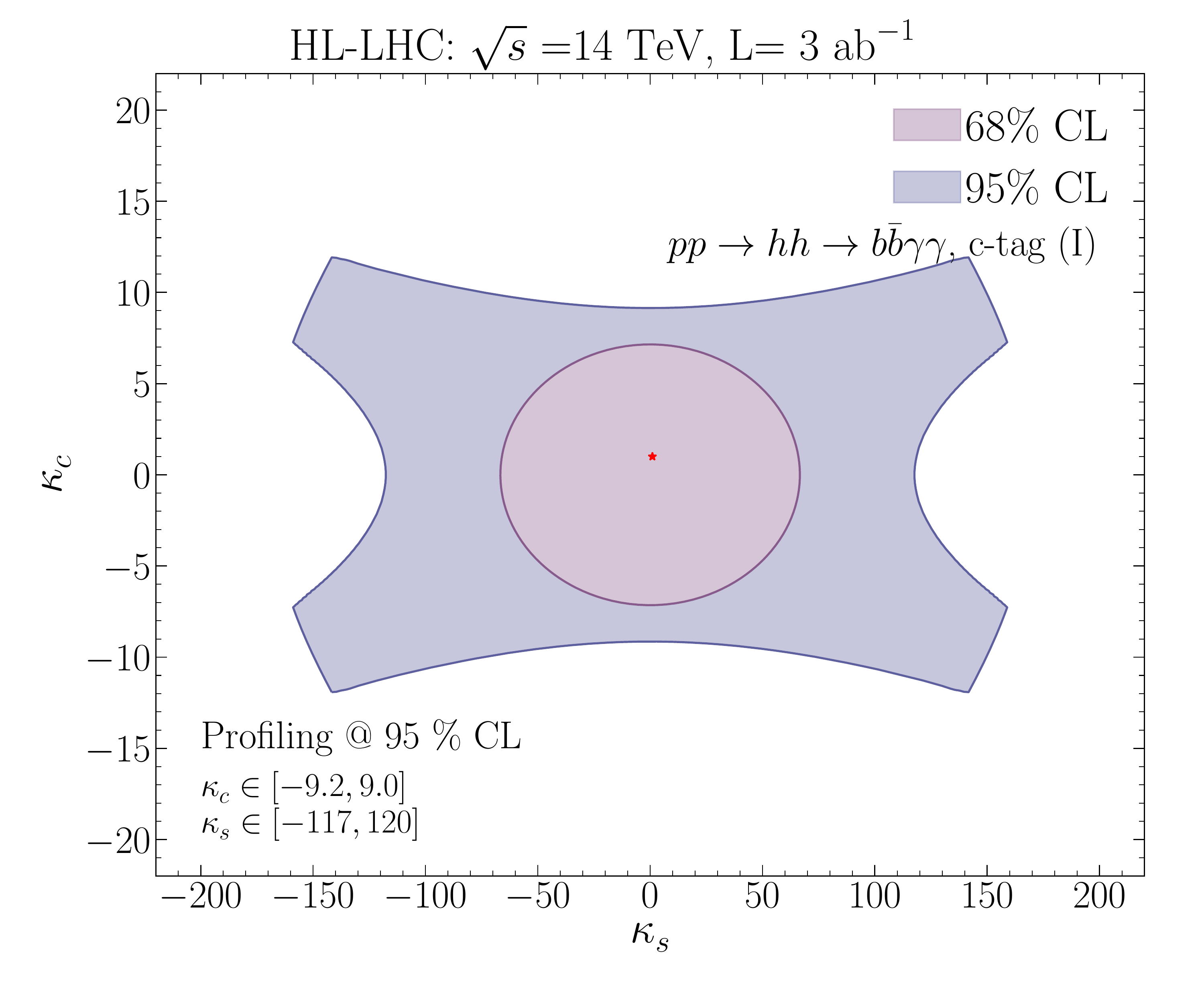}
   \includegraphics[width = 0.49\textwidth]{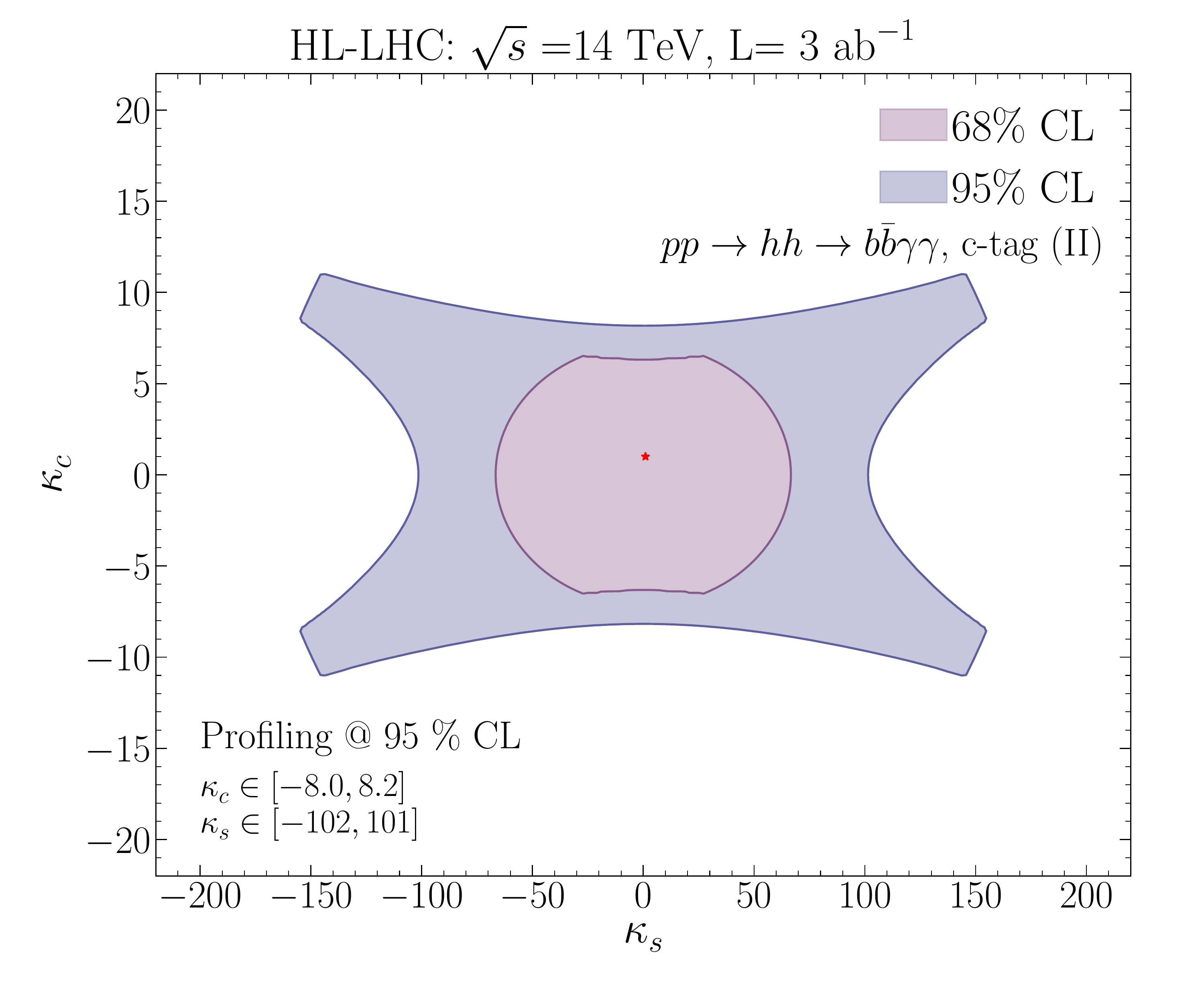}
   \centering
   \includegraphics[width = 0.49\textwidth]{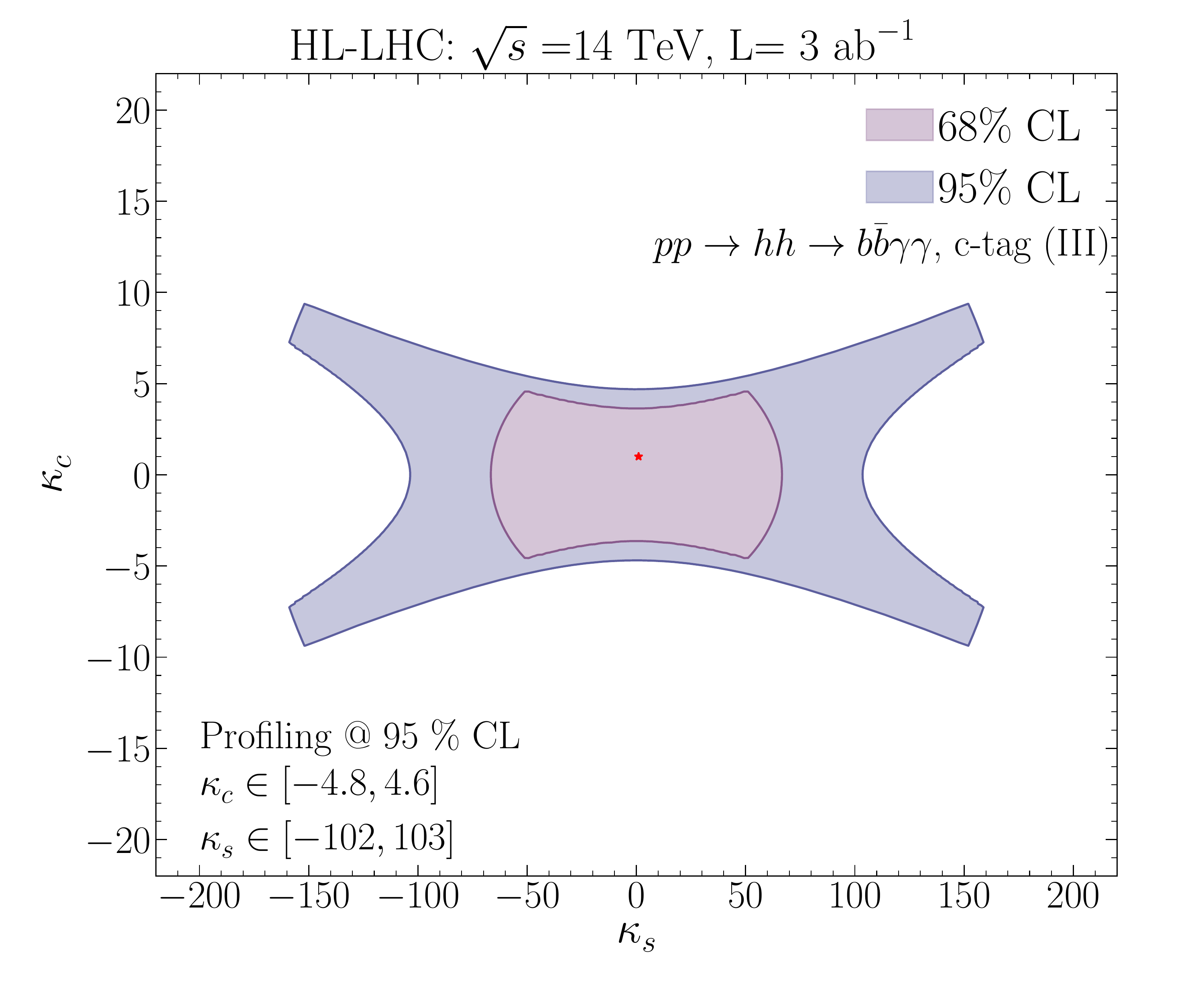}
  \caption{The expected sensitivity likelihood contours at 68\% CL and 95\% CL for an integrated luminosity $L=3000\text{ fb}^{-1}$ for modified second generation quark Yukawa couplings, using the c-tagging I (\textit{upper pannel, left}), II (\textit{upper pannel, right}) and III (\textit{lower pannel}) working points. }
  \label{bounds_2ndgen1}
\end{figure}

 Fitting  signal strengths with varying $\kappa_c$,  $\kappa_s$ for charm and bottom final states (\textit{cf.} eq.~\eqref{modelmu}) for constructing the likelihood $\mathscr{L}(\kappa_c,\kappa_s)$, we can set limits from the anticipated charm tagging working points as shown in fig.~\ref{bounds_2ndgen1}.
These projected limits are an improvement compared to the current direct bound and prospects for HL-LHC, particularly for charm quark Yukawa modifications~\cite{Perez:2015aoa,Perez:2015lra}.
 Again, it should be kept in mind that the bounds on $\kappa_q$ do not just correspond to the scaling of the Yukawa coupling, but also to the new coupling $g_{hhq\bar{q}}$ arising in SMEFT.
%%%%%%%%%%%%%%%%%%%%%%%%%%%
\subsection{Bounds with trilinear coupling scaling}
Since we expect that most of the UV-complete models will modify the trilinear Higgs coupling by a scaling $ \kappa_\lambda=\lambda_{hhh}/\lambda_{hhh}^{SM}$, we have investigated the light quark Yukawa bounds along with a modified trilinear Higgs self-coupling.

\begin{figure}[!t]
\centering
  \includegraphics[width = 0.75\textwidth]{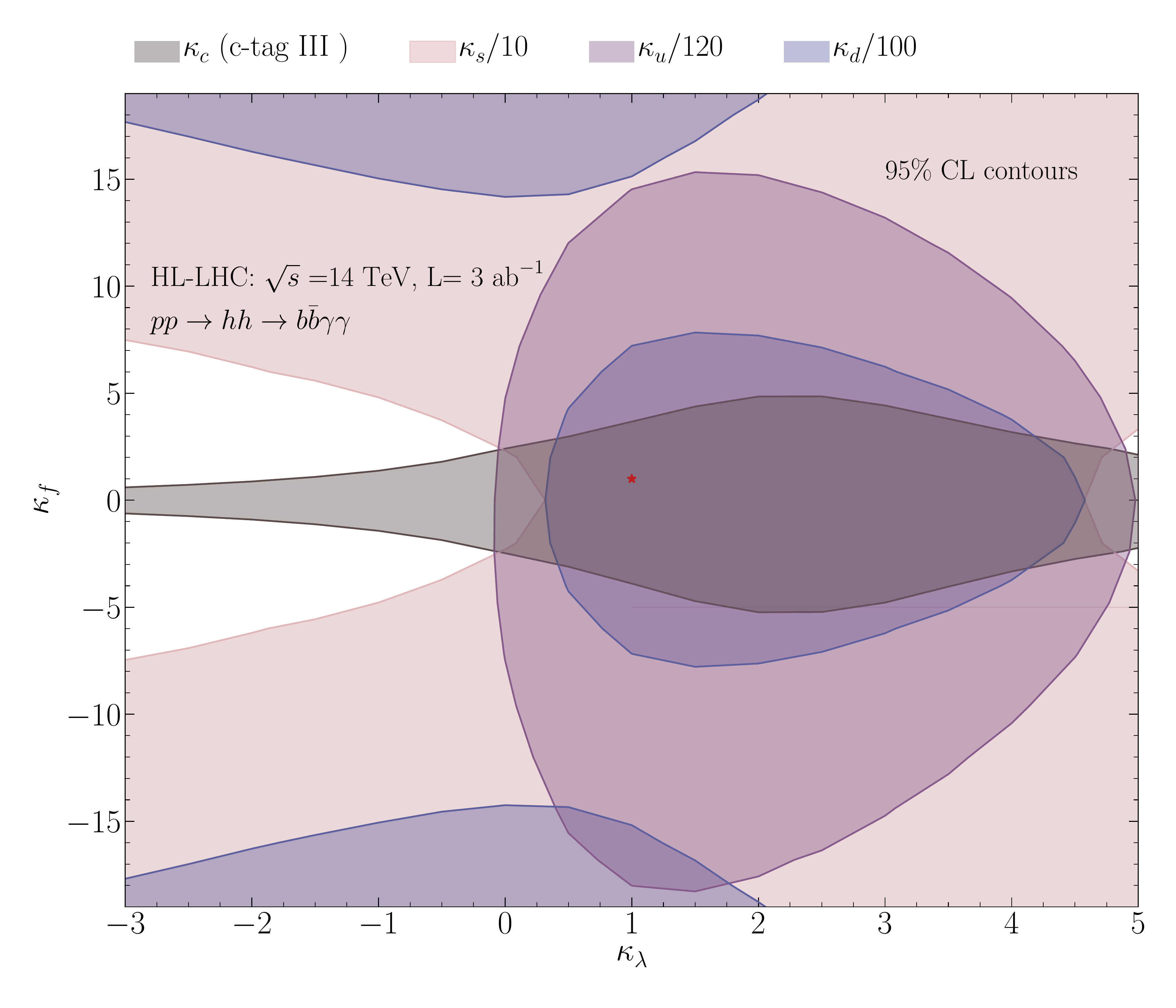}
  \caption{The expected sensitivity likelihood contours at 95\% CL for an integrated luminosity $L=3000\text{ fb}^{-1}$ for modified Higgs trilinear coupling $\kappa_\lambda$ vs the light quark Yukawa couplings scalings $\kappa_q$.}
  \label{klkf}
\end{figure}

The likelihood contours obtained in fig.~\ref{klkf} assume that a single flavour coupling modifier $ \kappa_f$ is not correlated to the others, nor with the trilinear coupling scaling. The correlated case in terms of a two Higgs doublet model has been discussed in ref.~\cite{Bauer:2017cov}.
The modification of  $\kappa_\lambda $ enhances the contributions to the $s$-channel qqA and triangle ggF diagrams, where the first interferes destructively with the $hh \bar q q$ diagram, and the latter with the spin-0 box form factor for $\kappa_{\lambda}>0$. Therefore, we observe that if we let  $\kappa_\lambda $  be in the range of $\kappa_{\lambda} \in \, [1,4]$ the sensitivity for $\kappa_q$ becomes worse.

%%%%%%%%%%%%%%%%%%%%%%%%%%%%%%%%%%%%%%%%%%%%%%%%
\section{Conclusion \label{sec:conclusion}}
The couplings to the first and second generation fermions remain among the less well measured couplings of the Higgs boson.
In this paper we investigated the possibility of measuring light quark Yukawa couplings in Higgs pair production.
For enhanced Yukawa couplings of the first generation quarks, we found that limits can be set when considering
quark annihilation with subsequent decay of the Higgs boson pair to $b\bar{b}\gamma\gamma$. In an effective theory description with dimension 6 operators that modify the quark Yukawa couplings,
there exists also a coupling of two Higgs bosons to two fermions. This coupling increases the Higgs pair production cross section and hence allows to set bounds on the light quark Yukawa coupling modifications.
For the HL-LHC we found a sensitivity of $|\kappa_u| \lesssim 1170 $ and $|\kappa_d| \lesssim 850 $, \textit{cf}.~fig.~\ref{bounds_1stgen}, which is comparable to the sensitivity of other channels that can directly probe the light quark Yukawa couplings though being weaker than the results from a global fit. Further improvements could be possible with a more dedicated analysis.
We note though that the bounds we find stem mostly
due to the diagram involving the coupling of two Higgs bosons to two quarks, as we showed explicitly also by considering a non-linear effective
theory in which the coupling of one and two Higgs boson to fermions are uncorrelated.
This channel can hence also be used to distinguish between a linear vs non-linear Higgs EFT hypothesis in the light quark sector.
The LHC experiments should hence consider the Higgs pair production process in addition to other channels for probing the light quark Yukawa couplings. 
\par
For the second generation quarks we found that at the HL-LHC in the di-Higgs channel we will be able to set competitive bounds on the
charm Yukawa coupling if final states with tagged charm quarks are considered.
We were in particularly considering the final state $c\bar{c}\gamma\gamma$, in which we found a sensitivity of $|\kappa_c| \lesssim 5$ and $|\kappa_s | \lesssim 100 $, \textit{cf.}~fig.~\ref{bounds_2ndgen1}, where the first prospective limit is comparable to the prospects from charm tagging in the $Vh$ channel \cite{Perez:2015lra}.

%%%%%%%%%%%%%%%%%%%%%%%%%%%%%%%%%%%%%%%%%%%%%%%%
\section*{Acknowledgments}
We thank S.~Bar Shalom and A.~Soni for clarifications regarding \cite{Bar-Shalom:2018rjs}.
We thank L.~Di Luzio for useful comments on the manuscript.
LA thanks M.~Schmelling and H.~Dembinski for their help improving the statistical methods used in this work.
RG is supported by the ``Berliner Chancengleichheitsprogramm''. RCL acknowledges support by an IPPP summer studentship.

%%%%%%%%%%%%%%%%%%%%%%%%%%%%%%%%%%%%%%%%%%%%%%%%
\appendix
%%%%%%%%%%%%%%%%%%%%%%%%%%%%%%%%%%%%%%%%%%%%%%%%
\section{Parameter values as used in the analysis \label{app:pythia}}
In this appendix, we give the input parameters for masses, widths, and couplings as used in the \textsc{Pythia} simulation, see table~\ref{pp}. The collider input is given in table~\ref{cp} and the parton shower parameters in table~\ref{pyp}.
\begin{table}[!htpb]
    \centering
    \begin{tabular}{ccc}
        \toprule
        Parameter	     & value                    & notes  \\
        \midrule
         $ m_h$          & \SI{125.25}{\GeV}      & \\
         $\Gamma_h$      &\SI{0.013}{\GeV}         &  SM value, changes with $\kappa_f$\\
         $v$             & \SI{246.2}{\GeV}      &   \\
         $m_W$           & \SI{80.397}{\GeV}       &  \\
         $m_Z$           & \SI{91.1876}{\GeV}       &   \\
         $\Gamma_W$      & \SI{2.0886}{\GeV}        & \\
         $\Gamma_Z$      & \SI{2.4958}{\GeV}        & \\
         $m_t$           & \SI{173.21}{\GeV}      & pole mass \\
         $m_b$           & \SI{4.18}{\GeV}        & \multirow{2}{*}{$\bar{\text{MS}}$ mass at $\mu = 2$ \si{\GeV} }\\
         $m_c$           & \SI{1.27}{\GeV}        &  \\
         $m_s$           & \SI{96.0}{\MeV}       & \\
         $m_u$           & \SI{2.20}{\MeV}        & \\
         $m_d$           & \SI{4.70}{\MeV}        & \\
         $\alpha_s(m_Z)$ & $0.118$                 & \\
         $(hc)^2$            &\SI{3.894e11}{\femto\barn \per \GeV\squared} & conversion from $\GeV^2$ to $\femtobarn$ \\
         \bottomrule
    \end{tabular}
    \caption{The input parameters used in this work, all taken from the PDG \cite{Tanabashi:2018oca}. }
    \label{pp}
\end{table}

\begin{table}[!htpb]
    \centering
    \begin{tabular}{ccc}
        \toprule
        Parameter	& value & description  \\
        \midrule
         PDG ID's of initial states  & (2212,2212)  &$pp$ collision\\
         $\sqrt{s}$      & \SI{14}{\TeV} &  centre of mass energy\\
         $L$             & \SI{3}{\per\atto\barn} &  integrated luminosity \\
         LHAPDF ID     & $262000$   &  NNPDF30        \\
         \bottomrule
    \end{tabular}
    \caption{The collider parameters used in this work for the HL-LHC. }
    \label{cp}
\end{table}

\begin{table}[!htpb]
    \centering
    \begin{tabular}{ccc}
        \toprule
        Parameter	& value & description  \\
        \midrule
         \texttt{  MSTU(112)} & 5   &  5-flavour scheme. \\
         \texttt{  MSTP(48) } & 1  & Top quark decay before fragmentation. \\
         \texttt{  MSTP(61)}  & 1 & Turn on initial state radiation.\\
         \texttt{  MSTP(71)}  & 1 & Turn on final state radiation.\\
          \texttt{MSTJ(41)}  & 1 & Turn off QED bremsstrahlung.\\
          \texttt{MSTP(81)}  & 1 & Multiple interaction on. \\
          \texttt{MSTP(111)}  & 2 & Allow fragmentation and decay. \\
          \texttt{MSTP(42)}   & 0 & Turn off  shell boson production. \\
         \bottomrule
    \end{tabular}
    \caption{The values of \textsc{Pythia} 6.4 parameters that are changed compared to the default values.}
    \label{pyp}
\end{table}
\bibliographystyle{utphys.bst}
\bibliography{bibliography}

\end{document}